\documentclass[12pt]{article}
\usepackage{amsthm,amsmath, amssymb, array,
                     bm,booktabs,color,float,
                     natbib,graphicx, multirow,url}
\usepackage{enumerate}
\usepackage[total={6.4in,8.6in}]{geometry}
\usepackage[colorlinks,citecolor=blue, linkcolor = blue,urlcolor=blue]{hyperref}
\usepackage{pdflscape}

\newtheorem{theorem}{Theorem}

\newcommand{\blind}{0}

\begin{document}

\def\spacingset#1{\renewcommand{\baselinestretch}%
{#1}\small\normalsize} \spacingset{1}


\if0\blind
{
  \title{
   \bf On Estimating Optimal Regime for Treatment Initiation Time Based on Restricted Mean Residual Lifetime}
  \author{
   Xin Chen,  Rui Song, Jiajia Zhang, Swann Arp Adams, Liuquan Sun,\\ and 
  Wenbin Lu}  		
\date{}		
  \maketitle
} \fi

\if1\blind
{
  \bigskip
  \bigskip
  \bigskip
  \begin{center}
    {\Large{\bf  On Estimating Optimal Regime for Treatment Initiation Time Based on Restricted Mean Residual Lifetime}}
   \end{center}
  \medskip
} \fi

\bigskip
\begin{abstract}
When to initiate treatment on patients is an important problem in many medical studies such as AIDS and cancer.
In this article, we formulate the treatment initiation time problem for time-to-event data 
and propose an optimal individualized regime that determines the best treatment initiation time for individual patients based on their characteristics. 
	Different from existing optimal treatment regimes where treatments are undertaken at a pre-specified time,  
	here new challenges arise from the complicated missing mechanisms in treatment initiation time data 
	and the continuous treatment rule in terms of initiation time. 
	To tackle these challenges,
	we propose to use restricted mean residual lifetime as a value function  
	to evaluate the performance of different treatment initiation regimes,
	and develop a nonparametric estimator for the value function, which is consistent even when treatment initiation times are not completely observable and their distribution is unknown.
	We also establish the asymptotic properties of the resulting estimator in the decision rule and its associated value function estimator. In particular, the asymptotic distribution of the estimated value function is nonstandard, which follows a weighted chi-squared distribution. 
	The finite-sample performance of the proposed method is evaluated by simulation studies and is further illustrated with an application to a breast cancer data.
	\end{abstract}

\noindent%
\vfill

\newpage
\spacingset{1.5} 
\section{Introduction}
\label{sec:intro}

Finding the optimal time to initiate treatment is a critical issue in many medical studies including AIDS and cancer. 
For example,
in the treatment of patients with tuberculosis and newly identified 
infection with human immunodeficiency virus (HIV), 
antiretroviral therapy (ART) must be started during the treatment for tuberculosis \citep{abdool2010timing}, 
but the optimal timing to initiate ART on patients who are receiving tuberculosis therapy is a challenging question to address. 
Starting ART in the early stages of tuberculosis treatment would increase the pill burden, the potential drug toxicity and the risk of tuberculosis associated immune reconstitution inflammatory syndrome (IRIS), 
while delaying ART initiation may also result in a higher risk of HIV-related complications and death \citep{havlir2011timing,yang2014impact}.
Another example is the initiation of adjuvant therapy for patients diagnosed with breast cancer. 
In practice, adjuvant chemotherapy or radiotherapy 
is routinely recommended to breast cancer patients after definitive surgery and within 24 weeks from the surgery \citep{Lohrisch2006impact}.
However,
the optimal time to initiate adjuvant therapy during the 24 weeks after surgery is controversial.
Due to the heterogeneity in patients and diseases, 
several retrospective studies evaluating the role of early or delayed initiation of adjuvant chemotherapy reported conflicting results \citep{Yu2017Influence}. 
Similar treatment initiation time problems also arise in finding 
the timing of neurosurgery for medically refractory epilepsies
\citep{ Sugano2015epilepsy} 
and the timing for cardiovascular surgeries
\citep{Jung2019progression}.

The goal of this article is to develop a method to search for the optimal individualized treatment initiation regime (OTIR) that selects the best treatment initiation time based on each individual's characteristics.
{	Specifically, we focus on the case where 
	the outcome of interest is time-to-event data, 
	such as the death time of patients,
	and initiating treatment at the proper time within a pre-specified time range may decrease the hazard rate of the failure event.
	In the example of breast cancer patients, 
	the pre-specified time range could be the 24 weeks after surgery.
	In the area of precision medicine, 
	although numerous efforts have been made in finding optimal individualized treatment regime (OTR) for discrete choices of treatment options \citep{
		watkins1992q, blatt2004learning,murphy2003optimal,qian2011performance, zhang2012robust, zhao2012estimating}, 
	the optimal individualized treatment initiation problem for time-to-event data has been seldom studied,
	and the estimation of OTIR has to tackle several new challenges.}

One main challenge in the estimation of OTIR rises from the missing of the treatment initiation time.
In clinical studies,
although every patient would be assigned with a treatment initiation time,
the value of the assigned treatment initiation time would be missing 
if the patient does not survive to the assigned treatment initiation time.   
Therefore, 
the potential outcome of a given treatment initiation time assignment regime can not be evaluated directly.

Another challenge in the searching for OTIR comes from the continuity of treatment initiation time.
In the treatment initiation time data, since the set of decision options is a time period containing an infinite number of time points, 
the relationship among the treatment options, covariates and outcome of interest
could be too complicated to be correctly specified, 
and thus, the regression-based estimation methods 
may not be suitable. 
Besides, 
when treatment initiation time follows a continuous distribution,
the probability that the observed treatment initiation time  
exactly matches a given regime is zero.
Thus,
existing value search methods  \citep{zhang2012robust, zhao2012estimating},
which evaluate each treatment regime based on samples whose treatment option exactly follows this regime,
can not be directly applied to treatment initiation time data. 
Lastly,
since in practice, when to initiate treatment is usually decided by physicians based upon patients' status,
there could exist unknown dependence between assigned treatment initiation times and covariates,
and statistical inference should consider such dependence.

In the literature,
several works related to the estimation of OTIR have been conducted.
In estimating the effect of treatment initiation time on an outcome measured at a fixed duration after initiation,
\cite{lok2012impact} 
discretized the treatment initiation time into multiple treatment points and 
developed structural nested mean models 
to deal with the non-random assignment of treatment initiation time in observational data.
\cite{zhao2011Reinforcement}
presented an adaptive reinforcement learning approach to discover 
the optimal individualized treatment regimen 
that selects the optimal time to initiate second-line therapy in a specially designed clinical reinforcement trial. 
\cite{hu2018modeling}
proposed a structural proportional hazards model to evaluate the effect of treatment initiation time on the survival time in the absence of baseline covariates.
For the case where treatment initiation is not assigned randomly,
they fit a semi-parametric model on the distribution of the assigned treatment initiation time and adopted inverse probability weighting techniques to deal with the missing of treatment initiation time.

{
	In this article, 
	we formulated the treatment initiation time decision problem in a meaningful and practical framework so as to overcome the aforementioned challenges and estimate the OTIR via a value search method.}  
Specifically,
we proposed a new value function, 
which is constructed on a restricted mean residual lifetime, 
to evaluate the performance of the treatment initiation regime 
and 
developed a nonparametric kernel-based estimation method for the value function.
The proposed method has three important advantages.
First, it does not posit any specific model on how the failure time depends on the treatment initiation time and covariates 
and  is thus more robust than regression-based methods.
Second, the obtained estimates are consistent even when treatment initiation times are not completely observable,  and their distribution is unknown.
Third, the estimation procedure allows treatment initiation times to depend on covariates arbitrarily, and thus, 
this method can be applied to datasets from both clinical trials and observational studies.

In the remainder of this article, 
Section \ref{sec:data} describes a breast cancer study that motivates the treatment initiation time problem.
Section \ref{sec:method} {formulates the optimal treatment initiation problem and} 
presents the details of the proposed OTIR estimation method. 
Section \ref{sec:prop} establishes the asymptotic properties of the resulting estimator in the decision rule and its associated value function estimator. In particular, since kernels are included in the estimation procedure, the asymptotic distribution of the estimated value function is nonstandard, which follows a weighted chi-squared distribution.
Section \ref{sec:simu} evaluates the performance of the proposed method via simulation studies, followed by an application to a breast cancer dataset for further illustration in Section \ref{sec:app}. 
Section \ref{sec:dis} provides some concluding remarks. 

\section{Data}
\label{sec:data}

\begin{figure}\small
	\graphicspath{{figures/}}
	\begin{center} 
	\begin{tabular}{cc}
	(a) & (b) \\
	\includegraphics[width=6 cm, height= 5 cm]{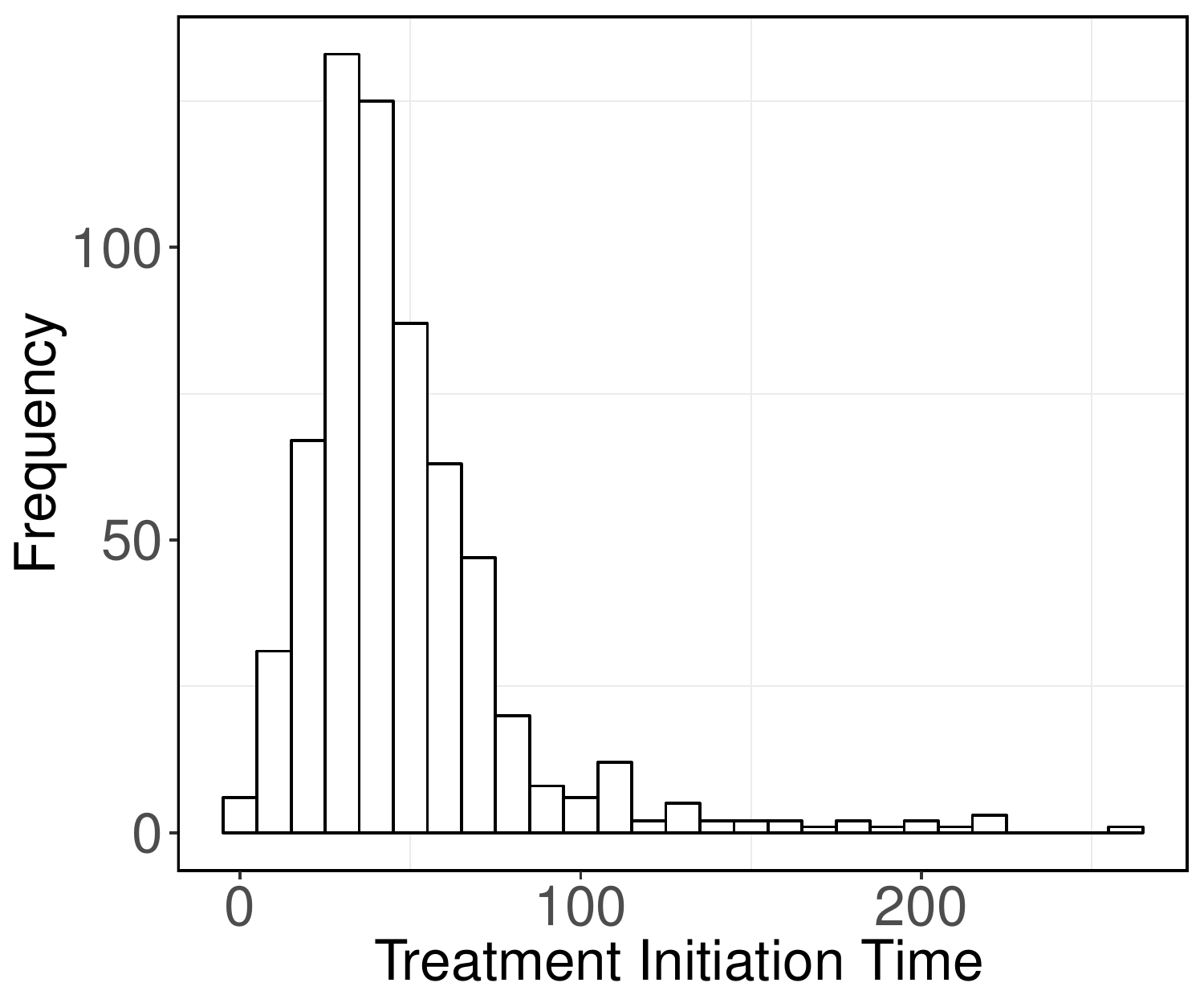} &
	\includegraphics[width=6 cm, height= 5 cm]{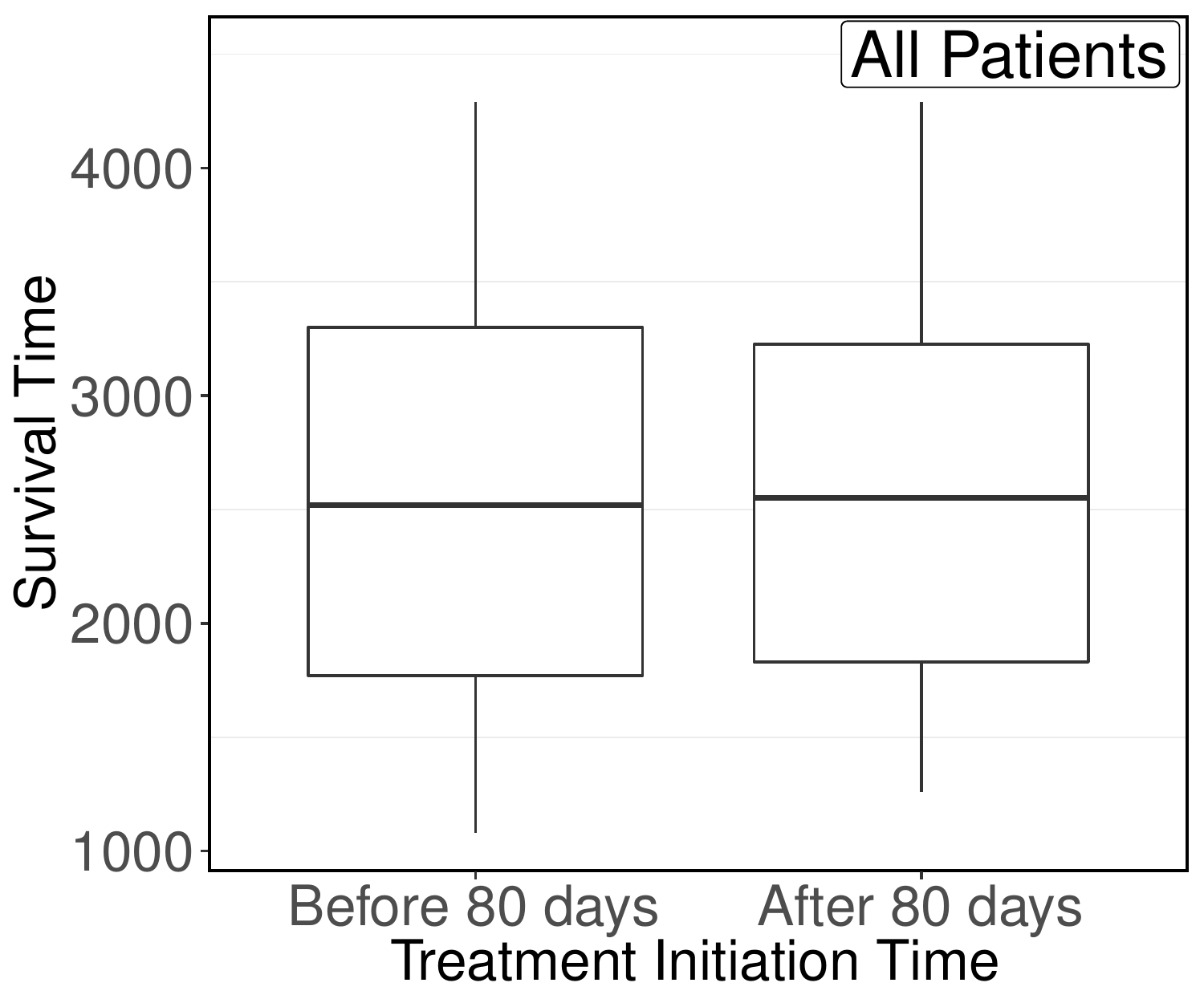} \\
	(c) & (d) \\
\includegraphics[width=6 cm, height= 5 cm]{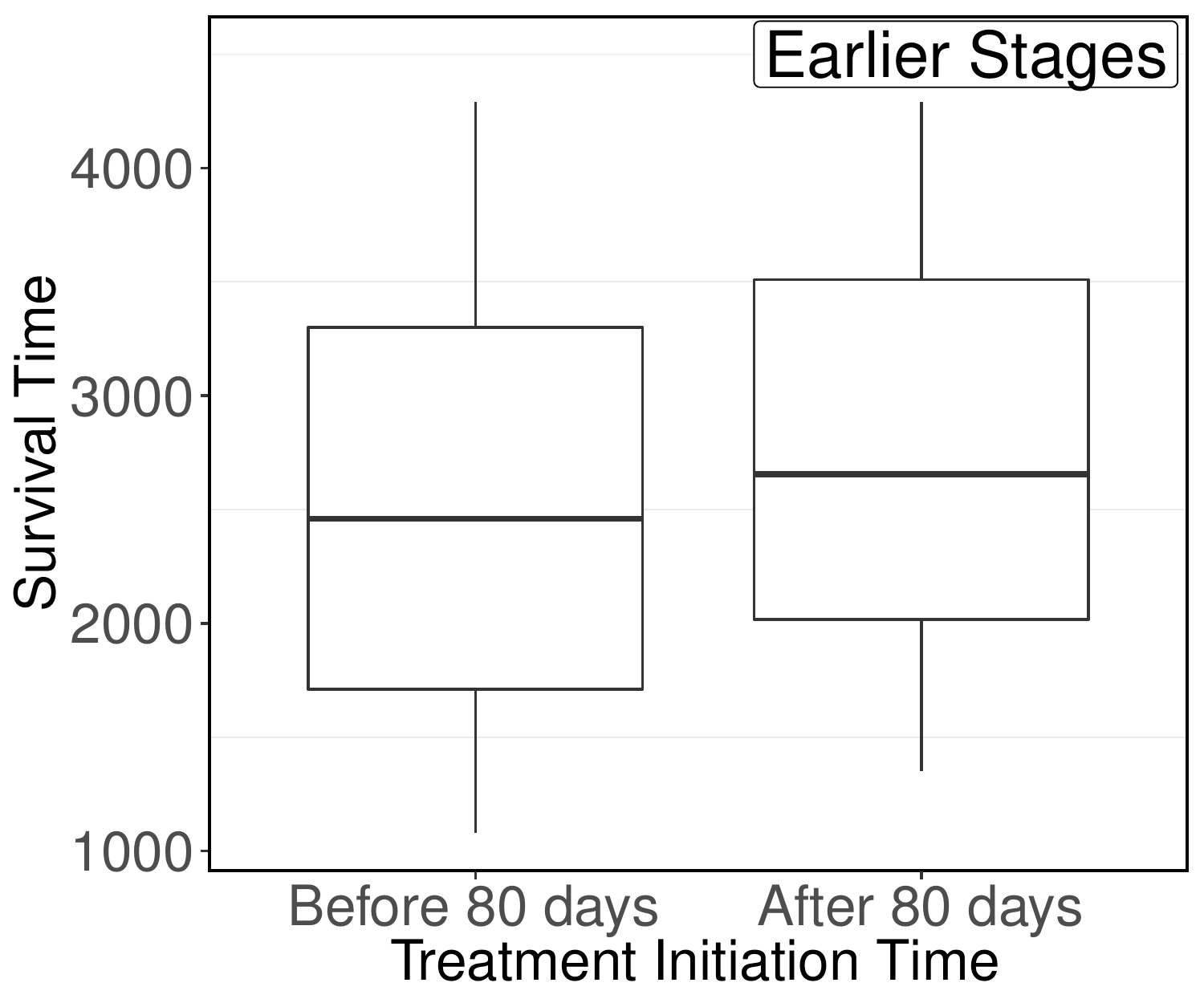} & 
\includegraphics[width=6 cm, height= 5 cm]{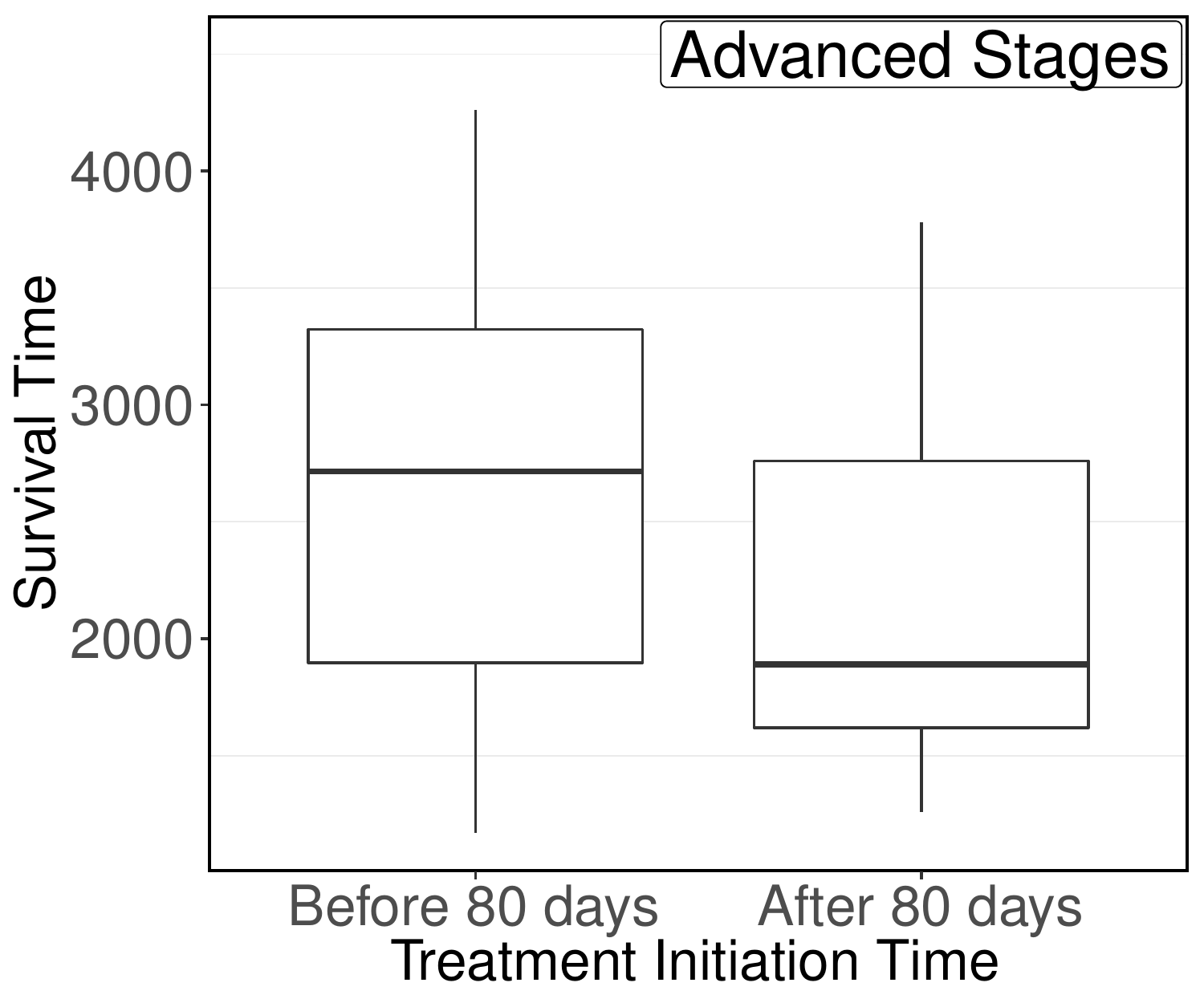} 
	\end{tabular}
	\end{center}
	\caption{
	Analysis of the breast cancer dataset, including (a) the distribution of adjuvant therapy initiation times, and the comparison of initiating adjuvant therapy within 80 days since surgery and  after 80 days since surgery among (b) all patients, (c) patients in earlier stages,  and (d) patients in advanced stages.
	}
	\label{fig:Distri of A}
\end{figure}

\begin{figure}
	\graphicspath{{figures/}}
	\begin{center}
		\includegraphics[width=8 cm, height= 6 cm]{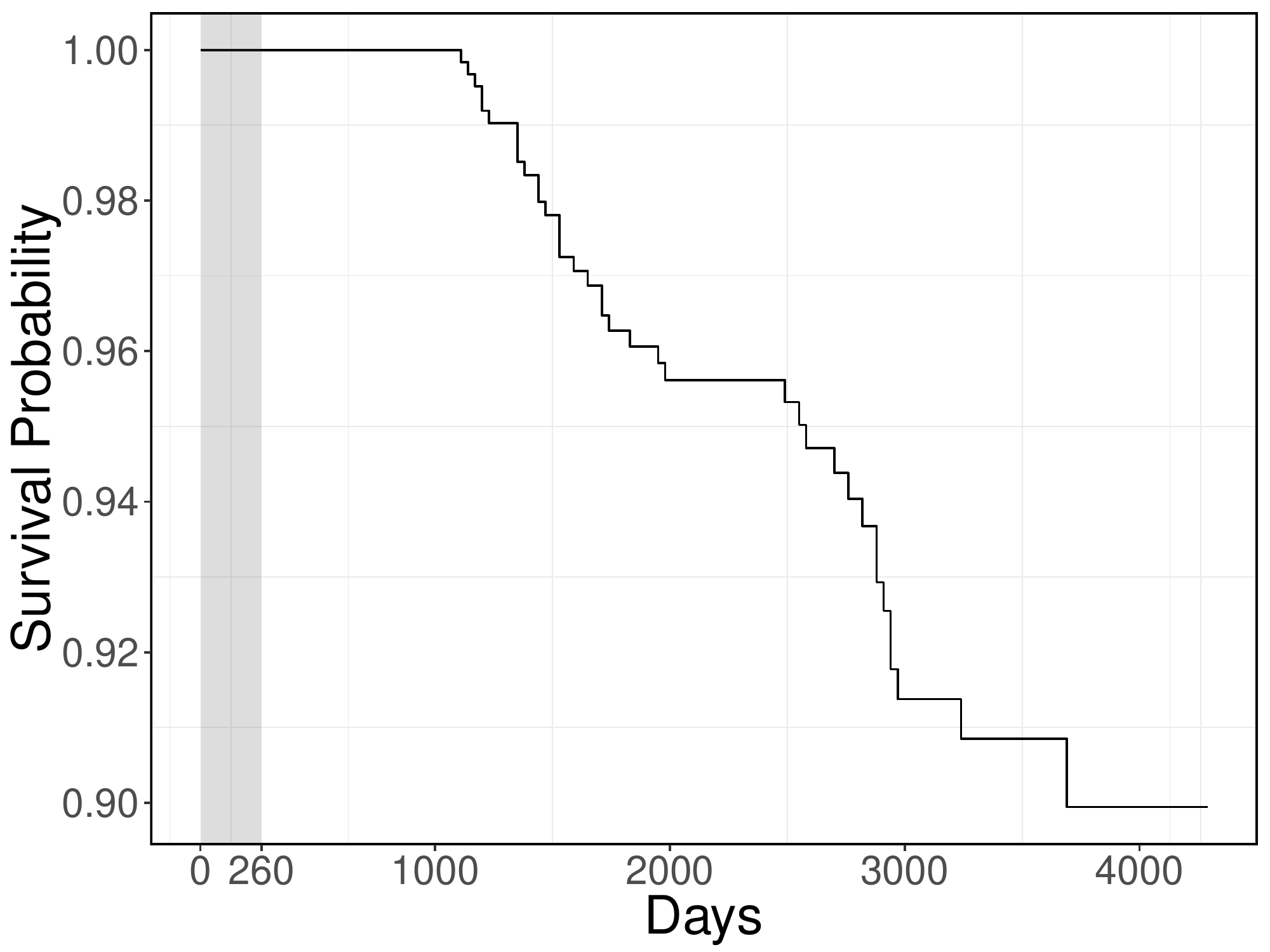}
	\end{center}
	\caption{
		The Kaplan-Meier estimates of the survival function in the breast cancer dataset.
		The shadow part marks the the range of the patients' treatment initiation times, which is  a relatively short time interval compared to their survival times.
	}
	\label{fig:KM}
\end{figure}

This research is motivated by a breast cancer dataset 
linked between the South Carolina Central Cancer Registry (SCCCR) and the South Carolina Revenue and Fiscal Affairs Office (RFA). 
The SCCCR is a population-based cancer surveillance system  that collects, processes, analyzes, and publishes cancer incidence for the state of South Carolina.
In the linked dataset, 
a total of 629 diagnosed breast cancer patients who received breast cancer surgery after the age of 45 years old 
initiated adjuvant chemotherapy or radiotherapy therapy at a specific time after surgery.
Among these patients, 
$19.9\% 
$ were in stage 0 (non-invasive breast cancers), 
$54.2\%$ 
were in stage 1 (localized only),
and 
$25.9\%$ 
 were in stage 2 (regional by direct extension only), stage 3 (regional lymph nodes involved only), stage 4 (regional by both direct extension and lymph) or stage 7 (distant sites/nodes involved).
The patients' age at surgery ranges from 45 to 62 years, 
and the initiation time of the adjuvant therapy is observed on all the patients.
Figure \ref{fig:Distri of A}(a)  shows the distribution of adjuvant therapy initiation time since surgery (in days), 
where the average duration from breast cancer surgery to adjuvant therapy is 49 days, and 
the majority of patients (619 of 629 patients) started adjuvant therapy within 24 weeks since surgery. 
The dataset also includes the time from surgery to death or the loss of follow-up (in days) for each patient, 
where the censoring rate of survival time is about $94\%$ and the observed survival time ranges from 1110 to  3690 days.
Figure \ref{fig:KM} shows the Kaplan-Meier estimates of the survival function, where the range of treatment initiation time was marked in shadow.
As shown in the plot, all the patients in this dataset initiated adjuvant therapy within a relatively short time interval compared to their survival times.


For patients diagnosed with breast cancer, 
numerous randomized trials have demonstrated that 
adjuvant chemotherapy or radiotherapy after the definitive surgeries 
could decrease the risk of death caused by breast cancer versus those without adjuvant therapy.
However, due to the heterogeneity in patients and their responses to the treatment, 
finding the optimal time from surgery to the initiation of adjuvant therapy is still challenging \citep{
Yu2017Influence,Riba2018surgical}.
In Figure \ref{fig:Distri of A},
we also compared the censored survival times of patients who initiated adjuvant therapy within 80 days since surgery and of patients who initiated adjuvant therapy after 80 days since surgery 
among all the patients, patients in earlier stages (stage 0 and 1), and patients in advanced stages (stage 2, 3, 4, and 7), respectively. 
Although on average, delaying the initiation of adjuvant therapy makes no significant differences on the patients' survival time distribution (as shown in Figure \ref{fig:Distri of A}b),
it is also clear that, among patients in advanced stages, 
those initiating adjuvant therapy within 80 days tend to survive longer than
those initiating adjuvant therapy after 80 days (Figure \ref{fig:Distri of A}d),  
whereas the situation is reversed for patients in earlier stages (Figure \ref{fig:Distri of A}c).
Such an observation indicates that a treatment initiation time that is beneficial for some group of patients may not be a good choice for the others, and 
an individualized treatment initiation regimen that is based on patients' characteristics is desired.

\section{ Methodology}
\label{sec:method}


\subsection{Definitions}
\label{ssec:def}

Consider a study with 
$n$ patients who started treatment at various time points and were followed up until the event of interest or censoring occurs.
Let $[0,a_0]$ be the pre-specified time range for initiating the treatment 
and let $\tau$ be the maximum follow-up time.
In medical practices, $[0,a_0]$ is usually a relatively short time range compared to the whole follow-up time $\tau$.
For example, 
in the breast cancer data studies, 
patients are recommended to start the adjuvant therapy within 168 days from the surgery, while their maximum follow-up time could be more than 10 years.

For the $i$th patient, $i=1, \ldots,n $, 
let $\bm{X}_i \in \mathcal{X}$ denote the $p$-dimensional vector of baseline covariates, 
$A_i^*$ denote the assigned treatment initiation time, 
$T_i$ denote the event time of interest, and
$C_i \leq \tau $ denote the censoring time.
Since $T_i$ could be censored by $C_i$, 
define $\tilde T_i= \min( T_i, C_i)$ and $\Delta_i=I(T_i \leq C_i)$.
Moreover, since 
$A_i^*$ could be unobserved 
if the patient does not survive beyond the assigned treatment time,
we define $A_i= A_i^* I\{ A_i^* \leq \tilde T_i\} + \infty I\{ A_i^* > \tilde T_i\} $ 
as the observed treatment initiation time.
Then the observed data consist of $\{ (\bm{X}_i,  A_i, \tilde T_i,\Delta_i), i = 1, \ldots, n \}$, which are
independent and identically distributed across $i$.

A treatment initiation regime $d(\bm{x})$ is a deterministic function
that maps the value of covariate $\bm{x} \in \mathcal X$ to a treatment initiation time $a\in[0,a_0]$.
Let $T^*(a)$ denote the potential survival time of a patient if he/she were assigned to
start the treatment at $a$
and  let $m(a, \bm{x})= \int_{a_0}^{\tau} pr\{ T^*(a) \geq t \mid T^*(a)\geq a_0, \bm{X}=\bm{x} \} dt $
denote the restricted mean residual lifetime of $T^*(a)$ at $a_0$ for patients with the covariate value $\bm{x}$. 
Then we propose to evaluate the performance of a treatment regime $d$ 
by  the average of restricted mean residual lifetime among all the patients:
\begin{align}
V(d)
&=E[ m\{d(\bm{X}), \bm{X} \} ] 
= \iint_{a_0}^{\tau} pr[ T^*\{ d(\bm{x})\} \geq t \mid T^*\{ d(\bm{x})\} \geq a_0, \bm{X}=\bm{x} ]  dt dF_X(\bm{x}) \,,
\label{Vd}
\end{align}
where $F_X(\bm{x})$ is the cumulative distribution function of $\bm{X}$.
For simplicity, we assume the probability density function of $\bm{X}$ also exists and denote it as $f_X(\bm{x})$. 
Now,
given a collection of treatment regimes $\mathcal{D}$ that are of interest,
the optimal treatment regime in $\mathcal{D}$
could be defined as 
\begin{align}
d^{opt}
=\arg  \max_{d\in \mathcal{D}} V(d)
\,.
\label{d_opt}
\end{align}

{\it Remark 1.}
Note that $T^*(a)$ is defined as the potential outcome if patients were assigned to initiate treatment at $a$.
Under this definition, 
if a doctor assigns a patient to initiate treatment at time $a$, 
then regardless of whether the patient dies before $a$, 
$T^*(a)$ would still be the potential outcome of the already assigned treatment initiation time $a$.
However, 
it should be noted that,
such treatment assignment could only have effects on a patient's survival time when $T^*(a)> a$.
In other words, 
if a patient does not survive to the assigned treatment initiation time, 
then he or she would not actually initiate the treatment, 
and thus the assigned treatment initiation time $a$ has no effect on the patient's survival time $T^*(a)$.

In the definition of $V(d)$, we restricted  $T^*\{d(\bm{x})\} > a_0$, 
which means we considered the restricted mean residual lifetime instead of restricted mean survival time.
Under this restriction,
$T^*\{d(\bm{x})\} > d(\bm{x})$ holds for any $\bm{x}$ and $d$,
which ensures that
all the samples used to evaluate the treatment initiation regime $d$ did initiate the treatment. 
It is noted that, for any $0\leq t_0 < a_0$, 
if we replace the restriction $T^*\{d(\bm{x})\} > a_0$ in $V(d)$ with 
$T^*\{d(\bm{x})\} > t_0$, 
there always exists $(\bm{x}, d)$ such that $t_0 < T^*\{d(\bm{x})\} < d(\bm{x})$.
We will show later that 
for such $(t_0, \bm{x}, d)$, the conditional survival probability 
$pr[ T^*\{ d(\bm{x})\} \geq t \mid T^*\{ d(\bm{x})\} \geq a_0, \bm{X}=\bm{x} ]$, $t > t_0$, is hard to estimate, because
in observational studies,
the treatment assignment time $a$ for $T< a$ is usually unobservable.


\subsection{An Example}
\label{ssec:example}

In this section, we will provide an example, under which  
the optimal treatment initiation regime $d^{opt}$ defined in $\eqref{d_opt}$ 
does select the optimal treatment initiation time.
Consider 
a class of hazard models for the potential survival time $T^*(a)$ conditional on covariate $\bm{X}=\bm{x}$:
\begin{equation}
\lambda(t; a, \bm{x})= \lambda_0(t) \exp\left[   \mu_0(\bm{x})+  I(t \geq a)Q\{ a- d_0(\bm{x})\} H_0(\bm{x}) \right], 
\label{model}
\end{equation}
where
$\lambda_0(\cdot)$ is the baseline hazard function, $\mu_0(\cdot) $ is an unspecified function for baseline covariate effects, 
$H_0(\cdot)$ is an unspecified non-negative function,
$d_0(\bm{x}) \in [0, a_0]$ is a given function of $\bm{x}$,
and
$Q(\cdot)$ is an unspecified  differentiable function with a unique minimum value $Q(0)<0$.
This model indicates that, 
given covariate $\bm{x}$, the ratio between 
hazard rate of patients who are receiving treatment at time $t$ 
and that of patients who have not started treatment at time $t$ 
equals to $\exp\left[ Q\{ a- d_0(\bm{x}) \} H_0(\bm{x})\right] $ 
and is minimized at $a= d_0(\bm{x})$.
Thus, $ d_0(\bm{x})$ is the optimal treatment initiation time for patient with covariate $\bm{X}=\bm{x}$ 
in the sense that it leads to the largest reduction in the patient's hazard rate after treatment.

On the other hand, we claim that $d_0$ is also the maximizer of the proposed value function $V(d)$. 
Here for  briefness, we provide an intuitive interpretation. 
Consider a patient with covariate value $\bm{x}$ and baseline hazards $\lambda_0(t) = \lambda_0$, $\mu_0(\bm{x}) = 0$ and $H_0(\bm{x}) = 1$. 
Let the dotted line in Figure \ref{Sect22} represent the mapping   
$ t \rightarrow \lambda_0 \exp\left[    Q\{ t - d_0(\bm{x})\} \right]$ on $[0,a_0]$.
Then, the solid line shows the hazard rate of the patient if the treatment is initiated at $a$,
and the dashed line shows the hazard rate of the patient if the treatment is initiated at $d_0(\bm{x})$.
Furthermore, let $S(a; \bm{x})$ 
denote the area of shadow part when $a$ ranges from $0$ to $a_0$. 
By calculation, the restricted mean residual lifetime at $a_0$ satisfies  
$ m(a, \bm{x}) = (\tau - a_0) [1- \exp\{ - S(a; \bm{x}) \} ]/ S(a; \bm{x})$, 
which implies that the optimal treatment initiation time which maximizes  $m(a, \bm{x})$ also minimizes the area of shadow part. 
From Figure \ref{Sect22}, it can be seen that the area of shadow part is minimized when $a = d_0(\bm{x})$.
Therefore, the proposed value function, which is constructed on the mean residual lifetime, is  maximized at $d_0$.

\begin{figure}
	\graphicspath{{figures/}}
	\begin{center}
	\includegraphics[width= 12 cm, height= 6 cm]{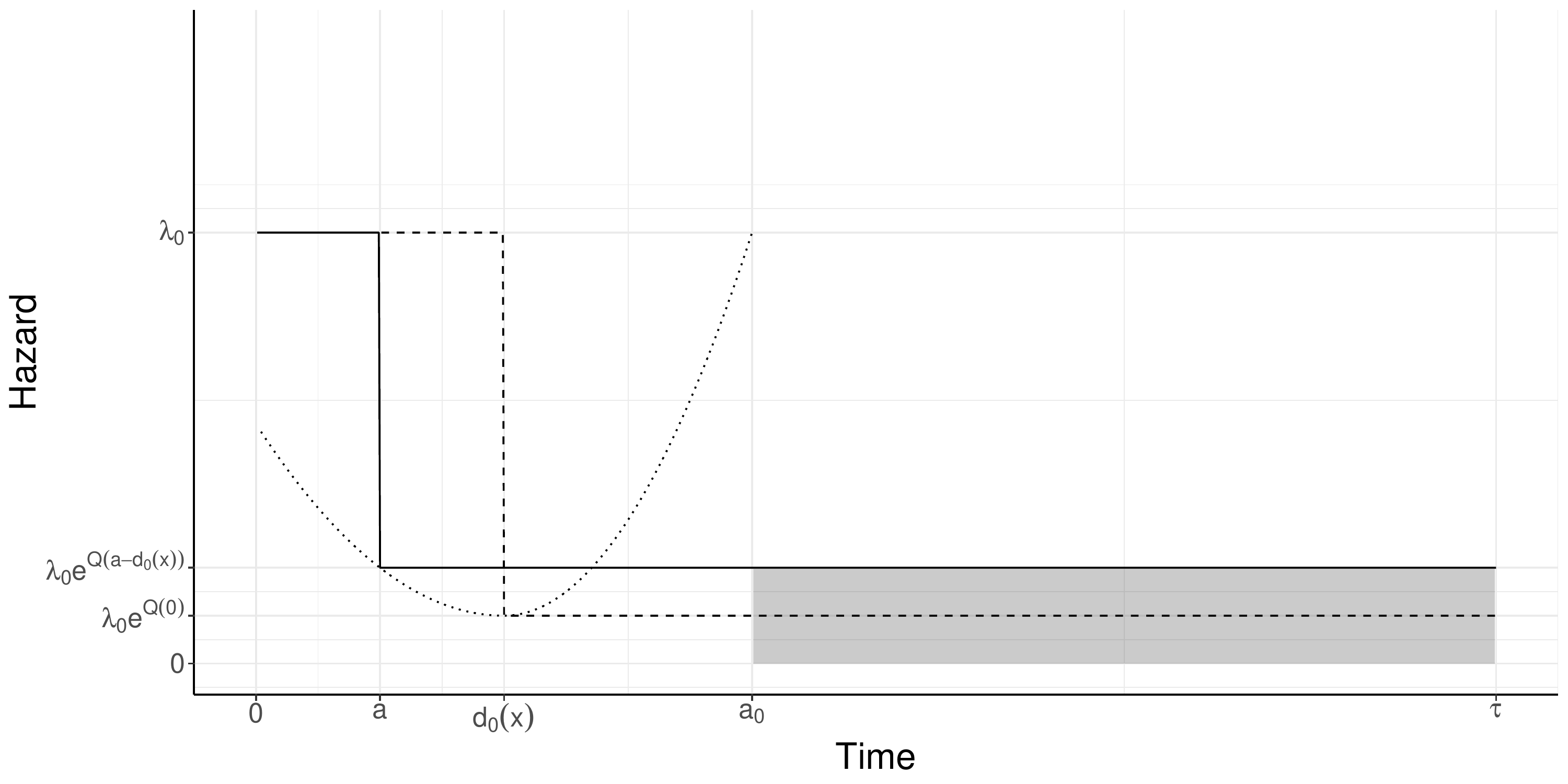}
	\end{center}
	\caption{ 
	The effect of treatment initiation time $a$ on the restricted mean residual lifetime $m(a, \bm{x})$ for a patient with covariate $\bm{X} = \bm{x}$ when 
	$\lambda_0(t) = \lambda_0$, $\mu_0(\bm{x}) = 0$ and $H_0(\bm{x}) = 1$.
         The hazard function $\lambda(t; a, \bm{x})$ is plotted in solid line,
         and the value of restricted mean residual lifetime $m(a, \bm{x})$ 
	 is monotone increasing with the area of shadow part. 
		\label{Sect22}
	}
\end{figure}

Besides, we noted that, 
although the restricted mean survival time has been widely used
to evaluate treatment effect in survival analysis. 
Here under model $(\ref{model})$, 
the maximizer of the restricted mean survival time 
does not equal to $d_0$.
For an intuitive interpretation, 
we still take Figure \ref{Sect22} as an example. 
Since the area under the solid curve on $[0,t]$  
represents the cumulative hazard $\Lambda(t; a, \bm{x})$ 
and that under the dashed curve represents $\Lambda\{ t; d_0(\bm{x}), \bm{x}\}$, 
it is not hard to obtain from Figure \ref{Sect22} that, 
if $a$ satisfies $Q\{a - d_0(\bm{x})\}<0$ and $a < d_0(\bm{x})$, 
we have $\Lambda(t; a, \bm{x}) < \Lambda\{ t; d_0(\bm{x}), \bm{x}\}$ for all $a < t \leq d_0(\bm{x}) $.  
Moreover, 
if the difference between $Q\{a - d_0(\bm{x})\}$ and $Q(0)-Q\{a - d_0(\bm{x})\}$ is large enough, 
the inequality may hold for all $ a < t \leq \tau$, 
and thus,  
the restricted mean survival time 
$\int_0^{\tau} \exp\{- \Lambda(t; a, \bm{x}) \} d t $  is not maximized at $a = d_0(\bm{x})$. 
In general, it can be proved that for any $0\leq t_0 < a_0$, the maximizer of the restricted mean residual lifetime, $\iint_{t_0}^{\tau} pr[ T^*\{ d(\bm{x})\} \geq t \mid T^*\{ d(\bm{x})\} \geq  t_0,  \bm{X}=\bm{x} ]  dt dF_X(\bm{x})$, does not equal to $d_0$.  

%
\subsection{ Estimation procedure}
\label{ssec:est}

To ensure that the proposed value function 
can be estimated using observed data, 
the following assumptions are required:
(A1) (consistency assumption) $T=T^*(A^*)$;
(A2) (no unmeasured confounder assumption)
$ \{ T^*(a), 0 \leq a \leq a_0 \} \perp A^* \mid \bm{X} $;
(A3) (conditionally independent censoring assumptions) $C \perp T \mid (A^*, \bm{X})$ and $ C \perp A^* \mid \bm{X}$.

Under these assumptions,
the value function $V(d)$ can be written as 
\begin{align*}
V(d)
=&
\iint_{a_0}^{\tau} 
\frac{ pr\{ \tilde T\geq t  \mid A^*=d(\bm{x}), \bm{X}=\bm{x}\} }{pr\{ \tilde T\geq a_0\mid A^*=d(\bm{x}), \bm{X}=\bm{x}\}}
\frac{pr( C \geq  a_0\mid \bm{X}=\bm{x})}{pr( C \geq t \mid \bm{X}=\bm{x} ) } f_X(\bm{x}) dt  d\bm{x} \,.
\end{align*}
Since $\tilde T \geq a_0$ implies  $A^* = A$,
we have 
\begin{align*}
V(d)
=&
\iint_{a_0}^{\tau} 
\frac{ pr\{ \tilde T\geq t ,     A^*= A \mid  A^*=d(\bm{x}), \bm{X}=\bm{x}\}
       }{pr\{ \tilde T\geq a_0, A^*= A \mid  A^*=d(\bm{x}), \bm{X}=\bm{x}\}
       }
\frac{pr( C \geq  a_0\mid \bm{X}=\bm{x})}{pr( C \geq t \mid \bm{X}=\bm{x} ) } f_X(\bm{x}) dt  d\bm{x} \\
=&
\iint_{a_0}^{\tau} 
\frac{ pr\{ \tilde T\geq t      \mid  A^*= A, A^*=d(\bm{x}), \bm{X}=\bm{x}\}
       }{pr\{ \tilde T\geq a_0 \mid  A^*= A, A^*=d(\bm{x}), \bm{X}=\bm{x}\}
       }
\frac{pr( C \geq  a_0\mid \bm{X}=\bm{x})}{pr( C \geq t \mid \bm{X}=\bm{x} ) } f_X(\bm{x}) dt  d\bm{x} \\
=&
\iint_{a_0}^{\tau} 
\frac{ pr\{ \tilde T\geq t       \mid   A=d(\bm{x}), \bm{X}=\bm{x}\} }{
	pr\{ \tilde T\geq a_0  \mid  A=d(\bm{x}), \bm{X}=\bm{x}\}}
\frac{pr( C \geq  a_0\mid \bm{X}=\bm{x})}{pr( C \geq t \mid \bm{X}=\bm{x} ) } f_X(\bm{x}) dt  d\bm{x}   \,.
\end{align*}

{
	Here, $(\tilde T, A, X)$ are observable, 
	$pr\{ \tilde T\geq t       \mid   A=d(\bm{x}), \bm{X}=\bm{x}\}$ can be directly estimated by smoothing techniques such as kernel smoothing, local polynomial fitting, and spline methods, 
	and thus $V(d)$ is estimable.
	For example, 
	let $\hat f_X(\cdot)$ denote the kernel density estimator of $f_X(\cdot)$, 
	and let $\hat S_C( t \mid \bm{x} )$ be an estimator for the conditional survival function 
	$S_C( t \mid \bm{x} ) = pr( C \geq t\mid \bm{X}=\bm{x} )$, 
	then a kernel-based estimator of $V(d)$ could be }
\begin{align*}
\hat V(d)
=&
\iint_{a_0}^{\tau} 
\frac{ 
	\sum_{i=1}^n I(\tilde T_i \geq t     ) K_{\bm{h_1}} (\bm{X}_i - \bm{x}) 
	K_{h_2}[g( A_i) - g\{ d(\bm{x}) \}  ] }{
	\sum_{i=1}^n I(\tilde T_i \geq a_0) K_{\bm{h_1}} (\bm{X}_i - \bm{x}) 
	K_{h_2}[g( A_i) - g\{ d(\bm{x}) \}  ]  }
\frac{\hat S_C( a_0 \mid \bm{x} )}{\hat S_C( t \mid \bm{x} )} 
\hat f_X(\bm{x}) dt  d\bm{x}\,,
\end{align*}
where
\begin{align*}
K_{\bm{h_1}}(\bm{X}_i - \bm{x}) 
= \prod_{j = 1}^p \frac{1}{h_1^{(j)}}  K\{ \frac{X_i^{(j)}- x^{(j)} }{h_1^{(j)}} \} \,,
\,~
K_{h_2}[  g( A_i) - g\{ d(\bm{x}) \}  ] 
= \frac{1}{h_2}K[ \frac{g( A_i) - g\{ d(\bm{x}) \} }{h_2}] \,,
\end{align*}
$K(\cdot)$ is a kernel function, 
$\bm{h_1} \in \mathbb{R}^p$ and $h_2\in \mathbb{R}$ are kernel bandwidths,
$x^{(j)}$ ($i = 1, \ldots, p$) denotes the $j$th component of any $p$-dimensional vector $\bm{x}$,
and $g$ is a monotonic increasing transform function that maps interval $(0,a_0)$ into real line.

There are several approaches to obtain $\hat S_C(t\mid \bm{x})$.
For example, we may construct a semi-parametric model on the censoring time $C$ conditional on $\bm{X}$, 
and obtain a model-based estimator of $pr( C \geq t\mid \bm{X}=\bm{x})$. 
Alternatively, for more robust estimation,
we could  use the kernel-based local Kaplan-Meier method 
\citep{dabrowska1989uniform}  
which estimates the conditional survival function nonparametrically.
In some applications  
such as clinical studies with satisfactory follow-up,
there is no obvious evidence that 
censoring events are related to treatment or covariates.
For such cases,
it is reasonable to make
an independent censoring assumption (A4): $C \perp (A^*,\bm{X})$,
and estimate $pr( C \geq t\mid \bm{X}=\bm{x})=pr(C \geq t)$ by the standard Kaplan-Meier estimator.

For simplicity,
from now on we only consider the case where independent censoring assumption (A4) holds
and let $S_C(t)=pr(C \geq t)$ denote the survival function of the censoring time $C$.
Let $\hat S_C(t)$ be the Kaplan-Meier estimator for $ S_C(t)$. Then, 
under assumptions (A1)-(A4), 
the value of treatment regime $d$ can be estimated by
\begin{align*} 
\hat V(d)
=&
\int 
\frac{ 
	\sum_{i=1}^n W_n( \tilde T_i) I(\tilde T_i \geq a_0) K_{\bm{h_1}} (\bm{X}_i - \bm{x}) K_{h_2}[  g( A_i) - g\{ d(\bm{x}) \}  ] 
}{
	\sum_{i=1}^n I(\tilde T_i \geq a_0)  K_{\bm{h_1}} (\bm{X}_i - \bm{x}) K_{h_2}[  g( A_i) - g\{ d(\bm{x}) \}  ] }
\hat f_X(\bm{x}) d\bm{x} \,,
\end{align*}
where
\begin{align*}
W_n(t)  
=&
\int_{a_0}^{t} 
\frac{\hat S_C(a_0)}{\hat S_C(u)}du
= 
\int_{a_0}^t 
\prod_{a_0 < v \leq u} \left\{1- \frac{ \sum_{j=1}^n I(\tilde T_j =v, \Delta_j=0) 
}{ \sum_{j=1}^n I(\tilde T_j \geq v)   }\right\}^{-1}
du   \,,
\end{align*}
for all $t \geq a_0$.

{\it Remark 2.}
Under the independent censoring assumption, 
the integrand of our value function can be written as 
$ 
pr( \tilde T \geq t \mid  \tilde T \geq a_0, A^* =a, \bm{X}) S_C(a_0)\{S_C(t)\}^{-1}
f_X(\bm{x}) 
$.
Here, $pr( \tilde T \geq t \mid  \tilde T \geq a_0, A^* =a, \bm{X})$
is a conditional survival function 
that focuses on the patients who are alive and uncensored at $a_0$.
Since $ \tilde T \geq a_0$ implies $\tilde T \geq A^*$, 
all these patients will have observed treatment initiation times $A =A^*$, 
and thus 
$pr( \tilde T \geq t \mid  \tilde T \geq a_0, A^* =a, \bm{X}) = 
  pr( \tilde T \geq t \mid  \tilde T \geq a_0, A =a, \bm{X})
$ can be directly estimated by the observed data.
Therefore,   
although in practice, $A^*$ could be censored by both $C$ and $T$ in some complicated mechanism, 
our estimation procedure 
avoids the possible problems associated with the missingness of $A^*$.
On the other hand, 
since the other part of the integrand $S_C(a_0)\{S_C(t)\}^{-1} f_X(\bm{x}) $ does not  concern with the unobservable $A^*$, 
 the corresponding estimators $\hat S_C(t)$ and $\hat f_X(\bm{x})$ can be directly  obtained based on all the samples. 
Therefore, the information for the patients who do not survive to $a_0$
will also be utilized in the estimation of our value function.


\subsection{Optimal treatment initiation regime}
\label{ssec:otit}

Now we search for the optimal treatment initiation regime 
in a collection of treatment regimes $\mathcal{D}$ indexed by finite-dimensional parameters, which is 
$\mathcal D =\{ d_{\beta}: d_{\beta}( \bm{x})= \phi({ \bm{\tilde x}} ^ T \bm{\beta}) \} $.
Here, 
$\bm{\tilde x} =(1,\bm{x}^T)^T$, $\bm{\beta}$ is a $(p+1)$-dimensional parameter 
and $\phi$ is a fixed function mapping $(-\infty, +\infty)$ to $(0,a_0)$.
In practice, $\phi$ could be defined as a logistic link function 
$\phi(u)= a_0 \exp(u) / \{1+\exp(u)\}$ or  a normal link  function 
$\phi(u)= a_0 \Phi(u) $, 
where $\Phi$ is the cumulative distribution function of the standard normal distribution.

Let $d^{opt}=\arg \max_{d_{\beta} \in \mathcal{D}} V(d_{\beta})$ 
denote the optimal treatment regime among  $\mathcal{D}$
and let 
\begin{align*}
M(\bm{\beta}) =V(d_{\beta})
=  \iint_{a_0}^{\tau} 
\frac{ pr\{ \tilde T\geq t     \mid   A=\phi(\bm{ \tilde x} ^ T \bm{\beta}), \bm{X}=\bm{x}\} }{
	pr\{ \tilde T\geq a_0\mid A=\phi(\bm{ \tilde x} ^ T \bm{\beta}), \bm{X}=\bm{x}\} }
\frac{S_C( a_0)}{S_C(t)} f_X(\bm{x}) dt  d\bm{x},
\end{align*}
denote the value of the treatment initiation regime indexed by $\bm{\beta}$.
Define
$\bm{\beta}^{opt}=\arg \max_{ \bm{\beta} \in \mathbb{R}^{p+1} } M(\bm{\beta})\,,$
then we have $d^{opt}(x)= \phi( \bm{\tilde x}^ T \bm{\beta}^{opt}) $.

Following the estimation procedure in section \ref{ssec:est}, we can estimate 
$M(\bm{\beta})$ by
\begin{align*} 
M_n(\bm{\beta})
=&
\int 
\frac{ 
	\sum_{i=1}^n W_n( \tilde T_i) I(\tilde T_i \geq a_0) K_{\bm{h_1}}(\bm{X}_i - \bm{x}) K_{h_2}[  g( A_i)-g\{ \phi(\bm{ \tilde x} ^ T \bm{\beta}) \}  ] 
}{
	\sum_{i=1}^n I(\tilde T_i \geq a_0)  K_{\bm{h_1}}(\bm{X}_i - \bm{x}) K_{h_2}[  g( A_i)-g\{ \phi(\bm{ \tilde x} ^ T \bm{\beta}) \}  ] }
\hat f_X(\bm{x}) d\bm{x}.
\end{align*}
Let  $\bm{\hat \beta}^{opt}=\arg \max_{\beta} M_n(\bm{\beta})$ be the  estimate for  $\bm{\beta}^{opt}$.
Then $\hat d^{opt}(\bm{x})= \phi(\bm{\bm{\tilde x}}^ T \bm{\hat \beta}^{opt})$
is an estimator for the optimal treatment initiation regime $d^{opt}$, 
and 
$M_n(\bm{\hat \beta}^{opt})$ estimates the value function of the optimal treatment initiation regime, $V(d^{opt})$.

\section{Asymptotic properties}
\label{sec:prop}

This section provides the asymptotic properties of $\bm{\hat \beta}^{opt}$ and $M_n(\bm{\hat \beta}^{opt})$. Their proofs are given in the  Appendix.
For simplicity, we consider the case where 
$A^*$ and  $\bm{X}$ are continuous with joint probability density function $f_{(A^*, X)} (a, \bm{x})$ on $ [0,a_0]\times \mathcal{X}$.
Define $S_T(t; a, \bm{x}) = pr( T \geq t \mid A^* = a, \bm{X} = \bm{x})$.  
We assume the following conditions:
\begin{enumerate}
\item[C1] 
There exists a bounded set $\mathcal{B}$ of $\bm{\beta}$ satisfying 
	$\bm{\beta}^{opt} \in \mathcal{B}$ and 
	$\sup \{ M(\bm{\beta}):  \bm{\beta} \in \mathcal{B}, || \bm{\beta} - \bm{\beta}^{opt} ||_2 \geq \epsilon \}  
	< M(\bm{\beta}^{opt})$ for any $\epsilon >0 $.
	\item[C2] 
	There exist two positive constants $\epsilon_c$ and $\epsilon_t$, such that  
	$ pr( C \geq \tau \mid \bm{X}=\bm{x}) \geq \epsilon_c$ and 
	$ \inf_{a\in [0,a_0]} S_T(a_0; a, \bm{x}) \geq \epsilon_t$ 
	for all $x \in \mathcal{X}$. 	
	\item[C3]
	For each fixed $t$, $S_T(t; a, \bm{x})$ and  $f_{(A^*, X)} (a, \bm{x})$ are thrice-differentiable functions of $(a, \bm{x})$ with 
	partial derivatives uniformly bounded  on $t\in [0,\tau]$, $a\in[0,a_0]$ and $\bm{x}\in \mathcal{X}$.
	Link function $\phi(u)$ is twice-differentiable with derivative 
	$\dot \phi(u) = \rm d \phi(u) / \rm du$ bounded on $\mathbb{R}$.
	Transform function $g(a)$ is four times differentiable for $a \in [0, a_0]$.
	\item[C4] 
	$K(u)$ is a twice differentiable symmetric kernel function with derivative $\dot K(u)$ and second derivative $\ddot K(u)$. 
	Define $\kappa_{i,j}= \int u^i K(u)^j du$ and 
	$\dot \kappa_{i,j}= \int u^i \dot K(u)^j du$ for any $i \geq 0$ and $j \geq 0$.
	Then 
	$u^4K(u) \rightarrow 0$ as $|u| \rightarrow \infty$,
	$\int |\dot K(u)| du < \infty$,
	$\sup_u K(u) \leq K_{max} < \infty$,
	$\sup_u \dot K(u) \leq \dot K_{max} < \infty$ and 
	$\kappa_{2,1}$, $\kappa_{0,2}$ and $\dot \kappa_{0,2}$ are bounded.	
	\item[C5] 
	$n |\bm{h_1}| h_2^3 \rightarrow \infty$, $  n |\bm{h_1}| h_2 ( ||\bm{h_1}||_2^2+h_2^2)^2 \rightarrow 0$
	as $n\rightarrow \infty$.
	Here $|\bm{h_1}| = \prod_{j=1}^p h_1^{(j)}$, $||\bm{h_1}||_2^2 = \sum_{j=1}^n \{ h_1^{(j)}\}^2$ for  
	$\bm{h_1} = (h_1^{(1)},\ldots, h_1^{(p)})$.	
	\item[C6] 
	Let $D(\bm{\beta})$ denote the Hessian matrix of $M(\bm{\beta})$.
	There exists a small neighborhood 
	$N_{\delta} = \{ \bm{\beta} \in \mathcal{B}: || \bm{\beta} - \bm{\beta}^{opt} ||_2 \leq \delta \}$, 
	such that $-D(\bm{\beta})$ is positive definite for any $\bm{\beta} \in N_{\delta}$. 
\end{enumerate}

Condition C1 is proposed 
to prove the strong consistency for $\bm{\hat \beta}^{opt}$. 
For a compact set $\mathcal{B}$ and continuous function $M(\bm{\beta})$, 
the uniqueness of $\bm{\beta}^{opt}$ as a maximizer of $M(\bm{\beta})$ also implies this condition.
In Condition C2, the positiveness of $S_C(\tau)$ is commonly assumed in survival analysis to ensure the uniform convergence of $\hat S_C(t)$ on $ t\in [0, \tau]$. 
Also, Condition C2 implies that 
$ \inf_{a\in [0,a_0]} pr( A^* \leq T, A^* \leq C \mid  A^* =a, X) > 0 $ almost surely, 
which indicates any candidate treatment initiation time on $[0,a_0]$ has a potential chance to be observed in our setting.
The boundedness of derivatives in Condition C3 is posited for 
the uniform convergence of $M_n(\bm{\beta})$ and its derivatives on $\mathcal{B}$. 
In addition, it is not hard to verify that both logistic link function and normal link function satisfy the existence and boundedness assumption for $\dot \phi$.
Conditions C4 and C5 are posited for the asymptotic properties of the Kernel estimators.
It is not hard to verify that most of the kernel functions, 
including Gaussian and all the bounded symmetric kernels,
satisfy condition $4$.
Condition C5 also provides a guide to select bandwidth $\bm{h_1}$ and $h_2$.
For example, 
when $p=1$ and $h_1=h_2$, an appropriate bandwidth should range from 
$n^{-1/4}$ to $n^{-1/6}$.
Condition C3 implies that $M(\bm{\beta})$ is twice-differentiable on $\mathcal{B}$, 
and thus, the Hessian matrix $D(\bm{\beta})$ defined in Condition C6 exists for any $\bm{\beta} \in \mathcal{B}$.
Moreover, by Condition C6, 
$D(\bm{\beta})$ is invertible in a neighborhood of $\bm{\beta}^{opt}$, 
which is needed for deriving the asymptotic distribution of $\bm{\hat \beta}^{opt}$.

Based on the above conditions, 
we can establish the asymptotic properties of the proposed estimators.

\begin{theorem}
	Under Conditions C1--C6, 
	$\sup_{\bm{\beta} \in \mathbb{R}^{p+1}} |  M_n(\bm{\beta}) - M(\bm{\beta}) |$ converges to zero almost surely, 
	and $\bm{\hat \beta}^{opt}$ is a consistent estimator for $\bm{\beta}^{opt}$.
\end{theorem}
	
\begin{theorem}
	Under Conditions C1--C6,
	$ (n |\bm{h_1}| h_2^3)^{1/2} ( \bm{\hat \beta}^{opt} - \bm{\beta}^{opt})$
	converges in distribution to a normal variable with mean zero and variance
	$\Sigma = 
	D^{-1}(\bm{\beta}^{opt}) \Sigma_U D^{-1}(\bm{\beta}^{opt})\,.
	$
	Here $D(\bm{\beta})$ is a $p\times p$ matrix defined in condition $6$,
	$D^{-1}(\bm{\beta})$ denotes the inverse of matrix $D(\bm{\beta})$, 
	and $\Sigma_U$ is a $p\times p$ matrix with expression given in 
	the Appendix.
\end{theorem}

Let $\det(\Sigma)$ denotes the determinant of matrix $\Sigma$ and 
let $ a_i (i = 1,\ldots, s)$ denote the distinct eigenvalues of $-\Sigma_U D^{-1}(\bm{\beta}^{opt})$
satisfying  $ \det\{  \lambda I + \Sigma_U D^{-1}(\bm{\beta}^{opt}) \}= \prod_{i=1}^s (\lambda - a_i)^{r_i}$ and $\sum_{i=1}^s r_i = p$. 
Then the asymptotic distribution of the estimated value,  $M_n(\bm{\hat \beta}^{opt})$, of the derived optimal treatment initiation regime is stated in the following theorem.

\begin{theorem}
	Under Conditions C1--C6,
	$
	n |\bm{h_1}| h_2^3 \{ M_n(\bm{\hat \beta}^{opt}) - M(\bm{\beta}^{opt})\}
	$
	converges in distribution to 
	$ \sum_{i=1}^s a_i \chi^2(r_i)/2$, 
	where 
	$ \chi^2(r_i),i = 1, \ldots s$ are mutually independent chi-square distributions with degree of freedom $r_i$.
\end{theorem}

Theorem 3 shows that the asymptotic distribution of the estimated value function is a weighted chi-squared distribution.
To illustrate this, we provide some heuristic arguments below. Consider a general case where the M-estimation function $M_n$ and its derivative 
$U_n(\bm{\beta}) = \rm{d} M_n(\bm{\beta})/ \rm{d} \bm{\beta} $ satisfy 
$
c_n \{ M_n( \bm{\beta}) - M(\bm{\beta})\} \rightsquigarrow N\{ 0,\sigma_{M}^2(\bm{\beta})\}
$ 
and
$d_n \{ U_n( \bm{\beta}^{opt}) - U(\bm{\beta}^{opt})\} \rightsquigarrow N(0,\sigma_{U}^2)$.   
Since $\bm{\hat \beta}^{opt}$ is the maximizer of $M_n$,
it can be obtained from Taylor expansion that 
$ d_n^2\{ M_n(\bm{\hat \beta}^{opt}) - M_n(\bm{\beta}^{opt})\}$  converges to a weighted chi-square distribution.
Then by rewriting
\[
M_n(\bm{\hat \beta}^{opt}) - M(\bm{\beta}^{opt}) =
\{ M_n(\bm{\hat \beta}^{opt}) - M_n(\bm{\beta}^{opt})\} +  \{ M_n( \bm{\beta}^{opt}) - M(\bm{\beta}^{opt})\},
\]
we can conclude that 
if $d_n^2 / c_n \rightarrow 0$,
$c_n \{ M_n(\bm{\hat \beta}^{opt}) - M(\bm{\beta}^{opt})\} \approx c_n  \{ M_n( \bm{\beta}^{opt}) - M(\bm{\beta}^{opt})\}$ 
converges to a normal distribution; 
while if $d_n^2 / c_n \rightarrow \infty$, 
$d_n^2 \{ M_n(\bm{\hat \beta}^{opt}) - M(\bm{\beta}^{opt})\} \approx 
d_n^2\{ M_n(\bm{\hat \beta}^{opt}) - M_n(\bm{\beta}^{opt})\}$ converges to  a weighted chi-square distribution. 
In our estimation, 
since treatment initiation time $A$ follows a continuous distribution and 
$M_n(\bm{\beta})$ contains kernel term  $K_{h_2}[  g( A_i)-g\{ \phi(\bm{ \tilde x} ^ T \bm{\beta}) \}  ]$,
this leads to
$d_n = ( n |\bm{h_1}|h_2^3)^{1/2}$, 
$c_n = ( n |\bm{h_1}|h_2)^{1/2}$ 
and $d_n^2 / c_n = (n |\bm{h_1}| h_2^5)^{1/2} \rightarrow 0$ 
as $n \rightarrow \infty$ under condition $5$.
Thus, the asymptotic distribution of $ M_n(\bm{\hat \beta}^{opt})$ is a weighted chi-square distribution.
In contrast, if $A$ follows a discrete distribution, 
$M_n(\bm{\beta})$ will not include the kernel term  $K_{h_2}[  g( A_i)-g\{ \phi(\bm{ \tilde x} ^ T \bm{\beta}) \} ]$ 
and $c_n = d_n$. In that case, $ M_n(\bm{\hat \beta}^{opt})$ would converge to a normal distribution as $d_n^2 / c_n = d_n  \rightarrow \infty $, as widely studied in the literature for the value search estimators 
\citep{zhang2012robust,fan2017concordance,jiang2017estimation}.

In addition, since the analytic forms of the asymptotic variances of the parameter and value estimators are too complicated due to various kernel estimations, direct estimation of these asymptotic variances in finite samples is difficult. 
Thus, in this article, we use the bootstrap method to obtain the variance estimators.
In particular, 
the confidence interval for the value function is constructed based on the empirical distribution of bootstrapped estimators,  while for $\bm{\hat \beta}^{opt}$, a normal-based confidence interval is adopted. 

\section{Simulations}
\label{sec:simu}

Now we conduct simulations to assess the finite-sample
performance of the proposed method.
Let covariate $\bm{X}=(X_1, X_2)^ T$ be a 2-dimensional random vector
with $X_1$  generated from a discrete Bernoulli distribution with mean 0.5 and
$X_2$  generated from a continuous normal distribution with mean 0 and variance 1.
Let $A^* \in [0,a_0]$  
be the assigned treatment initiation time which may depend on $\bm{X}$.
Given  covariate value $\bm{x}=(x_1,x_2)^ T$ and $A^*=a$, 
the failure event  $T$ is generated by one of the following hazard models:
\begin{enumerate}
	\item[$(m1)$] 
	$\lambda(t; a, \bm{x})= \lambda_0 \exp\left[   I(t \geq a)Q\{ a-\phi(\bm{\tilde x}^ T\bm{\beta}_0)\}  \right] $,
	\item[$(m2)$]
	$ \lambda(t; a, \bm{x})= \lambda_0 \exp\left[   I(t \geq a)Q\{ a-\phi(\bm{\tilde x}^ T\bm{\beta}_0)\} (1+ x_1) \right] $, 
	\item[$(m3)$]
	$\lambda(t; a, \bm{x})= \lambda_0 \exp\left[   \log \{(1+x_2^2)/2\} +  I(t \geq a)Q\{ a-\phi(\bm{\tilde x}^ T\bm{\beta}_0)\} \right] $. 
\end{enumerate}
Here, model $(m1)$ is the basic case.
Model $(m2)$ allows the effect of the optimal  treatment initiation $ Q(0) (1+ x_1) $ to depend on covariates,
which indicates that patients are heterogeneous even when the optimal treatment initiation time is adopted. 
Under model $(m3)$, the hazard rate before initiating the treatment is also a function of covariates $\bm{X}$.
For all these three models, 
we set  $Q(u) = 2 (u^2-1)$,  
$\bm{\beta}_0=(0, 0.5,0.5)^ T$, 
$\lambda_0=0.2$ or $0.3$ and $a_0=2$ or $3$.
Let $\mathcal{U}(a, b)$ denote a uniform distribution on $[a,b]$ and 
let $\mathcal{B}(a, b)$ denote a beta distribution with mean $1/(a+b)$ and variance $ab/\{(a+b)^2(a+b+1)\}$. 
We consider the following two scenarios for the treatment initiation time distribution: 
\begin{enumerate}
	\item[$(a1)$] independent case: $A^* \sim \mathcal{U} (0,a_0)$;
	\item[$(a2)$] dependent case :  $A^* = a_0 B_1 I(X_1+X_2<0) + a_0 B_2 I(X_1+X_2 \geq 0)$ with $B_1 \sim \mathcal{B}(1,2)$ and $B_2 \sim \mathcal{B}(2,1)$.
\end{enumerate}
Lastly, 
let $C=\min( C^*, \tau)$ be the censoring time for $T$, 
where $\tau=30$ is the endpoint of the study and 
$C^*$ is generated from uniform distribution $\mathcal{U}(0,100)$.
Two sample sizes ($n=600$ and $1000$) are considered.

In total, we generated the data in 36 scenarios. 
Among these scenarios, 
the censoring rate of event time $T$ ranges from $10.5\%$ to $44.2 \%$,
and the observable rate of treatment initiation time $A^*$ (i.e. $A^* \leq T $ and $A^* \leq C$) ranges from
$61.7\%$ to $85.4\%$. 
For each scenario, 
we applied the proposed method to estimate  $\bm{\beta}^{opt} = (\beta_1, \beta_2, \beta_3)$ and $V_0=V(d^{opt})= M(\bm{\beta}^{opt})$, 
and calculated the variance of $\bm{\hat \beta}^{opt}$ and $\hat M_n(\bm{\hat \beta}^{opt})$ by bootstrapped samples. In our implementation, we take monotonic transform function $g(u) = \Phi^{-1}(u/a_0)$,
where $\Phi$ denotes the cumulative distribution function of the standard normal distribution. Since $X_1$ is a binary covariate, we use the indicator function instead of a kernel for stratification. 
For continuous variables, $X_2$ and $A$,  
a Gaussian kernel is used, and 
the bandwidth is selected as 
$h_1= \gamma_1 n^{-1/5} \rm{sd}( X_2)$ and 
$h_2= \gamma_2 n^{-1/5} \rm{sd}(A)$,
where $\gamma$ is a constant,
$\rm{sd}(X_2)$ is the sample standard deviation of $X_2$ and 
$\rm{sd}(A)$ is the sample standard deviation of the observed treatment initiation time. 
In our numerical studies, 
$\gamma_1 = \gamma_2 = 1$ generally gives good results for all scenarios. 
For a better performance, 
$(\gamma_1, \gamma_2)$ can also be selected 
by a cross-validation procedure.
For example, 
we can divide the data into $K$ equal sized subsamples 
and consider a finite set of candidate values for $(\gamma_1, \gamma_2)$.
For each pair $(\gamma_1, \gamma_2)$, 
let $\hat d^{opt}_{-k}$ $(1\leq k \leq K)$ be the estimated OTIR obtained under  
$K-1$ subsamples excluding the $k$th one,
and 
let $\hat V_k (\hat d^{opt}_k)$ be the estimated value of $\hat d^{opt}_{-k}$ where $\hat V_k$ is obtained under the $k$th subsample.
Then by searching over the candidate values of $(\gamma_1, \gamma_2)$, 
we can select the tuning parameter as the pair maximizing $K^{-1} \sum_{k =1}^K \hat V_k (\hat d^{opt}_k)$.
To do the optimization, 
we adopted the Nelder-Mead algorithm \citep{nelder1965simplex},
which can be implemented by the R function $optim$. All the initial values are set as zero.
Since kernel estimation and bootstrapping could be computationally expensive, 
the simulations were carried out  with an  AMD EPYC 7452 32-Core processor, 
and processing of one data set with $n = 600$ samples takes about 115-155 seconds.

Table $\ref{tab1}$-$\ref{tab3}$ summarizes the simulation results for $\bm{\hat \beta}^{opt}$ and $\hat M_n(\bm{\hat \beta}^{opt})$ based on 500 replications. 
For each scenario, 
we report 
the censoring rate of event time ($CR$), 
the observable rate  of treatment initiation time ($OR$),
the bias of the estimators ($Bias$), 
the standard deviation of the estimators ($SD$), 
the mean of estimated standard errors ($SE$), 
and the empirical coverage probability of $95\%$ confidence intervals ($CP$).
We also reported the true values of $V_0$ for each scenario in the parentheses.
From the results, we can see that under all cases,
the proposed estimators for $\bm{\beta}^{opt}$ and $V_0$ are nearly unbiased, 
and the estimated standard errors are close to the standard deviation of the estimators.
Moreover, 
the empirical coverage probabilities of $95\%$ confidence intervals are close to the nominal level for both parameters and value estimators.  
Both bias and standard deviation of the estimators get smaller 
when the sample size increases from $n=600$ to $n=1000$ as expected.

To demonstrate the effect of the proposed treatment initiation regimes on individual level, 
Table $\ref{tab1}$ also presents the percent of individuals with improved counterfactual outcomes  
under the optimal individualized treatment initiation regime (PI$_i$), 
and under the optimal constant treatment initiation regime (PI$_c$).
From the results, it can be concluded that 
around $78.5\%-89.6\%$ of the individuals would achieve better outcomes  if  $\hat d^{opt}$ had been followed by the entire population, 
and 
the optimal individualized treatment initiation regime $\hat d^{opt} $ does perform better than the optimal constant regime.

\section{Application}
\label{sec:app}

Now we apply our method to the breast cancer dataset linked between the SCCCR and the South Carolina RFA 
and aim to choose the optimal initiation regime of adjuvant therapy for these breast cancer patients.  
Let $A$ be the initiation time (in days) of the adjuvant chemotherapy or radiotherapy since surgery and let $T$ be the patients' survival time (in days). 
For the selection of $a_0$, 
since patients are usually recommended to start adjuvant therapy within 24 weeks from the surgery \citep{Lohrisch2006impact}, we set $a_0 = 168$ (in days),
and exclude patients who started the adjuvant therapy after 24 weeks from the surgery.
As shown in Figure \ref{fig:Distri of A},  the majority of patients did start adjuvant therapy within 168 days since surgery. 
Also, since $pr(C \geq \tau)>0$ is required to ensure the uniform convergence of $\hat S_C(t)$ (as discussed in section \ref{sec:prop}),  we choose $\tau = 3720$ such that about $85\%$ of patients'  event times (censoring or failure) are less than $\tau$.
In our analysis,  we consider two covariates $\bm{X}=(X_1, X_2)$, where $X_1$ is the age of the patient at surgery standardized to mean 0 and variance 1, 
and $X_2$ is an indicator of breast cancer stage.
Specifically, we define $X_2=0$  for patients with breast cancer stage 0 or stage 1 (localized only); and define $X_2 = 1$ for  patients with breast cancer stages 2, 3, 4 or 7. 
In medical practice, stages 0 and 1 are earlier stages of breast cancer.  Thus, we refer $X_2=0$ as earlier stage and $X_2=1$ as advanced stage. 
Since $X_1$ and $A$ are continuous,  
a Gaussian kernel is used in the estimation procedure, and 
the kernel bandwidths are selected as 
$h_1 =  n^{-1/5} \rm{sd}( X_1)$ and 
$h_2 =  n^{-1/5} \rm{sd}(A)$ separately. 
For a better performance, it would also be worthwhile to develop some data-driven bandwidth selection algorithms based on the empirical bias bandwidth selection (EBBS) method. 

\begin{figure}
	\graphicspath{{figures/}}
	\begin{center}
		\includegraphics[width= 12 cm, height=6cm]{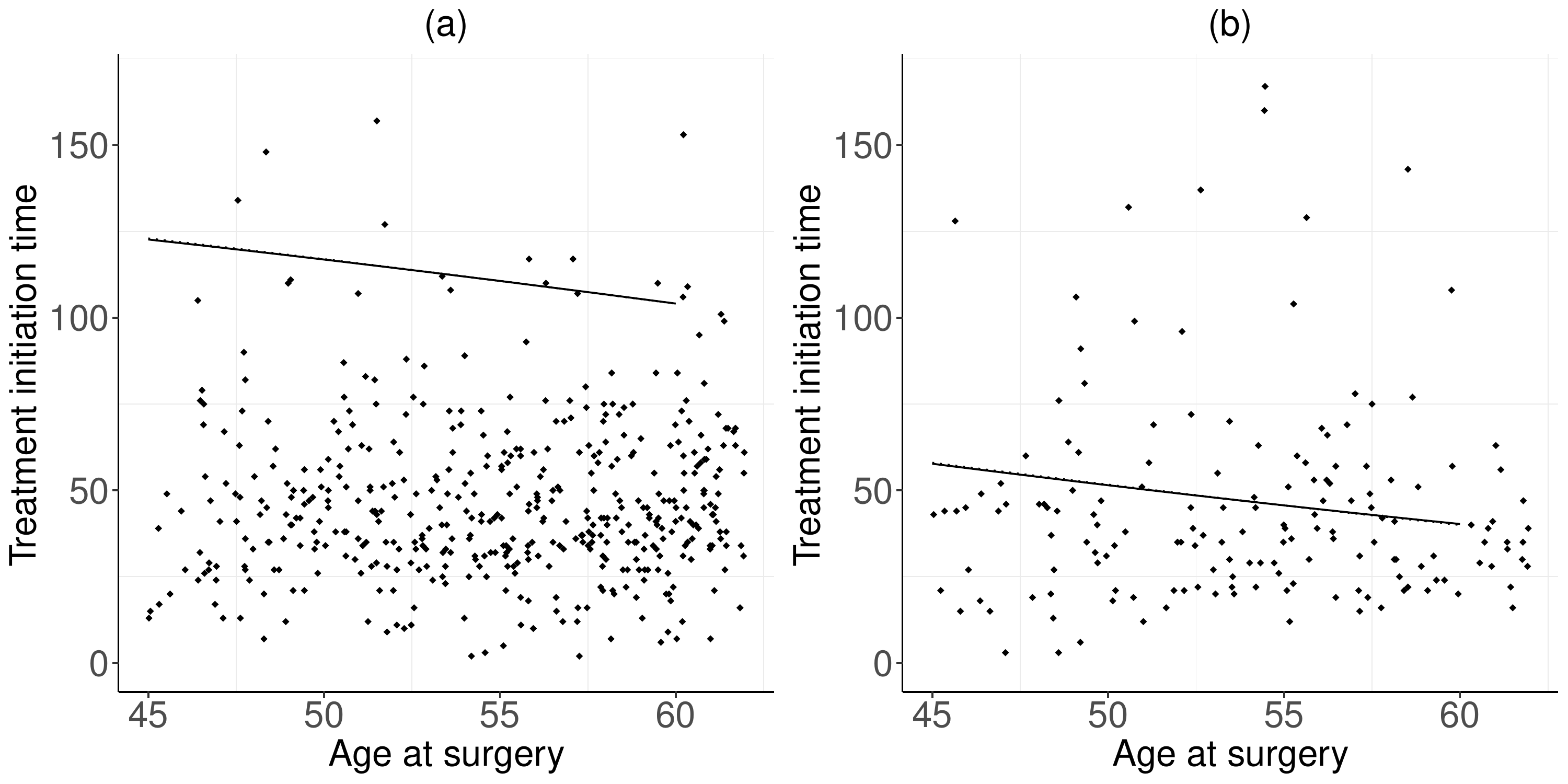} \\[2mm]
		\includegraphics[width=12 cm, height= 6 cm]{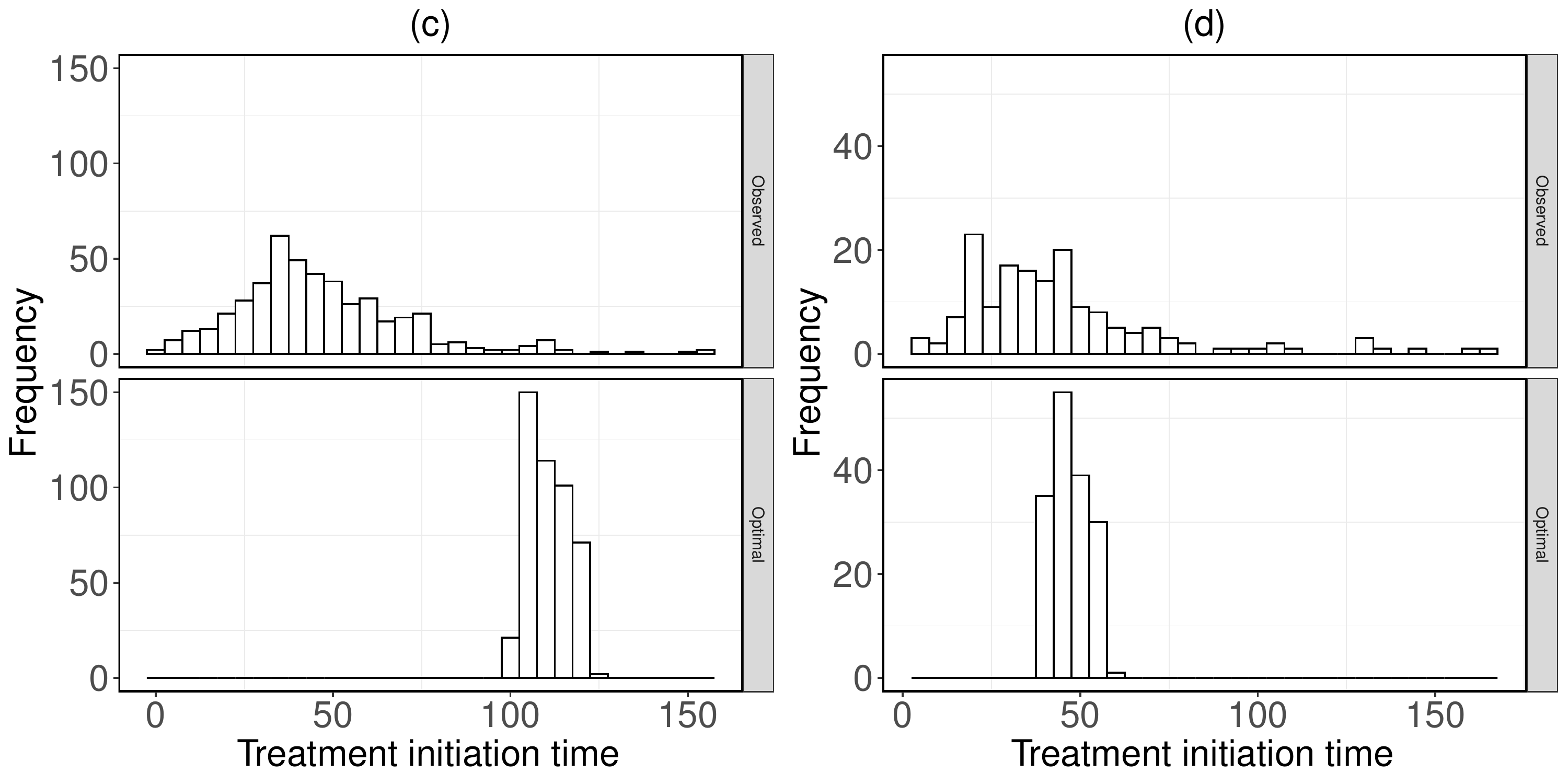}\\
	\end{center}
	\caption{
		The optimal treatment initiation time 
		estimated based on the logistic link function (solid line), 
		and the normal link function (dashed line), and 
		the observed treatment initiation time (points) for patients with (a) earlier stages and (b) advanced stages.
		The distribution of observed treatment initiation times (upper panel) and 
		estimated optimal treatment initiation times (lower panel)
		for patients with (c) earlier stages and (d) advanced stages.}
	\label{fig:app_est}   
\end{figure}

Figures  \ref{fig:app_est}(a) and \ref{fig:app_est}(b) present the estimated optimal treatment initiation regime
among two classes of decision rules:
\begin{align*}
\mathcal D_{\text{logistic}} 
=&
\{ d_{\beta}: d_{\beta}( \bm{x})=  \frac{ a_0 \exp( \bm{\tilde x}^ T \bm{\beta})}{ 1+ \exp( \bm{\tilde x}^ T \bm{\beta}) }, x  \in \mathbb{R}\times\{0,1\}, \bm{\beta} \in \mathbb{R}^3 \}
\\
\mathcal D_{\text{normal}}  =&
\{ d_{\beta}: d_{\beta}( \bm{x})= a_0 \Phi( \bm{\tilde x}^ T \bm{\beta}) , x   \in \mathbb{R}\times\{0,1\}, \bm{\beta} \in \mathbb{R}^3 \} \,.
\end{align*}
It can be seen from the plots that the optimal treatment initiation regimes obtained based on the 
logistic link function and the normal link function
recommend similar treatment initiation times for patients. This shows certain robustness of the proposed method to the choice of link functions in the considered treatment initiation regimes. 

For simplicity, from now on we only present analysis results under the logistic link function.
By calculation, 
the estimate of $\beta$ is $\bm{\hat \beta}^{opt}=(0.669,- 0.154, -1.643)$ 
with the standard error $SE = (0.288, 0.163, 0.476)$,  
and the $p$-values are $0.020$, $0.348$ and $0.001$, respectively.
Further illustration is presented in Figure \ref{fig:app_est}.
Specifically, 
Figure \ref{fig:app_est}$(a)$ and \ref{fig:app_est}$(b)$ plot the observed treatment initiation times $A_i$ and the estimated optimal treatment initiation times $\hat{d}^{opt}(\bm{X}_i)$ over patients' age. 
As shown in the plots, 
although there is no significant difference in the observed treatment initiation times among patients with different ages, 
the estimated optimal treatment initiation regime suggests moderate delay  in treatment initiation for younger patients. 
Figure \ref{fig:app_est}$(c)$ and \ref{fig:app_est}$(d)$ 
compare the distribution of the observed treatment initiation times 
and the distribution of the estimated optimal treatment initiation times 
for patients with earlier stages 
and advanced stages. 
It can be concluded from the plots that, 
according to the estimated optimal treatment initiation regime, patients with advanced breast cancer stages ($X_2 = 1$) should initiate adjuvant therapy
earlier than those with earlier breast cancer stages. 
{
All these results are consistent with the $p$-values.
}


We have also compared the performance of the estimated optimal treatment initiation regime $\hat{d}^{opt}$ with that of the observed treatment initiation regime $d_A(\bm{X}_i) = A_i$.
On one hand, 
we let $\hat{V}(d)$ denote the estimated value function under a given regime $d$. Using the proposed kernel estimation method, we can calculate $\hat{V}(\hat{d}^{opt})  = 4134.528$ and $\hat{V}(d_{A}) = 3408.576$. Therefore,  
the increase in the value function comparing the  estimated optimal treatment initiation regime with the observed treatment initiation time is $\hat{V}_{\text{diff}} = \hat{V}(\hat{d}^{opt}) - \hat{V}(d_{A}) = 725.952$, 
 which suggests a nearly 2-year improvement in expected overall restricted survival time for breast cancer patients when patients follow $\hat{d}^{opt}$. 
 We can further  obtain the empirical distribution of $\hat{V}_{\text{diff}}$ 
by bootstrapping samples for 500 times.
Specifically, 
for each bootstrapped sample, 
we calculate $\hat{V}^*$, $\hat{d}^{opt*}$, and  let  $\hat{V}_{\text{diff}}^* = \hat{V}^*(\hat{d}^{opt*}) - \hat{V}^*(d_{A}^*)$.
Then based on the 500 bootstrapped $\hat{V}_{\text{diff}}^*$, 
we can obtain a quantile-based $95\%$ confidence interval of $\hat{V}_{\text{diff}}$ as $(308.128, 1489.948)$.
This again suggests a significant improvement in the overall restricted mean survival time.
On the other hand, 
 to compare the performances of  $\hat{d}^{opt}$ and  $d_A$ on individual level, 
 we calculate the percent of individuals with improved counterfactual outcomes under $\hat{d}^{opt}$ compared to $d_A$. 
 The obtained PI$_i$ is $0.798$,  which indicates that $79.8\%$ of the breast cancer patients would achieve larger restricted mean survival time if $\hat d^{opt}$ had been followed by 
 all the breast cancer patients.

\begin{figure}
	\graphicspath{{figures/}}
	\begin{center}
		\includegraphics[width=10 cm, height=6cm]{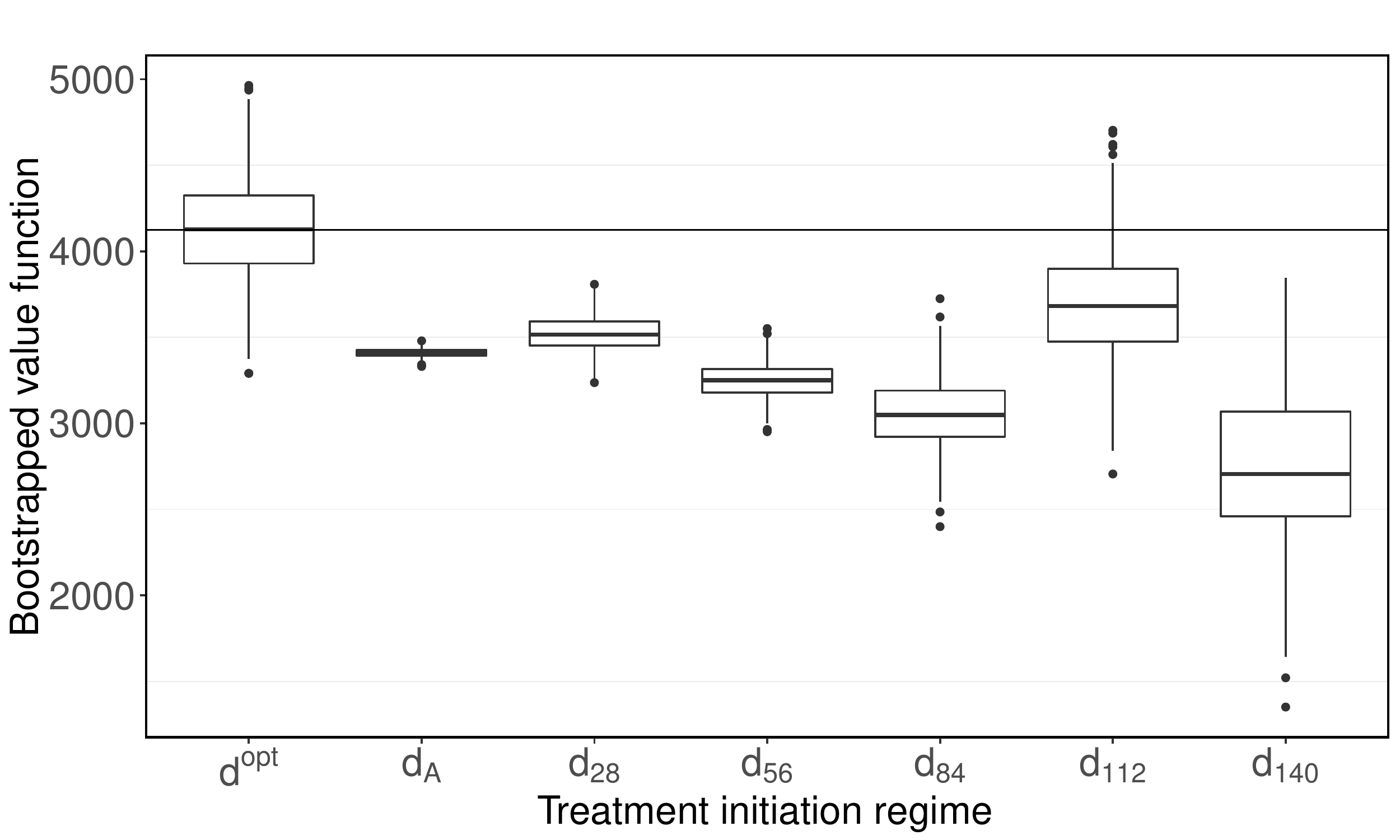}
	\end{center}
	\caption{
		Distribution of the estimated value function $\hat{V}(d)$ under treatment initiation regimes $\hat{d}^{opt}$, $d_A$ and $d_a$ for $a = 28, 56, 84, 112, 140$.
	}
	\label{fig:com_boxplot}
\end{figure}

Lastly, we compare the derived
optimal treatment initiation regime with
some fixed treatment initiation regimes $d_a(\bm{x}) \equiv a$ where $a$ is a constant taking value in $[0,a_0]$.
Specifically, given a  regime $d$, 
we calculate the value functions $\hat{V}^*(d)$ based on 500 bootstrapped samples,
and compare the empirical distributions of $\hat{V}^*(d)$ under
$\hat{d}^{opt}$,  
$d_{28}(\bm{x}) \equiv 28$, 
$d_{56}(\bm{x}) \equiv 56$,   
$d_{84}(\bm{x}) \equiv 84$, 
$d_{112}(\bm{x}) \equiv 112$, 
and 
$d_{140}(\bm{x}) \equiv 140$. 
For completeness, we also include the results for $d_A$. 
The results are plotted in Figure \ref{fig:com_boxplot}.
Based on the plot, the values of the estimated optimal treatment initiation regime $\hat{d}^{opt}$ usually are much larger than those under
$d_a$ and $d_A$, which indicates that 
patients following $\hat{d}^{opt}$ tend to have better treatment effect than those following constant regime $d_a$ for $a \in [0,a_0]$.

\section{Discussion}
\label{sec:dis}

In this article, we 
presented a formulation of treatment initiation time decision problem 
and proposed a new value search approach  
to find the optimal individualized treatment initiation time regime for censored time-to-event data.
Different from existing value search methods,
our value function is constructed on the restricted mean residual lifetime at the endpoint of treatment initiation interval $a_0$.
The proposed value function can be estimated consistently even when the treatment initiation times are not completely observable, and their distribution is unknown.
As a matter of fact, if the value function is constructed based on the restricted mean survival time or restricted mean residual lifetime at some time point $t < a_0$,
the estimation procedure could be challenging due to the missing in treatment initiation time. 

In the estimation procedure, we focus on patients whose survival times are longer than the maximum treatment initiation time $a_0$.
Such a value function may cause some selection bias when there is a proportion of patients in critical conditions with potentially short life expectancy. 
To deal with this issue, we
propose a refined two-step estimation procedure, 
by first identifying a subset of patients who may be in critical conditions and need to start the treatment early on 
based on their estimated optimal treatment initiation time obtained in the first step.  
More details are given in the Web Appendix of Supplementary Materials.


For simplicity, this paper only considered the covariate-independent censoring case $(A4)$. 
For the case with the conditionally independent censoring given in assumption $(A3)$,
 we may constructing a Cox model on the censoring time $C$ conditional on $X$, 
or we can estimate the  conditional survival function of censoring times by  kernel conditional Kaplan-Meier estimator \citep{dabrowska1989uniform}.
The associated asymptotic properties of the proposed estimators can also be derived 
but will be more involved. 
Also, 
since the  kernel conditional Kaplan-Meier estimator needs $nh_1^{p+4} \rightarrow 0$ 
to obtain the convergence rate $\{ n h_1^{p}/(log h_1^{-p}) \}^{1/2}$ \citep{dabrowska1989uniform}, 
the convergence rate of the OTIR estimator could be slower 
especially when the dimension  of covariates $p$ is high.
Moreover, it is worthwhile considering the cases where the censoring time is  also affected by the assigned treatment initiation time. 
For such cases, if the dependence between the censoring time and the assignment treatment initiation time can be fully captured by the observed covariates, ( i.e. $C \perp A^* | \bm{X}$), 
then our method is still valid.
Otherwise, finding the 
optimal treatment initiation regime would be an open problem that warrants future research.

\bibliographystyle{biom}
\bibliography{OTIT-ref}

\newpage
\renewcommand{\baselinestretch}{1.2}
{\small
	\begin{table} [!ht]
		\tabcolsep 5pt 
		\caption{ \label{tab1} Simulation results  when sample size $n=600$ and $a_0 = 3$.
}
		\vspace*{-12pt}
		\begin{center}
			\def\temptablewidth{1 \textwidth}
			{\rule{\temptablewidth}{1pt}}
			\begin{tabular*}
				{\temptablewidth}{@{\extracolsep{\fill}}cccrccccrccc}
         \hline
         & &  &\multicolumn{4}{c}{$A^*$ independent with $X$} &&
               \multicolumn{4}{c}{$A^*$ dependent with $X$} \\
        \cline{4-7}\cline{9-12}
$model$ &$\lambda_0$ & estimate &  Bias       &    SD     &    SE     &   CP      &&  Bias     &SD&SE&CP\\
				\hline 
$m_1$	  &	$0.2$	& $ \beta_1$	  &  $0.021   $ &  $0.232$  &  $0.215$  &  $0.934$  &&  $0.033$  &  $0.218$  &  $0.216$  &  $0.928$ \\
				&		    &	$\beta_2$	    &  $ 0.064 	$ &  $0.345$  &  $0.342$  &  $0.960$  &&  $0.020 $  &  $0.317$  &  $0.325$  &  $0.966$ \\
				&		    &	$\beta_3$	    &  $ 0.014 	$ &  $0.180$  &  $0.186$  &  $0.970$  &&  $-0.001$  &  $0.171$  &  $0.176$  &  $0.958$ \\
				&		    &	$V_0(19.155)$	&  $ -0.465 $ &  $0.974$  &  $0.971$  &  $0.966$  &&  $-0.667$  &  $0.820$  &  $0.841$  &  $0.968$ \\
        \cline{4-7}\cline{9-12}
  			&       &               &  \multicolumn{4}{c}{CR $:0.172$ \ PI$_{i} :0.881$}&& \multicolumn{4}{c}{CR $:0.208$ \ PI$_{i} :0.856$} \\         
				&       &               &  \multicolumn{4}{c}{OR $:0.741$ \ PI$_{c}:0.789$} && \multicolumn{4}{c}{OR $:0.715$ \ PI$_{c}:0.737$}\\         
\hline				
$m_1$		&	$0.3$	&	$\beta_1$	    &  $ 0.033 	$ &  $0.238$  &  $0.223$  &  $0.922$  &&  $0.033 $  &  $0.221$  &  $0.221$  &  $0.932$ \\
				&		    &	$\beta_2$	    &  $ 0.048 	$ &  $0.347$  &  $0.372$  &  $0.954$  &&  $0.039 $  &  $0.324$  &  $0.347$  &  $0.964$ \\
				&		    &	$\beta_3$	    &  $ 0.003 	$ &  $0.185$  &  $0.193$  &  $0.964$  &&  $-0.026$  &  $0.168$  &  $0.186$  &  $0.968$ \\
				&		    &	$V_0(16.400)$	&  $ -0.476 $ &  $1.060$  &  $1.094$  &  $0.948$  &&  $-0.688$  &  $0.896$  &  $0.948$  &  $0.966$ \\
        \cline{4-7}\cline{9-12}
  			&       &               &  \multicolumn{4}{c}{CR $:0.105$ \ PI$_{i}:0.883$} && \multicolumn{4}{c}{CR $:0.126$ \ PI$_{i}:0.857$} \\         
				&       &               &  \multicolumn{4}{c}{OR $:0.650$ \ PI$_{c}:0.792$} && \multicolumn{4}{c}{OR $:0.617$ \ PI$_{c}:0.744$}\\         
				\hline                                                                                                                    
$m_2$	  &	$0.2$	&	$\beta_1$	    &  $0.027   $ &  $0.234$  &  $0.217$  &  $0.926$  &&  $0.035 $  &  $0.216$  &  $0.214$  &  $0.922$ \\
				&		    &	$\beta_2$	    &  $0.057   $ &  $0.367$  &  $0.345$  &  $0.938$  &&  $0.036 $  &  $0.313$  &  $0.331$  &  $0.950$ \\
				&		    &	$\beta_3$	    &  $0.022   $ &  $0.176$  &  $0.183$  &  $0.970$  &&  $-0.006$  &  $0.166$  &  $0.173$  &  $0.972$ \\
				&		    &	$V_0(22.432)$	&  $-0.228  $ &  $0.857$  &  $0.861$  &  $0.968$  &&  $-0.453$  &  $0.725$  &  $0.745$  &  $0.964$ \\
        \cline{4-7}\cline{9-12}
  			&       &               &  \multicolumn{4}{c}{CR $:0.245$ \ PI$_{i}:0.880$} && \multicolumn{4}{c}{CR $:0.296$ \ PI$_{i}:0.856$} \\         
				&       &               &  \multicolumn{4}{c}{OR $:0.741$ \ PI$_{c}:0.787$} && \multicolumn{4}{c}{OR $:0.715$ \ PI$_{c}:0.734$}\\         
\hline
$m_2$		&	$0.3$	&	$\beta_1$	    &  $0.038   $ &  $0.230$  &  $0.221$  &  $0.920$  &&  $0.036 $  &  $0.228$  &  $0.220$  &  $0.930$ \\
				&		    &	$\beta_2$	    &  $0.056   $ &  $0.355$  &  $0.340$  &  $0.936$  &&  $0.044 $  &  $0.310$  &  $0.328$  &  $0.960$ \\
				&		    &	$\beta_3$	    &  $0.015   $ &  $0.180$  &  $0.184$  &  $0.966$  &&  $-0.024$  &  $0.159$  &  $0.173$  &  $0.960$ \\
				&		    &	$V_0(20.747)$	&  $-0.287  $ &  $0.970$  &  $0.994$  &  $0.970$  &&  $-0.540$  &  $0.845$  &  $0.862$  &  $0.972$ \\
        \cline{4-7}\cline{9-12}
  			&       &               &  \multicolumn{4}{c}{CR $:0.170$ \ PI$_{i}:0.884$}&& \multicolumn{4}{c}{CR $:0.207$ \ PI$_{i}:0.860$} \\         
				&       &               &  \multicolumn{4}{c}{OR $:0.650$ \ PI$_{c}:0.791$}&& \multicolumn{4}{c}{OR $:0.617$ \ PI$_{c}:0.743$}\\         
				\hline                                                                                                                    
$m_3$	  &	$0.2$	&	$\beta_1$	    &  $0.013  $  &  $0.241$  &  $0.233$  &  $0.948$  &&  $0.025 $  &  $0.231$  &  $0.231$  &  $0.956$ \\
				&		    &	$\beta_2$	    &  $0.048  $  &  $0.366$  &  $0.368$  &  $0.962$  &&  $0.022 $  &  $0.349$  &  $0.357$  &  $0.966$ \\
				&		    &	$\beta_3$	    &  $-0.005 $  &  $0.187$  &  $0.208$  &  $0.970$  &&  $-0.037$  &  $0.193$  &  $0.198$  &  $0.946$ \\
				&		    &	$V_0(19.697)$	&  $-0.516 $  &  $0.925$  &  $0.928$  &  $0.950$  &&  $-0.682$  &  $0.792$  &  $0.814$  &  $0.960$ \\
 			         \cline{4-7}\cline{9-12}
        &       &               &  \multicolumn{4}{c}{CR $:0.217$ \ PI$_{i} :0.870$}&& \multicolumn{4}{c}{CR $:0.254$ \ PI$_{i}:0.849$} \\         
				&       &               &  \multicolumn{4}{c}{OR $:0.793$ \ PI$_{c}:0.789$} && \multicolumn{4}{c}{OR $:0.775$ \ PI$_{c}:0.744$}\\         
\hline
$m_3$	  &	$0.3$	&	$\beta_1$	    &  $0.022  $  &  $0.243$  &  $0.229$  &  $0.924$  &&  $0.030 $  &  $0.233$  &  $0.233$  &  $0.952$ \\
				&		    &	$\beta_2$	    &  $0.047  $  &  $0.376$  &  $0.359$  &  $0.960$  &&  $0.014 $  &  $0.350$  &  $0.356$  &  $0.950$ \\
				&		    &	$\beta_3$	    &  $-0.032 $  &  $0.214$  &  $0.217$  &  $0.960$  &&  $-0.079$  &  $0.194$  &  $0.214$  &  $0.942$ \\
				&		    &	$V_0(17.268)$	&  $-0.398 $  &  $1.033$  &  $1.022$  &  $0.960$  &&  $-0.624$  &  $0.890$  &  $0.898$  &  $0.974$ \\
        \cline{4-7}\cline{9-12}
  			&       &               &  \multicolumn{4}{c}{CR $:0.146$ \ PI$_{i} :0.871$} && \multicolumn{4}{c}{CR $:0.170$ \ PI$_{i} :0.851$} \\         
			&       &               &  \multicolumn{4}{c}{OR $:0.717$ \ PI$_{c} :0.793$} && \multicolumn{4}{c}{OR $:0.692$ \ PI$_{c}:0.754$}\\         
 				\hline
			\end{tabular*}
		\end{center}
	\end{table}
}

\newpage
\renewcommand{\baselinestretch}{1.2}
{\small
	\begin{table} [!ht]
		\tabcolsep 5pt 
		\caption{ 
			\label{tab2} Simulation results  when sample size $n=600$ and $a_0 = 2$.}
		\vspace*{-12pt}
		\begin{center}
			\def\temptablewidth{1 \textwidth}
			{\rule{\temptablewidth}{1pt}}
			\begin{tabular*}
				{\temptablewidth}{@{\extracolsep{\fill}}cccrcccrccc}
				\hline
				& & &\multicolumn{4}{c}{$A^*$ independent with $X$} &
				\multicolumn{4}{c}{$A^*$ dependent with $X$} \\
				\cline{4-7}\cline{8-11}
				$model$ &$\lambda_0$& estimate &Bias&SD&SE&CP&Bias&SD&SE&CP\\
				\hline 
				$m_1$	&	$0.2$	&	$\beta_1$	&$	0.019 	$&$	0.389 	$&$	0.358 	$&$	0.940 	$&$	0.049 	$&$	0.384 	$&$	0.387 	$&$	0.970 	$\\
				&		&	$\beta_2$	&$	0.046 	$&$	0.581 	$&$	0.583 	$&$	0.956 	$&$	-0.015 	$&$	0.541 	$&$	0.588 	$&$	0.968 	$\\
				&		&	$\beta_3$	&$	0.031 	$&$	0.301 	$&$	0.294 	$&$	0.968 	$&$	-0.030 	$&$	0.293 	$&$	0.290 	$&$	0.942 	$\\
				&		&	$V_0(19.631)$	&$	0.054 	$&$	0.915 	$&$	0.924 	$&$	0.962 	$&$	0.013 	$&$	0.814 	$&$	0.821 	$&$	0.952 	$\\[2mm]
				&	$0.3$	&	$\beta_1$	&$	0.033 	$&$	0.353 	$&$	0.329 	$&$	0.938 	$&$	0.060 	$&$	0.376 	$&$	0.360 	$&$	0.962 	$\\
				&		&	$\beta_2$	&$	0.053 	$&$	0.523 	$&$	0.575 	$&$	0.956 	$&$	0.011 	$&$	0.544 	$&$	0.557 	$&$	0.960 	$\\
				&		&	$\beta_3$	&$	0.017 	$&$	0.279 	$&$	0.281 	$&$	0.962 	$&$	-0.019 	$&$	0.285 	$&$	0.284 	$&$	0.944 	$\\
				&		&	$V_0(16.728)$	&$	0.007 	$&$	0.986 	$&$	0.991 	$&$	0.940 	$&$	-0.073 	$&$	0.813 	$&$	0.873 	$&$	0.968 	$\\
				\hline
				$m_2$	&	$0.2$	&	$\beta_1$	&$	0.021 	$&$	0.378 	$&$	0.352 	$&$	0.938 	$&$	0.062 	$&$	0.371 	$&$	0.381 	$&$	0.966 	$\\
				&		&	$\beta_2$	&$	0.016 	$&$	0.592 	$&$	0.569 	$&$	0.950 	$&$	-0.006 	$&$	0.575 	$&$	0.613 	$&$	0.966 	$\\
				&		&	$\beta_3$	&$	0.025 	$&$	0.314 	$&$	0.296 	$&$	0.956 	$&$	-0.036 	$&$	0.305 	$&$	0.295 	$&$	0.944 	$\\
				&		&	$V_0(23.121)$	&$	0.231 	$&$	0.786 	$&$	0.825 	$&$	0.940 	$&$	0.114 	$&$	0.712 	$&$	0.742 	$&$	0.946 	$\\[2mm]
				&	$0.3$	&	$\beta_1$	&$	0.046 	$&$	0.362 	$&$	0.333 	$&$	0.926 	$&$	0.060 	$&$	0.365 	$&$	0.359 	$&$	0.954 	$\\
				&		&	$\beta_2$	&$	0.004 	$&$	0.540 	$&$	0.552 	$&$	0.954 	$&$	0.019 	$&$	0.554 	$&$	0.547 	$&$	0.960 	$\\
				&		&	$\beta_3$	&$	0.031 	$&$	0.278 	$&$	0.277 	$&$	0.966 	$&$	-0.024 	$&$	0.299 	$&$	0.278 	$&$	0.948 	$\\
				&		&	$V_0(21.340)$	&$	0.152 	$&$	0.931 	$&$	0.908 	$&$	0.936 	$&$	0.043 	$&$	0.767 	$&$	0.806 	$&$	0.962 	$\\
				\hline 
				$m_3$	&	$0.2$	&	$\beta_1$	&$	-0.017 	$&$	0.406 	$&$	0.386 	$&$	0.944 	$&$	0.017 	$&$	0.392 	$&$	0.420 	$&$	0.956 	$\\
				&		&	$\beta_2$	&$	0.010 	$&$	0.646 	$&$	0.641 	$&$	0.956 	$&$	-0.015 	$&$	0.581 	$&$	0.652 	$&$	0.976 	$\\
				&		&	$\beta_3$	&$	-0.001 	$&$	0.330 	$&$	0.317 	$&$	0.962 	$&$	-0.101 	$&$	0.291 	$&$	0.317 	$&$	0.950 	$\\
				&		&	$V_0(20.219)$	&$	0.003 	$&$	0.871 	$&$	0.901 	$&$	0.962 	$&$	-0.092 	$&$	0.788 	$&$	0.811 	$&$	0.952 	$\\[2mm]
				&	$0.3$	&	$\beta_1$	&$	0.047 	$&$	0.388 	$&$	0.353 	$&$	0.938 	$&$	0.051 	$&$	0.384 	$&$	0.383 	$&$	0.958 	$\\
				&		&	$\beta_2$	&$	0.004 	$&$	0.548 	$&$	0.582 	$&$	0.956 	$&$	-0.001 	$&$	0.578 	$&$	0.586 	$&$	0.958 	$\\
				&		&	$\beta_3$	&$	-0.021 	$&$	0.300 	$&$	0.310 	$&$	0.982 	$&$	-0.088 	$&$	0.301 	$&$	0.311 	$&$	0.936 	$\\
				&		&	$V_0(17.662)$	&$	-0.030 	$&$	0.951 	$&$	0.948 	$&$	0.950 	$&$	-0.085 	$&$	0.811 	$&$	0.853 	$&$	0.974 	$\\
				\hline
			\end{tabular*}
		\end{center}
	\end{table}
}

\newpage
\renewcommand{\baselinestretch}{1.2}
{\small
	\begin{table} [!ht]
		\tabcolsep 5pt 
		\caption{ 
			\label{tab3} Simulation results  when sample size $n=1000$ and $a_0 = 3$.}
\begin{center}
\def\temptablewidth{1 \textwidth}
{\rule{\temptablewidth}{1pt}}
\begin{tabular*}
{\temptablewidth}{@{\extracolsep{\fill}}cccrcccrccc}
\hline
& & &\multicolumn{4}{c}{$A^*$ independent with $X$} &
\multicolumn{4}{c}{$A^*$ dependent with $X$} \\
\cline{4-7}\cline{8-11}
$model$ &$\lambda_0$& estimate &Bias&SD&SE&CP&Bias&SD&SE&CP\\
\hline 
$m_1$	&	$0.2$	&	$\beta_1$	&$	0.020 	$&$	0.190 	$&$	0.188 	$&$	0.926 	$&$	0.025 	$&$	0.176 	$&$	0.179 	$&$	0.948 	$\\
	&		&	$\beta_2$	&$	0.049 	$&$	0.307 	$&$	0.296 	$&$	0.952 	$&$	0.029 	$&$	0.248 	$&$	0.269 	$&$	0.960 	$\\
	&		&	$\beta_3$	&$	0.018 	$&$	0.164 	$&$	0.161 	$&$	0.948 	$&$	-0.006 	$&$	0.142 	$&$	0.145 	$&$	0.972 	$\\
	&		&	$V_0(19.155)$	&$	-0.387 	$&$	0.752 	$&$	0.791 	$&$	0.972 	$&$	-0.546 	$&$	0.674 	$&$	0.680 	$&$	0.952 	$\\[2mm]
	&	$0.3$	&	$\beta_1$	&$	0.022 	$&$	0.199 	$&$	0.189 	$&$	0.940 	$&$	0.037 	$&$	0.180 	$&$	0.179 	$&$	0.924 	$\\
	&		&	$\beta_2$	&$	0.055 	$&$	0.304 	$&$	0.298 	$&$	0.966 	$&$	0.013 	$&$	0.257 	$&$	0.278 	$&$	0.954 	$\\
	&		&	$\beta_3$	&$	0.012 	$&$	0.170 	$&$	0.164 	$&$	0.952 	$&$	-0.022 	$&$	0.141 	$&$	0.149 	$&$	0.960 	$\\
	&		&	$V_0(16.400)$	&$	-0.392 	$&$	0.864 	$&$	0.894 	$&$	0.962 	$&$	-0.598 	$&$	0.728 	$&$	0.759 	$&$	0.952 	$\\[2mm]
\hline 
$m_2$	&	$0.2$	&	$\beta_1$	&$	0.022 	$&$	0.196 	$&$	0.188 	$&$	0.916 	$&$	0.026 	$&$	0.177 	$&$	0.179 	$&$	0.948 	$\\
	&		&	$\beta_2$	&$	0.037 	$&$	0.324 	$&$	0.296 	$&$	0.930 	$&$	0.031 	$&$	0.270 	$&$	0.274 	$&$	0.970 	$\\
	&		&	$\beta_3$	&$	0.019 	$&$	0.163 	$&$	0.161 	$&$	0.948 	$&$	-0.002 	$&$	0.139 	$&$	0.146 	$&$	0.972 	$\\
	&		&	$V_0(22.432)$	&$	-0.164 	$&$	0.691 	$&$	0.696 	$&$	0.972 	$&$	-0.372 	$&$	0.571 	$&$	0.601 	$&$	0.972 	$\\[2mm]
	&	$0.3$	&	$\beta_1$	&$	0.024 	$&$	0.199 	$&$	0.191 	$&$	0.936 	$&$	0.038 	$&$	0.184 	$&$	0.180 	$&$	0.926 	$\\
	&		&	$\beta_2$	&$	0.049 	$&$	0.314 	$&$	0.289 	$&$	0.940 	$&$	0.033 	$&$	0.253 	$&$	0.266 	$&$	0.964 	$\\
	&		&	$\beta_3$	&$	0.006 	$&$	0.163 	$&$	0.158 	$&$	0.956 	$&$	-0.009 	$&$	0.134 	$&$	0.141 	$&$	0.970 	$\\
	&		&	$V_0(20.747)$	&$	-0.205 	$&$	0.814 	$&$	0.800 	$&$	0.966 	$&$	-0.455 	$&$	0.686 	$&$	0.691 	$&$	0.966 	$\\[2mm]
\hline
$m_3$	&	$0.2$	&	$\beta_1$	&$	0.004 	$&$	0.195 	$&$	0.201 	$&$	0.964 	$&$	0.007 	$&$	0.188 	$&$	0.192 	$&$	0.958 	$\\
	&		&	$\beta_2$	&$	0.027 	$&$	0.323 	$&$	0.313 	$&$	0.952 	$&$	0.025 	$&$	0.273 	$&$	0.291 	$&$	0.958 	$\\
	&		&	$\beta_3$	&$	-0.007 	$&$	0.161 	$&$	0.172 	$&$	0.956 	$&$	-0.039 	$&$	0.147 	$&$	0.158 	$&$	0.944 	$\\
	&		&	$V_0(19.697)$	&$	-0.526 	$&$	0.715 	$&$	0.749 	$&$	0.960 	$&$	-0.635 	$&$	0.636 	$&$	0.654 	$&$	0.936 	$\\[2mm]
	&	$0.3$	&	$\beta_1$	&$	0.011 	$&$	0.207 	$&$	0.195 	$&$	0.934 	$&$	0.017 	$&$	0.191 	$&$	0.191 	$&$	0.938 	$\\
	&		&	$\beta_2$	&$	0.040 	$&$	0.313 	$&$	0.305 	$&$	0.954 	$&$	0.016 	$&$	0.281 	$&$	0.288 	$&$	0.962 	$\\
	&		&	$\beta_3$	&$	-0.024 	$&$	0.186 	$&$	0.180 	$&$	0.956 	$&$	-0.061 	$&$	0.162 	$&$	0.171 	$&$	0.930 	$\\
	&		&	$V_0(17.268)$	&$	-0.441 	$&$	0.809 	$&$	0.828 	$&$	0.966 	$&$	-0.618 	$&$	0.684 	$&$	0.722 	$&$	0.950 	$\\
\hline
\end{tabular*}
\end{center}
\end{table}
}

\newpage
\renewcommand{\baselinestretch}{1.2}
{\small
	\begin{table} [!ht]
		\tabcolsep 5pt \caption{ 
			\label{tab4} Estimated optimal treatment initiation regime for breast cancer data.}
		\vspace*{-12pt}
		\begin{center}
			\def\temptablewidth{0.6 \textwidth}
			{\rule{\temptablewidth}{1pt}}
			\begin{tabular*}
				{\temptablewidth}{@{\extracolsep{\fill}}cccc}
				\hline
				& $\beta_1$(Intercept) & $\beta_2$(Age) & $\beta_3$(Stage) \\
				\hline
				$Est$ &$0.669$&$-0.154$&$-1.643$\\
				$Sd$ &$0.288$&$0.163$&$0.476$\\
				$P value$&$0.020$&$0.348$&$0.001$\\
				\hline
			\end{tabular*}
		\end{center}
	\end{table}
}

\appendix
\newpage
This supplementary material is organized as follows.

In Web Appendix A, we present some discussions. 
Specifically, 
Web Appendix A1 presents an example model that motivated us to construct the value function based on restricted mean residual lifetime.
Web Appendix A2 shows the calculation procedure of PI$_i$ and PI$_c$ in the simulation and application studies;
Web Appendix A3 discusses how to develop a data-driven bandwidth selection algorithm based on  the empirical bias bandwidth selection (EBBS) method;
Web Appendix A4 discusses how to estimate OTIR  for other possible value functions;
Web Appendix A5 proposes a refined OTIT for patients in critical conditions.

In Web Appendix B, we present the technical proofs for the asymptotic properties, 
including the proofs of Theorems 1-3 and the proofs of Lemmas 1-4.

Additional simulation results are given in Web Appendix C.

\section*{Web Appendix A:  Some Discussions}

\section*{Web Appendix A1. An example}

In this section, we will provide an example, under which  
the optimal treatment initiation regime $d^{opt}$ defined in Section 3.1  
does select the optimal treatment initiation time.
Consider 
a class of hazard models for the potential survival time $T^*(a)$ conditional on covariate $\bm{X}=\bm{x}$:
\begin{equation}
\lambda(t; a, \bm{x})= \lambda_0(t) \exp\left[   \mu_0(\bm{x})+  I(t \geq a)Q\{ a- d_0(\bm{x})\} H_0(\bm{x}) \right], 
\label{model}
\end{equation}
where
$\lambda_0(\cdot)$ is the baseline hazard function, $\mu_0(\cdot) $ is an unspecified function for baseline covariate effects, 
$H_0(\cdot)$ is an unspecified non-negative function,
$d_0(\bm{x}) \in [0, a_0]$ is a given function of $\bm{x}$,
and
\textcolor{blue}{$Q(\cdot)$ is an unspecified  differentiable function with a unique minimum value $Q(0)<0$}.
This model indicates that, 
given covariate $\bm{x}$, the ratio between 
hazard rate of patients who are receiving treatment at time $t$ 
and that of patients who have not started treatment at time $t$ 
equals to $\exp\left[ Q\{ a- d_0(\bm{x}) \} H_0(\bm{x})\right] $ 
and is minimized at $a= d_0(\bm{x})$.
Thus, $ d_0(\bm{x})$ is the optimal treatment initiation time for patient with covariate $\bm{X}=\bm{x}$ 
in the sense that it leads to the largest reduction in the patient's hazard rate after treatment.

On the other hand, we claim that $d_0$ is also the maximizer of the proposed value function $V(d)$. 
A rigorous proof of this result is provided later, 
and here for  briefness, we provide an intuitive interpretation. 
Consider a patient with covariate value $\bm{x}$ and baseline hazards $\lambda_0(t) = \lambda_0$, $\mu_0(\bm{x}) = 0$ and $H_0(\bm{x}) = 1$. 
Let the dotted line in Figure \ref{Sect22} represent the mapping   
$ t \rightarrow \lambda_0 \exp\left[    Q\{ t - d_0(\bm{x})\} \right]$ on $[0,a_0]$.
Then, the solid line shows the hazard rate of the patient if the treatment is initiated at $a$,
and the dashed line shows the hazard rate of the patient if the treatment is initiated at $d_0(\bm{x})$.
Furthermore, let $S(a; \bm{x})$ 
denote the area of shadow part when $a$ ranges from $0$ to $a_0$. 
By calculation, the restricted mean residual lifetime at $a_0$ satisfies  
$ m(a, \bm{x}) = (\tau - a_0) [1- \exp\{ - S(a; \bm{x}) \} ]/ S(a; \bm{x})$, 
which implies that the optimal treatment initiation time which maximizes  $m(a, \bm{x})$ also minimizes the area of shadow part. 
From Figure \ref{Sect22}, it can be seen that the area of shadow part is minimized when $a = d_0(\bm{x})$.
Therefore, the proposed value function, which is constructed on the mean residual lifetime, is  maximized at $d_0$.

\begin{figure}
	\graphicspath{{figures/}}
	\begin{center}
	\includegraphics[width= 12 cm, height= 6 cm]{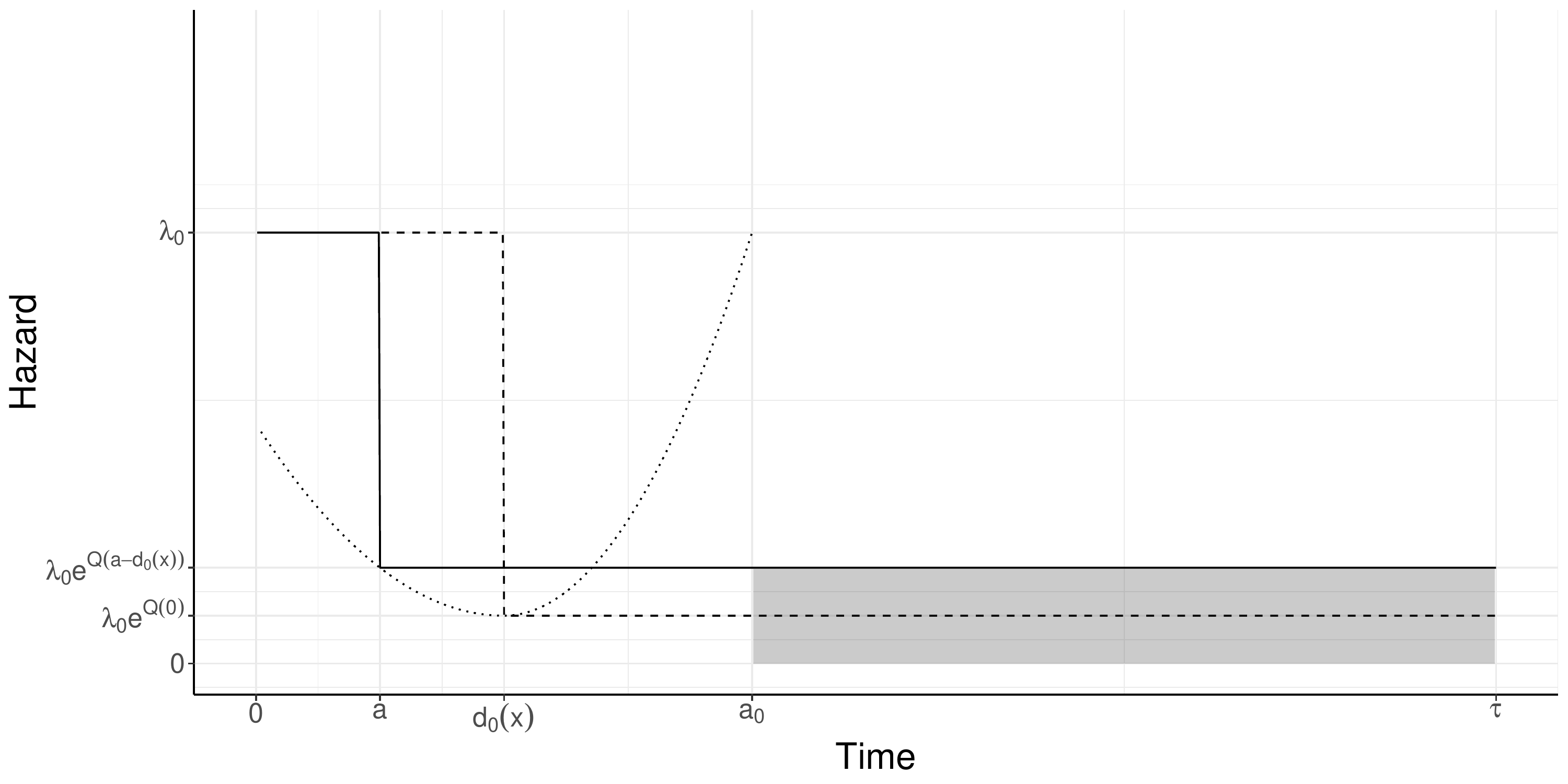}
	\end{center}
	\caption{ 
	The effect of treatment initiation time $a$ on the restricted mean residual lifetime $m(a, \bm{x})$ for a patient with covariate $\bm{X} = \bm{x}$ when 
	$\lambda_0(t) = \lambda_0$, $\mu_0(\bm{x}) = 0$ and $H_0(\bm{x}) = 1$.
         The hazard function $\lambda(t; a, \bm{x})$ is plotted in solid line,
         and the value of restricted mean residual lifetime $m(a, \bm{x})$ 
	 is monotone increasing with the area of shadow part. 
		\label{Sect22}}
\end{figure}

Besides, we noted that, 
although the restricted mean survival time has been widely used
to evaluate treatment effect in survival analysis. 
Here under model $(\ref{model})$, 
the maximizer of the restricted mean survival time 
does not equal to $d_0$.
For an intuitive interpretation, 
we still take Figure \ref{Sect22} as an example. 
Since the area under the solid curve on $[0,t]$  
represents the cumulative hazard $\Lambda(t; a, \bm{x})$ 
and that under the dashed curve represents $\Lambda\{ t; d_0(\bm{x}), \bm{x}\}$, 
it is not hard to obtain from Figure \ref{Sect22} that, 
if $a$ satisfies $Q\{a - d_0(\bm{x})\}<0$ and $a < d_0(\bm{x})$, 
we have $\Lambda(t; a, \bm{x}) < \Lambda\{ t; d_0(\bm{x}), \bm{x}\}$ for all $a < t \leq d_0(\bm{x}) $.  
Moreover, 
if the difference between $Q\{a - d_0(\bm{x})\}$ and $Q(0)-Q\{a - d_0(\bm{x})\}$ is large enough, 
the inequality may hold for all $ a < t \leq \tau$, 
and thus,  
the restricted mean survival time 
$\int_0^{\tau} \exp\{- \Lambda(t; a, \bm{x}) \} d t $  is not maximized at $a = d_0(\bm{x})$. 
In general, it can be proved that for any $0\leq t_0 < a_0$, the maximizer of the restricted mean residual lifetime, $\iint_{t_0}^{\tau} pr[ T^*\{ d(\bm{x})\} \geq t \mid T^*\{ d(\bm{x})\} \geq  t_0,  \bm{X}=\bm{x} ]  dt dF_X(\bm{x})$, does not equal to $d_0$.  

Lastly, we prove that $d_0$ is the maximizer of $V(d)$ under model $(\ref{model})$.
Define $\Lambda_0(t; a, x) = \int_0^t \lambda_0(u; a, x)du$, then 
the cumulative hazard at $ t\geq a$ takes the form
$$
 \Lambda(t; a, x) = \int_0^t \lambda(u; a, x)du
= \Lambda_0(a) \exp\{  \mu_0(x) \}  +
 \{\Lambda_0(t) - \Lambda_0(a) \} 
 \exp[ \mu_0(x) +  Q\{ a- d_0(x)\} H_0(x) ] \,,
$$ 
and the value function of treatment regime $d$ equals to
\begin{align*}
V(d) 
=&
\iint_{a_0}^{\tau} 
\frac{pr[ T^*\{d(x)\} \geq t | X= x ] }{pr[ T^*\{d(x)\} \geq a_0 | X= x ] } f_X(x) dt dx
\\
=&
\iint_{a_0}^{\tau} 
\exp[- \Lambda\{t; d(x),x\} + \Lambda\{a_0; d(x),x\}  ] f_X(x) dt dx
\\
=&
\iint_{a_0}^{\tau} 
\exp\Big(  -  \{ \Lambda_0(t) -\Lambda_0(a_0)\} 
\exp[ \mu_0(x) + Q\{d(x)-d_0(x)\}H_0(x) ]    \Big)
f_X(x) dt dx\,.
\end{align*}
Note that for any $x\in\mathcal{X}$ and $t \in [a_0,\tau]$,
we have 
$ \{ \Lambda_0(t) -\Lambda_0(a_0)\} \geq 0 $,
$H_0(x) \geq 0$  and 
 $Q\{d(x)-d_0(x)\} \geq Q(0) = Q\{d_0(x)-d_0(x)\}$.
Thus for any $d$, 
\begin{align*}
V(d) 
\leq &
\iint_{a_0}^{\tau} 
\exp\Big(  -  \{ \Lambda_0(t) -\Lambda_0(a_0)\} 
\exp[ \mu_0(x) + Q\{d_0(x)-d_0(x)\}H_0(x) ]    \Big)
f_X(x) dt dx
\\
= &
V(d_0)\,,
\end{align*}
which means $V(d)$ achieves its maximum value at $d_0$ under model $(\ref{model})$.

We can also prove by contradiction that $d_0$ is not the maximizer of $V^{t_0}(d)$  for any $0 \leq t_0 < a_0$. 
 For simplicity, we assume that $Q$ is a differentiable function  with derivative $\dot Q$.
Rewrite 
$V^{t_0}(d) =\int m_{t_0}\{d(x);x\} f_X(x) dx$,
where 
$m_{t_0}(a; x) = E\{ T^*(a) -t_0 | T^*(a) \geq t_0, X= x \} $.
If $d_0$ is the maximizer of $V^{t_0}(d)$, 
then for any $x$ with $f_X(x) >0$, 
$$
d_0(x) = \arg \max_a m_{t_0}(a; x)
= \arg \max_a 
\int_{t_0}^{\tau} \exp\{- \Lambda(t; a, x) + \Lambda(t_0; a, x)\}dt \,,
$$
and
$$
 \dot m_{t_0}(a; x) =  
 \int_{t_0}^{\tau} \exp\{- \Lambda(t; a, x) + \Lambda(t_0; a, x)\}
\frac{\partial}{\partial a}\{- \Lambda(t; a, x) + \Lambda(t_0; a, x)\}
dt$$
equals to zero at $a = d_0(x)$.
However, 
when $t_0 < a_0$, there always exists $x$ such that $t_0 < d_0(x)\leq a_0 $. 
For such $x$ and $a = d_0(x) > t_0$,
 it can be calculated under model $(3)$ that 
\begin{align*}
 \dot m_{t_0}\{d_0(x); x\}
=&
 \int_{t_0}^{\tau} \exp[- \Lambda\{ t; d(x), x\} + \Lambda\{t_0; d(x), x\} ]
\frac{\partial}{\partial a}\{- \Lambda(t; a, x) + \Lambda(t_0; a, x)\}|_{a=d_0(x)}
dt
\\
=&
\int_{d_0(x)}^{\tau} 
\exp[- \Lambda\{ t; d(x), x\} + \Lambda\{t_0; d(x), x\} ]
\Big[
\lambda_0(a) \big( 1 -  \exp[ Q\{ a- d_0(x)\} H_0(x) ] \big)
\\
&\times  \exp\{\mu_0(x)\}
+
\{\Lambda_0(t) - \Lambda_0(a) \} 
\exp[ \mu_0(x) +  Q\{ a- d_0(x)\} H_0(x) ] 
\\
&\times
\dot Q\{a-d_0(x)\}H_0(x)\Big]\Big|_{a=d_0(x)}dt
\\
=&
\int_{d_0(x)}^{\tau} 
\lambda_0\{d(x)\} \exp[- \Lambda\{ t; d(x), x\} + \Lambda\{t_0; d(x), x\} +\mu_0(x)]
[1 -  \exp\{Q(0) H_0(x)  \}]dt
\end{align*}
Since $Q(0)<0$ and $H_0(x)\geq 0$, 
it can be seen that $\dot m_{t_0}\{d_0(x); x\} <  0 $ for any $x$ satisfying
$t_0 < d_0(x)$ and $H_0(x) >0$. It contradicts with the conclusion that 
$d_0(x)$ maximizes $m_{t_0}(a; x)$.

\section*{Web Appendix A2. Calculation  of PI$_i$ and PI$_c$ }

To demonstrate the effect of the proposed treatment initiation regimes on individual level, 
we also calculated the percent of individuals with improved counterfactual outcomes  
under the optimal individualized treatment initiation regime (PI$_i$).  

Specifically, let 
$ \hat m\{ d(x), x\}  = 
\frac{ 
    \sum_{i=1}^n  W_n(\tilde T_i )  I(\tilde T_i \geq a_0)  K_{\bm{h_1}} (\bm{X}_i - \bm{x}) K_{h_2}[  g( A_i) - g\{ d(\bm{x}) \}  ] 
}{
  \sum_{i=1}^n I(\tilde T_i \geq a_0)  K_{\bm{h_1}} (\bm{X}_i - \bm{x}) K_{h_2}[  g( A_i) - g\{ d(\bm{x}) \}  ]  } $
 be the estimate of the counterfactual outcome under treatment regime $d$ for an individual with covariate $x$.
Then 
 $ \hat m\{ d(X_i), X_i\} > \hat m( A_i, X_i) $ denotes   
 an improved counterfactual outcome  on the $i$th individual if the regime $d$ had been followed.
Therefore,   
by calculating the percentage of individuals with $ \hat m\{ \hat d^{opt}(X_i), X_i\}  > \hat m( A_i, X_i) $ among patients with observed treatment initiation time, 
we found that around $78.5\%-89.6\%$ of the individuals in the simulation studies, and 
about $79.8\%$ of the breast cancer patients in the application study would achieve larger restricted mean survival time if $\hat d^{opt}$ had been followed by the entire population. 

Moreover, 
as a comparison, 
we also calculated the percentage of individuals with $\hat m\{ \hat d_c(X_i), X_i\}  > \hat m( A_i, X_i)$, where $\hat d_c$ is an estimate of the optimal constant regime $d_c = \arg \max _a E\{ m(a, X)\}$.  The simulation results are reported as the values of PI$_c$ in Table 1.
From the results, it can be seen that in all the scenarios, values of PI$_i$ are higher than those of  PI$_c$.
This indicates the estimated optimal treatment initiation regime $\hat d^{opt} $ does perform better than the optimal constant regime.

\section*{Web Appendix A3. A data-driven bandwidth selection algorithm based on EBBS method  }

When the patient observation is not evenly distributed across the covariate space, 
it would be worthwhile to develop some robust bandwidth selection methods that could provide data-driven bandwidth selection for different patient characteristics.
However, 
incorporating the existing methods  such as empirical bias bandwidth selection(EBBS) into our estimation procedure is non-trivial.

Firstly, the estimation of the proposed value function includes two kernel approximations, $K_{\bm{h_1}} (\bm{X}_i - \bm{x})$ and $K_{h_2}[g( A_i) - g\{ d(\bm{x}) \}  ]$, respectively.
Since $A_i$ is not always observed, 
it warrants future research to find the optimal local bandwidth  for the two kernels simultaneously,  
especially when the patient observation of $A_i$ is not evenly distributed across the time range $[0,a_0]$.

What's more, 
if we rewrite the estimated value function $M_n(\beta)$ as 
$\int \hat m( \bm{\beta}, \bm{x}; h_1,h_2) d \hat F_X(\bm{x}) $, 
then following the idea of EBBS \citep{ruppert1997}, 
for each given $x_0$, our target is to
estimate the mean square error of $ \hat m( \bm{ \hat \beta}, \bm{x}_0; h_1,h_2)$, 
which can be denoted by $MSE(h_1,h_2;x_0,\hat \beta)$,
 and the optimal  bandwidth 
 is the minimizer of $MSE(h_1,h_2;x_0,\hat \beta)$ among  all the candidate $(h_1,h_2)$.
However, 
note that  
the calculation of $ \hat \beta = \arg \max M_n(\beta) $  depends on 
$\{ h_1(x), h_2(x) \}$ for all $x\in\mathcal{X}$, 
the obtained  local bandwidth $\{\hat h_1(x_0), \hat h_2(x_0)\}$ for given $x_0$
will also depend on the value of $\{ h_1(x), h_2(x) \}$ for  $x \neq x_0$.
Therefore,
in order to find the optimal local bandwidth $\{\hat h_1(x), \hat h_2(x)\}$ for $x\in\mathcal{X}$ simultaneously, 
the development of some iterative algorithm would be required.

In the analysis of the breast cancer data, 
we developed a data-driven bandwidth selection algorithm under a naive setting, where 
the bandwidth in the kernel $K_{h_2}[g( A_i) - g\{ d(\bm{x}) \}]$ is fixed as 
$h_{2}= n^{-1/5} \rm{sd}(A)$, 
and the bandwidth in the kernel $K_{\bm{h_1}} (\bm{X}_i - \bm{x})$  is selected separately for each $x$  from a candidate set 
$\{ h_{1,1},h_{1,2},\dots,h_{1,21}\}$ with  
$h_{1l} = (0.78+0.02 \cdot l) n^{-1/5} \rm{sd}(X)$ for $ 1\leq l \leq 21$ .
Specifically, 
let $\hat \beta_{l}$ denote the obtained estimator when $h_1(x) = h_{1,l} $ for all $x\in \mathcal{X}$.
Then for given $x_0$, and for each $ 1\leq l \leq 21$,
we can estimate the $MSE(h_{1,l},h_2;x_0)$ by the EBBS method with $\hat \beta =\hat \beta_{l} $.
Following the smoothing technique introduced in \cite{ruppert1997}, 
a smoothed version of MSE, $SMSE(h_{1,l},h_2;x_0)$, can be obtained, 
and then $\hat h_1(x_0)$ is defined as the minimizer of $SMSE(h_{1,l},h_2;x_0)$ among all the candidate values $\{ h_{1,1},h_{1,2},\dots,h_{1,21}\}$. 
Lastly, using the smoothing technique again, the optimal local bandwidths $\tilde h_1(x)$ is finally obtained for  each $x \in \mathcal{X}$.
The analysis results under the selected local bandwidths are reported in the following table.

\begin{center}
 \renewcommand{\baselinestretch}{1}  \small 
{ Table R1. Analysis results for breast cancer data, including the 
the estimates for $\bm{\beta}$ ($Est$),
the standard deviations of $\hat \beta$ ($Sd$), and the $p$-values ($Pvalue$).
}\\
{\small		
\begin{tabular}{cccc}
\hline
& $\beta_1$(Intercept) & $\beta_2$(Age) & $\beta_3$(Stage) \\
				\hline
	\multicolumn{4}{c}{	global bandwidth }\\
$Est$    &$0.669$  &  $-0.154$&  $-1.643$ \\
$Sd$     &$0.272$  &  $0.164$ &  $ 0.468$  \\
$P value$&$0.014$  &  $0.350$ &  $<0.001$ \\
\hline
\multicolumn{4}{c}{ local bandwidth  }	 \\
$Est$    &$0.630$  &  $-0.078$&  $-1.605$ \\
$Sd$     &$0.238$  &  $0.171$ &  $ 0.458$  \\
$P value$&$0.008$  &  $0.649$ &  $<0.001$ \\
				\hline
			\end{tabular}}
		\end{center}

It can be seen that, for this breast cancer dataset, 
the estimates obtained by global bandwidths and  by local bandwidths are similar.

\section*{Web Appendix A4. Estimating the OTIT based on other value functions}

If the value function is constructed based on the restricted mean survival time or restricted mean residual lifetime at some time point $t < a_0$,
the estimation for the corresponding value function could be challenging due to the missing in treatment initiation time $A^*$.
To illustrate this,  
define 
\begin{align*}
V^t(d)
&
\equiv \iint_{t}^{\tau} pr[ T^*\{ d(\bm{x})\} \geq u \mid  T^*\{ d(\bm{x})\} \geq t, \bm{X}=\bm{x} ] f_X(\bm{x}) du d\bm{x} 
\end{align*}
where  $V^0(d)$ corresponds to the restricted mean survival time 
and $V^{a_0}(d)$ is the value function defined in $(1)$. 
Under assumptions (A1)-(A4),  we have 
\begin{align*}
V^t(d)
=&  
\iint_{t}^{\tau} \frac{ pr\{ \tilde T \geq u\mid   A^* = d(\bm{x}), \bm{X}=\bm{x} \} }{pr\{  \tilde T \geq t\mid  A^* = d(\bm{x}), \bm{X}=\bm{x} \}} \frac{S_C(t)}{S_C(u)}f_X(\bm{x}) du d\bm{x} \,.
\end{align*}
When $t > a_0$, as shown before, the integrand of $V^t(d)$ can be estimated consistently by the observed data for all $\bm{x}$ and $u \geq t$.
When $t < a_0$, 
there always exists some $(d, x)$  satisfying $t < d(\bm{x}) < a_0$.
Then for the samples with $t < \tilde T_i < d(\bm{x})$, 
whether or not $A_i^* = d(\bm{x})$ can not be determined by the observed data.
Thus, the conditional survival probability 
$pr\{ T \geq t \mid  A^* = d(\bm{x}),\bm{X}=\bm{x} \} $ can not be estimated unless further information about the distribution of $A^*$ is available.

To deal with this, one possible solution is assuming 
$T$ to be independent of $A^*$ given  $ \xi =0$,
then the distribution of $A^*$ conditioning on $\bm{X}$ 
can be estimated by the observed data,
and thus, the value function $V^t(d)$ for all $t\in [0,\infty)$ is estimable.
However, 
the corresponding estimates take a relatively complex form and 
could be computationally expensive.

\section*{Web Appendix A5.  A refined OTIT for patients in critical conditions}

In order to define a proper OTIR that can handle the challenge caused by the possible censoring in treatment initiation times,    
 we proposed a value function $V(d)$ based on  the restricted mean residual lifetime,
conditioning on the fact that counterfactual
survival times are longer than the maximum treatment initiation time. 
Such a value function may cause selection bias. 
However, as we have shown in the example given in Section 3.2., 
under a general class of hazard models, 
the maximizer of $V(d)$ could minimize patients' hazard rate post treatment.
 Therefore, our proposed OTIR can still provide an optimal rule for the whole population even there may be some selection bias in defining the value function.  
 
On the other hand, in practice, patients in critical conditions are less likely to survive longer than $a_0$, 
and thus would be eliminated in the estimation procedure of the OTIR. 
To deal with the potentially short life expectancy for patients in critical conditions, we may consider a refined estimation procedure. 
For patients in critical conditions, 
it is reasonable to assume that there exists some $a_1< a_0$ such that they should initiate treatment in $[0, a_1]$.
Then after obtaining the estimated OTIR $\hat d^{opt}(\bm{x})$ based on the originally defined restricted mean residual lifetime, we can diagnose the patients in critical conditions  by $ \mathcal{S}_{a_1} = \{i: \hat d^{opt}(\bm{X}_i) < a_1\} $.
Next, based on the samples in $ \mathcal{S}_{a_1}$ , we conduct the estimation procedure in Section 3 again with $a_0$ replaced by $a_1$, 
and obtain an updated OTIR estimator $\hat d^{opt}_{a_1}(\bm{x})$ for patients in critical conditions.
The updated OTIR is expected to be more accurate than the old OTIR for patients in critical conditions. 
To see this, we consider the following two cases for patients in critical conditions:

\includegraphics[width=10 cm, height=3 cm]{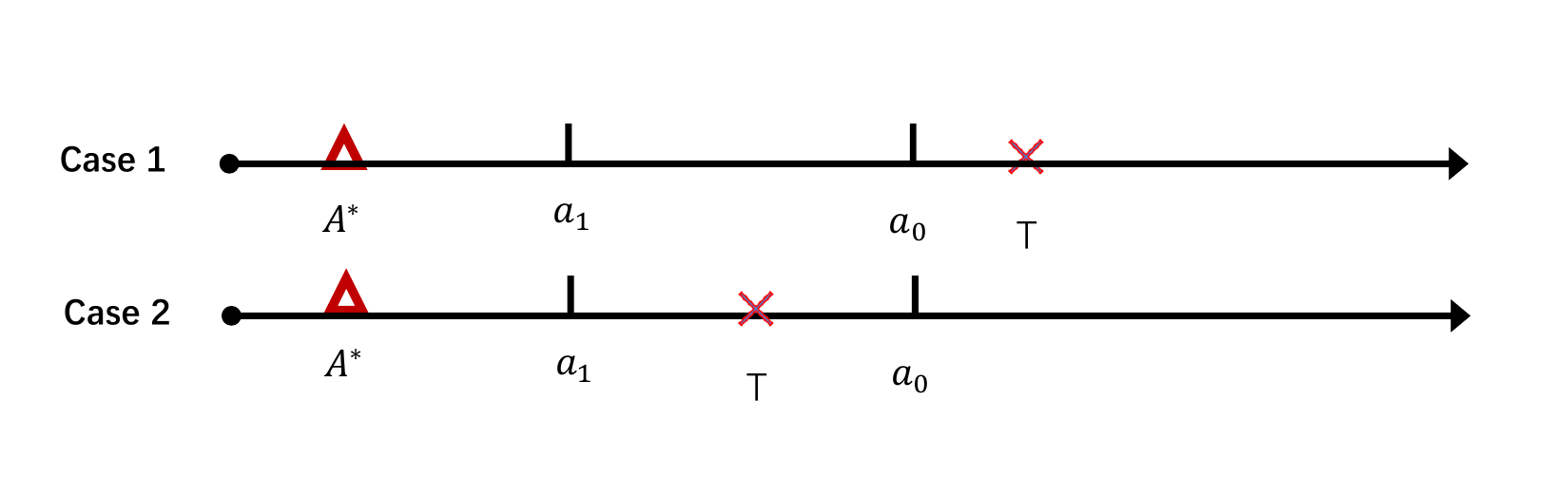}

Then, it is not hard to see that, 
compared with $\hat d^{opt}(\bm{x})$ which only focuses on samples in case 1, 
the refined estimator $\hat d^{opt}_{a_1}(\bm{x})$ utilizes the samples in both case 1 and 2, and thus could provide a more accurate estimator for the patient in critical conditions.

\newpage
\section*{Web Appendix B:  Proofs for the Asymptotic Properties}

\section*{B1.Proof of Theorems 1-3}

\subsection*{Notation and Lemmas}

For simplicity, we prove the theorems with $p = 1$.
All the arguments can be similarly generalized to $p \geq 2$.
Let $\rightsquigarrow $ denote convergence in distribution.
For $t \geq 0 $, $a\in[0,a_0]$ and $x \in \mathcal{X}$, define
\begin{align*}
S_n(t; a, {x}) 
=&
(n {h_1} h_2)^{-1} \sum_{i=1}^n  I(\tilde T_i \geq t)  K\{ (X_i-x){h_1}^{-1} \}
K[ \{ g(A_i^*)-g(a) \} h_2^{-1} ] \dot g(a)   \,,
\\
\dot S_n (t; a, {x})
=&
(n {h_1} h_2^2)^{-1} 
\sum_{i=1}^n  I(\tilde T_i \geq t)  
K\{ (X_i-x){h_1}^{-1} \} 
\dot K[ \{ g(A_i^*)-g(a) \} h_2^{-1} ]  \dot g(a)   \,,
\\
\Psi_n(a, {x})
=& 
(n {h_1} h_2)^{-1} \sum_{i=1}^n  \int_{a_0}^{\tau}  I(\tilde T_i \geq t) \{S_C(t)\}^{-1}  dt 
K\{ (X_i-x){h_1}^{-1} \}  K[ \{ g(A_i^*)-g(a) \} h_2^{-1} ] \dot g(a) 
\,,
\\
\dot \Psi_n(a, {x})
=& 
(n {h_1} h_2^2 )^{-1} \sum_{i=1}^n  \int_{a_0}^{\tau}  I(\tilde T_i \geq t) \{S_C(t)\}^{-1}  dt 
K\{ (X_i-x){h_1}^{-1} \} \dot K[ \{ g(A_i^*)-g(a) \} h_2^{-1} ] \dot g(a)  
\,.
\end{align*}
We first present several lemmas that are needed for the proof of asymptotic normality. 
The proofs of these lemmas are given in Web Appendix B2 of this Supplementary Material.

{\it Lemma 1.}
If Conditions C$1$--C$6$ hold,
then for any fixed $t \in [0,\tau]$, $a\in[0,a_0]$ and $x\in \mathcal{X}$,
$$
(n {h_1} h_2)^{1/2}\left\{ S_n(t; a, {x})  - \mu(t; a, {x}) \right\}
\rightsquigarrow 
\mathcal{N} \left\{ 0,  \dot g(a) \mu(t; a, {x}) \kappa_{0,2}^2 \right\} \,.
$$
$$
(n {h_1} h_2)^{1/2}
\left\{  \Psi_n(a, {x}) - \Psi(a, {x})
\right\}
\rightsquigarrow 
\mathcal{N} \left\{ 0,   \dot g(a) \omega( a, x) \kappa_{0,2}^2 \right\} \,,
$$
$$
(n {h_1} h_2^3)^{1/2}
\left[  \dot S_n(t; a, {x})+ \frac{\partial}{\partial a} \left\{\frac{\mu(t; a, {x})}{\dot g(a)} \right\}
\right] 
\rightsquigarrow
\mathcal{N}\{ 0, \dot g(a) \mu(t; a, {x})   \kappa_{0,2}\dot\kappa_{0,2}\} \,,
$$
$$
(n {h_1} h_2^3)^{1/2}
\left[  \dot \Psi_n(a, {x}) + \frac{\partial}{\partial a} \left\{\frac{\Psi( a, x)}{\dot g(a)} \right\} \right] 
\rightsquigarrow
\mathcal{N}\left\{  0,  \dot g(a) \omega(a,x) \kappa_{0,2}\dot\kappa_{0,2} \right\} \,,
$$
where 
$\mu(t; a, {x}) =                            S_T(t; a, {x}) f_{(A^*,X)}(a, {x}) S_C(t)$,
$\Psi(a, {x})    = \int_{a_0}^{\tau}  S_T(t; a, {x}) f_{(A^*,X)}(a, {x})  \rm{d}t$, 
$\dot g(a)$ is the derivative of $g(a)$ and 
$\omega(a, {x}) = 2 \int_{a_0}^{\tau}  S_T(t; a, {x}) f_{(A^*,X)}(a, {x}) 
\int_{a_0}^t \{ S_C(s) \}^{-1}   \rm{d} s  \rm{d} t  
$.
\\

{\it Lemma 2.}
Let $x_n$ denote a sequence of random vectors and 
let $y_n$ denote a sequence of random variables.
If $c_n ( x_n -  x_0) \rightsquigarrow \mathcal N(0, \Sigma_1) \,,$
$d_n ( y_n - y_0) \rightsquigarrow \mathcal N(0, \sigma^2_2) \,,$
$c_n/d_n \rightarrow 0 $ as $n \rightarrow \infty$, 
and $y_n$ and $y_0$ are bounded away from zero,
then
$c_n \left( x_n / y_n - x_0 / y_0 \right)
= c_n (x_n-x_0) / y_0 + o_p(1) 
$,  
$c_n \left( x_n y_n -x_0y_0 \right)
= c_n (x_n-x_0)y_0 + o_p(1) 
$.
\\

{\it Lemma 3.}
Let $\mathcal{G}(x)$  be a mean-zero Gaussian process with bounded, uniformly continuous
covariance function $r(x, y)$. If $a(x)$ is a $p$-dimensional piecewise smooth function 
and $\mathcal{G}(x)a(x)$ is integrable,
then $ Z= \int \mathcal{G}(x) a(x) d{x} $ is a Gaussian vector with mean zero and variance matrix
$\Sigma_Z= \iint r(x, y) a(x)a(y)^T d{x} dy $.
Moreover,
if $r(x, y)= 0 $ for all $x \neq y$, then 
$\Sigma_Z= \int r(x, x)a(x)a(x)^T d{x} $.
\\

Under condition 3,
$\mu(t; a, {x})$ is differentiable at $(a, {x}) \in [0,a_0]\times \mathcal{X}$ for each fixed $t \in[0,\tau]$.
Let $\dot \mu_a(t; a, {x})$ denote the first-order partial derivative of $\mu(t; a, {x})$ respect to $a$ 
and define
\begin{eqnarray*}
	G_n(a, {x})
	&=&
	\int_{a_0}^{\tau} 
	\left\{
	- \dot S_n(t; a, {x}) S_n(a_0; a, x) + S_n(t; a, {x}) \dot S_n(a_0; a, x)
	\right\}
	\frac{\hat S_C(a_0)}{\hat S_C(t)}  
	\frac{ \hat f_X({x}) \dot g(a)}{S_n(a_0; a, x)^2} dt \,,
	\\
	G(a, {x})
	&=&
	\int_{a_0}^{\tau} 
	\left\{
	\dot \mu_a(t; a, {x}) \mu(a_0; a, x) - \mu(t; a, {x}) \dot \mu_a(a_0; a, x)
	\right\}
	\frac{S_C(a_0)}{S_C(t)}  
	\frac{  f_X({x}) }{\mu(a_0; a, x)^2} dt\,.
\end{eqnarray*}
The asymptotic properties of $G_n$ is given in the following lemma.

{\it Lemma 4.}
If Conditions C1--C6 hold, 
then for any fixed $(a, {x}) \in [0,a_0]\times \mathcal{X}$,
$
(n {h_1} h_2^3)^{1/2} \{ G_n(a, {x}) - G(a, {x}) \}
$
converges in distribution to a normal variable with mean zero and variance  
$$
\sigma^2_G(a, {x})
=    \frac{ \omega(a, {x}) \mu(a_0; a, x) - \Psi(a, {x})^2  }{ \mu(a_0; a, x)^3} 
f_X(x)^2 \dot g(a)^3 S_C(a_0)^2 \kappa_{0,2} \dot \kappa_{0,2} \,.
$$
Moreover, the stochastic processes
$(n {h_1} h_2^3)^{1/2} [ G_n\{ \phi({\tilde x}^T {\beta}^{opt}),x\}  - G\{ \phi({\tilde x}^T {\beta}^{opt}),x\} ]$  
converges weakly to a Gaussian process $\mathcal G(x)$ 
with mean zero and covariance function 
$r(x, y)=I(x=y)\sigma^2_G\{ \phi({\tilde x}^T {\beta}^{opt}),x\}.$

\subsection*{Proof of Theorem 1}

By Condition C1 and Theorem 5.7 of 
\cite{van1998asymptotic}, 
it suffices to show 
$\sup_{{\beta} \in \mathcal{B}} | M_n({\beta}) - M({\beta}) | \rightarrow 0$ in probability.

Define 
$
R(x;{\beta})
=      [ S_T\{ a_0; \phi({\tilde x}^T {\beta}), x \} ]^{-1}
\int_{a_0}^{\tau} S_T\{ t; \phi({\tilde x}^T {\beta}), x \} dt 
=m\{ \phi({\tilde x}^T {\beta}), x \} 
$ and 
rewrite  $M_n({\beta}) - M({\beta})$ as
\begin{align*}
&
\iint_{a_0}^{\tau} 
\left[
\frac{ S_n\{ t; \phi({\tilde x}^T {\beta}), x \} }{S_n\{ a_0; \phi({\tilde x}^T {\beta}),x \} }
\frac{\hat S_C(a_0)}{ \hat S_C(t)}  \hat f_X(x) 
-
\frac{ \mu\{ t; \phi({\tilde x}^T {\beta}), x \} }{\mu\{ a_0; \phi({\tilde x}^T {\beta}), x \} }
\frac{S_C(a_0)}{S_C(t)}f_X(x) 
\right] dt d{x} 
\\
=&
\iint_{a_0}^{\tau} 
\left[
\frac{ S_n\{ t; \phi({\tilde x}^T {\beta}), x \} }{S_n\{ a_0; \phi({\tilde x}^T {\beta}),x \} }
\frac{\hat S_C(a_0)}{ \hat S_C(t)}  
-
\frac{ \mu\{ t; \phi({\tilde x}^T {\beta}), x \} }{\mu\{ a_0; \phi({\tilde x}^T {\beta}), x \} }
\frac{S_C(a_0)}{S_C(t)}
\right] \hat f_X(x)  dt d{x} 
\\
& +
\int   R(x;{\beta})  \left\{ \hat f_X(x)  - f_X(x) \right\} d{x} 
\\
\leq &
\int  \int_{a_0}^{\tau} 
\left|
\frac{ S_n\{ t; \phi({\tilde x}^T {\beta}), x \} }{S_n\{ a_0; \phi({\tilde x}^T {\beta}),x \} }
\frac{\hat S_C(a_0)}{ \hat S_C(t)}     -
\frac{ \mu\{ t; \phi({\tilde x}^T {\beta}), x \} }{\mu\{ a_0; \phi({\tilde x}^T {\beta}), x \} }
\frac{S_C(a_0)}{S_C(t)}
\right| \hat f_X(x) dt  d{x} 
\\
& +
\frac{1}{n{h_1}} \sum_{i=1}^n
\int R( X_i+u{h_1}; {\beta}) K(u) du- E\{ R(X;{\beta})\}
\end{align*}
\begin{align*}
\leq &
\sup_{t\in[0,\tau],x\in\mathcal{X}}
\left|
\frac{ S_n\{ t; \phi({\tilde x}^T {\beta}), x \} }{S_n\{ a_0; \phi({\tilde x}^T {\beta}),x \} }
\frac{\hat S_C(a_0)}{ \hat S_C(t)}     -
\frac{ \mu\{ t; \phi({\tilde x}^T {\beta}), x \} }{\mu\{ a_0; \phi({\tilde x}^T {\beta}), x \} }
\frac{S_C(a_0)}{S_C(t)}
\right| 
(\tau-a_0)  \int \hat f_X(x) d{x}
\\
& +
\frac{1}{n{h_1}} \sum_{i=1}^n
\int R(X_i+u{h_1}; {\beta}) K(u) du- E\{ R(X; {\beta})\}
\,.
\end{align*}

Under Condition C2, 
$\mu( a_0; a, x) $ and $ S_C(t) $ 
are uniformly bounded away from zero for $t\in[a_0,\tau]$,$a\in[0,a_0]$ and $ x \in \mathcal{X}$.
Thus, we only need to show that as $n \rightarrow \infty$,
$ \sup_{t\in[0,\tau]} |  \hat S_C(t)-  S_C(t) |  \rightarrow 0 $,   
$\sup_{t\in [a_0,\tau], a\in[0,a_0], x\in \mathcal{X}} 
| S_n(t; a, {x}) -  \mu(t; a, {x}) |   \rightarrow 0 $ and
$$
\sup_{{\beta} \in \mathcal{B}}
\left|
(n {h_1})^{-1} \sum_{i=1}^n
\int R(X_i+u{h_1};{\beta}) K(u) du- E\{ R(X;{\beta})\}
\right|
\rightarrow 0 
$$
almost surely.
The first result follows from the uniform convergence of Kaplan-Meier estimates.  
{
	For the second result, consider a class of functions 
	\begin{align*}
	\mathcal{F}_S 
	=  \{   &f_S(\tilde T, X, A^*; t, x, a) = 
	I(\tilde T \geq t)  K\{ {h_1}^{-1}(X-x) \} K[  h_2^{-1} \{ g(A^*)-g(a) \}] ( {h_1} h_2)^{-1};  \\
	&t\in[0,\tau],a\in[0,a_0], x\in \mathcal{X}  \} \,.
	\end{align*}
	Under Condition C5 and by the boundedness of $K(u)$ and $\int |\dot K(u)| du$ given in Condition C4, it can be calculated that the covering number of $\mathcal{F}_S$
	satisfies 
	$ 
	\log N( \epsilon, \mathcal{F}_S, L_1(\mathbb{P}_n)) \leq  O [ log\{ n^2 / ({h_1} h_2)^2 \} ] = o_P(n)
	$
	for any $\epsilon >0$. Thus, the second result follows from Theorem 2.4.3 of \cite{van1996weak}.
}

{ 
	For the third result, consider a class of functions 
	$$
	\mathcal{F}_R 
	= \{ f_R(X; {\beta}) = 
	\int   R(X+u{h_1};{\beta})  K(u) du; 
	{\beta} \in \mathcal{B} \} \,.
	$$
	For any fixed $\beta_1$, $\beta_2 \in \mathcal{B}$, 
	since $\int | K(u)| du = 1$ and $\sup_{u,{\beta}} |R(u;{\beta})| \leq \tau - a_0$,
	for any $\epsilon >0$, there exists $u_{min}$ and $u_{max}$  such that 
	\begin{align*}
	& | f_R(X;\beta_1) - f_R(X; \beta_2) |
	\\
	\leq & \frac{\epsilon}{3} + 
	K_{max} \int_{u_{min}}^{u_{max}} | R(X+u{h_1}; \beta_1) - R(X+u{h_1}; \beta_2) |  du 
	\\
	\leq & \frac{\epsilon}{3} 
	+ K_{max} \sum_{i = 1}^{I} 
	| R(X+u_i{h_1}; \beta_1) - R(X+u_i{h_1}; \beta_2) | ( u_i - u_{i-1}) 
	\\
	&+ K_{max} \sum_{i = 1}^{I} 
	\sup_{ u\in [u_{i-1},u_i] } | R(X+u{h_1}; \beta_1) - R(X+u_i{h_1}; \beta_1) |
	(u_i - u_{i-1})
	\\
	&+ K_{max} \sum_{i = 1}^{I} 
	\sup_{ u\in [u_{i-1},u_i] } |  R(X+u{h_1}; \beta_2) - R(X+u_i{h_1}; \beta_2) |
	(u_i - u_{i-1})
	\\
	\leq & \frac{\epsilon}{3} 
	+ K_{max}  \sup_{a\in [0,a_0], x\in \mathcal{X}} \left | \frac{\partial m(a, {x})}{ \partial a} \right |  (u_{max} - u_{min})
	\max_{1\leq i \leq I} \left | \phi\{ (1, X+u_i{h_1})^T  \beta_1\}  - \phi\{ (1, X+u_i{h_1})^T \beta_2\}  \right| 
	\\
	&+ 2 K_{max} \sup_{x \in\mathcal{X}, {\beta} \in \mathcal{B}} \left | \frac{\partial R(x; {\beta})}{ \partial x} \right |  {h_1}  (u_{max} - u_{min})
	\max_{1\leq i \leq I}  ( u_i - u_{i-1})\,.     
	\end{align*}
	Here
	$u_{min} = u_0 \leq u_1 \leq \ldots \leq u_{I} = u_{max}$ is a partition of $[u_{min},u_{max}]$. 
	By Condition C3 and the boundedness of $\mathcal{B}$, 
	both $\partial m(a, {x}) / \partial a$ and $\partial R(x; {\beta}) / \partial x$  exist 
	and are uniformly bounded on $a \in [0,a_0]$, $x \in \mathcal{X}$ and ${\beta} \in \mathcal{B}$.
	Denote $C_R = K_{max}  \sup_{a, x} \left | \partial m(a, {x}) / \partial a \right |  (u_{max} - u_{min})$, then 
	if the partition $\{ u_i; 1\leq i \leq I\}$ is dense enough, we have 
	\begin{align*}
	| f_R(X;\beta_1) - f_R(X; \beta_2) |
	\leq & \frac{2\epsilon}{3} 
	+ C_R
	\max_{1\leq i \leq I} \big | \phi\{ (1, X+u_i{h_1})^T  \beta_1\} 
	- \phi\{ (1, X+u_i{h_1})^T \beta_2\} 
	\big | \,.
	\end{align*}
}

{
	For each fixed $i$ and $u_i$ and by Lemma 2.6.15 and Lemma 2.6.17 of \cite{van1996weak}, 
	$ 
	\mathcal{F}_{\phi, i} = [ \phi\{ (1, X+u_i{h_1})^T  {\beta}\} ; {\beta} \in \mathcal{B} ]
	$
	is a VC class.
	Then by Theorem 2.6.7 of \cite{van1996weak}, 
	given any probability measure $Q$, 
	there exists $ \{ \mathcal{B}_{i, k} , 1\leq k \leq K_i \} $ satisfying 
	$ \cup_{1\leq k \leq K_i} \mathcal{B}_{i, k} \supseteq \mathcal{B}$ 
	and 
	$      \| \phi\{ (1, X+u_i{h_1})^T  \beta_1\}  - \phi\{ (1, X+u_i{h_1})^T \beta_2\}    \|_{Q} 
	<    \epsilon/ (3 C_R) $ 
	for any $\beta_1, \beta_2 \in \mathcal{B}_{i, k}$.
	Define 
	$ \mathcal{K} = \big \{ (k_1, \ldots, k_I); k_i \in \{1, \ldots, K_i \} , 1\leq i \leq I \big \} $ 
	and consider a finite set $\Big\{   \cap_{1\leq i \leq I}  \mathcal{B}_{i, k_i}; (k_1, \ldots, k_I)\in \mathcal{K}\Big\} $.
	For any  $\beta_1 ,\beta_2 \in  \cap_{1\leq i \leq I}  \mathcal{B}_{i,k_i}$,
	we have 
	$ \max_{1\leq i \leq I} \| \phi\{ (1, X+u_i{h_1})^T  \beta_1\}  - \phi\{ (1, X+u_i{h_1})^T \beta_2\}    \|_{Q} 
	<    \epsilon/ (6 C_R) $ 
	and 
	$\| f_R(X,\beta_1) - f_R(X, \beta_2)  \|_{Q}  <  \epsilon$. 
	This indicates that $N(\epsilon, \mathcal{F}_R, Q) $ is finite for any $\epsilon$ and $Q$, and thus, the third result follows from  Theorem 2.4.3 of \cite{van1996weak}.
}

\subsection*{Proof of Theorem 2}

Denote $U_n({\beta})= \rm{d} M_n({\beta}) /  \rm{d} {\beta}  $,
$U({\beta})= \rm{d} M({\beta}) / \rm{d} {\beta} $,
$D_n({\beta}) = \rm{d} U_n({\beta}) / \rm{d} {\beta}$,
and  $D({\beta}) =  \rm{d} U({\beta}) /\rm{d} {\beta}$.
By $U_n({\hat \beta}^{opt})=0$ and Taylor series expansion, 
\begin{equation}
U_n({\beta}^{opt}) 
=
- D_n( {\tilde \beta}) (  {\hat \beta}^{opt} - {\beta}^{opt} )  \,, 
\label{TaylorEx}
\end{equation}
where  ${\tilde \beta}$ is on the line segment between ${\hat \beta}^{opt}$ and ${\beta}^{opt}$.
To prove the asymptotic normality of $ ( n {h_1} h_2^3)^{1/2} ({\hat \beta}^{opt}-{\beta}^{opt})$,
it suffices to show that $ (n {h_1} h_2^3)^{1/2} U_n({\beta}^{opt})$ converges
in distribution to a normal variable, 
and 
$D_n({\beta})^{-1}$ exists and converges uniformly to a bounded matrix
$D({\beta})^{-1}$ on $\mathcal{B}$. 

We first derive the asymptotic properties of  $U_n({\beta}^{opt})$.
Since $U({\beta}^{opt}) = 0$ and 
$$U_n({\beta})= \int G_n\{  \phi({\tilde x}^T {\beta}),x \} \dot \phi( {\tilde x}^T {\beta}) {\tilde x} d{x} \,, 
U({\beta}) = \int G\{  \phi({\tilde x}^T {\beta}),x \} \dot \phi( {\tilde x}^T {\beta}) {\tilde x} d{x} \,,$$  
we have
\[
(n {h_1} h_2^3)^{1/2} U_n({\beta}^{opt})
= \int (n {h_1} h_2^3)^{1/2}  \left[ G_n\{  \phi({\tilde x}^T {\beta}^{opt}),x \} - G\{  \phi({\tilde x}^T {\beta}^{opt}),x \} \right] 
\dot \phi( {\tilde x}^T {\beta}^{opt}) {\tilde x} d{x} \,. 
\]
By Lemma $4$,
the stochastic processes
$(n {h_1} h_2^3)^{1/2} [ G_n\{ \phi({\tilde x}^T {\beta}^{opt}),x\}  - G\{ \phi({\tilde x}^T {\beta}^{opt}),x\} ]$  
converges weakly to a mean zero Gaussian process with
covariance function $r(x, y)$ satisfying
$r(x, x)= \sigma^2_G\{ \phi({\tilde x}^T {\beta}^{opt}),x\}$ and $r(x, y)= 0$ for all $x\neq y$.
Then, by Lemma $3$, 
we have 
\begin{equation}
(n {h_1} h_2^3)^{1/2} U_n({\beta}^{opt}) \rightsquigarrow N(0,\Sigma_U)  \label{Eq_U}
\end{equation}
where 
\begin{align*}
\Sigma_U
=& \int\sigma^2_G\{ \phi({\tilde x}^T {\beta}^{opt}),x\} 
\{\dot \phi( {\tilde x}^T {\beta}^{opt})\}^2 {\tilde x} {\tilde x}^T d{x} 
\\
=&\int   \left[  \frac{ \omega\{ \phi({\tilde x}^T {\beta}^{opt}),x\} }{ \mu^2\{a_0; \phi({\tilde x}^T {\beta}^{opt}),x\} } 
- \frac{\Psi^2\{ \phi({\tilde x}^T {\beta}^{opt}),x\} }{ \mu^3\{a_0; \phi({\tilde x}^T {\beta}^{opt}),x\} } \right]
\left\{ f_X(x)  S_C(a_0) \dot \phi( {\tilde x}^T {\beta}^{opt}) \right\}^2
\\
&\times
[\dot g\{ \phi({\tilde x}^T {\beta}^{opt}) \}]^3
\kappa_{0,2} \dot \kappa_{0,2} {\tilde x} {\tilde x}^T d{x} \,.
\end{align*}

Lastly we show the uniform convergence of $D_n({\beta})$ on $\mathcal{B}$.

Define
$
\ddot S_n (t; a, {x})
=
(n {h_1} h_2^3)^{-1} \sum_{i=1}^n  I(\tilde T_i \geq t)  
K\{ (X_i-x)/ {h_1}\} \ddot K [ \{ g(A_i^*)-g(a) \} / h_2 ] \dot g(a) \,,
$
then 
\begin{align*}
D_n({\beta})
=&
\iint_{a_0}^{\tau} H_{n}\{ t; \phi( {\tilde x}^T {\beta}), x\} 
[ \dot g\{ \phi({\tilde x}^T {\beta})\}  \dot \phi( {\tilde x}^T {\beta})]^2 
\hat S_C(a_0) \{ \hat S_C(t)\}^{-1} 
{\tilde x} {\tilde x}^T \hat f_X(x)            dt d{x}
\\
&+
\int G_n\{\phi({\tilde x}^T {\beta}), x \} 
\left( \ddot g\{ \phi({\tilde x}^T {\beta})\} [ \dot g\{ \phi({\tilde x}^T {\beta})\} ]^{-1} 
\{\dot  \phi( {\tilde x}^T {\beta}) \}^2 + \ddot \phi( {\tilde x}^T {\beta}) 
\right)
{\tilde x} {\tilde x}^T  d{x} 
\,,
\end{align*}
where
\begin{align*}
H_n(t;a, x)
=&
\frac{\ddot S_n(t;a, x)}{S_n(a_0;a, x)}-
\frac{2 \dot S_n(t;a, x) \dot S_n(a_0;a, x)}{\{S_n(a_0;a, x)\}^2}
+\frac{2 S_n(t;a,x)\{\dot S_n(a_0;a,x) \}^2}{ \{S_n(a_0;a, x)\}^3}
\\
&- \frac{S_n(t;a,x) \ddot S_n(a_0;a,x)}{ \{S_n(a_0;a,x)\}^2} \,.
\end{align*}
In the proof of Theorem 1, 
we have shown the uniform convergence of 
$S_n(t; a, {x})$ and $\hat S_C(t)$.
By similar arguments, we can also obtain that
\begin{align*}
& \sup_{t\in [a_0,\tau], a\in[0,a_0], x\in \mathcal{X}} 
\left| \dot S_n(t; a, {x})+ \frac{\partial}{\partial a} \Big\{\frac{\mu(t; a, {x})}{\dot g(a)} \Big\}
\right|  \rightarrow0 \,,
\\
& \sup_{t\in [a_0,\tau], a\in[0,a_0], x\in \mathcal{X}} 
\left|  \ddot S_n(t; a, {x})  -  \frac{\partial}{\partial a}\Big[ \frac{\partial}{\partial a} \big\{\frac{\mu(t; a, {x})}{\dot g(a)} \big\} \frac{1}{ \dot g(a)} \Big]  \right|  \rightarrow 0 
\end{align*}
almost surely.
Here, the limit of $\ddot S_n(t; a, {x})$ also equals to $B_2(t; a, {x})$  given in the proof of Lemma $1$.
Then, under Conditions C1--C6,
it can be calculated that
$D_n({\beta})$ converges uniformly to a 
non-zero bounded matrix 
\begin{align*}
D({\beta}) 
=& 
\iint_{a_0}^{\tau} 
\Bigg[ 
\frac{\ddot \mu_a \{ t; \phi( {\tilde x}^T{\beta}), x\} }{ \mu\{ a_0; \phi( {\tilde x}^T{\beta}), x\} }
- 2 \frac{ \dot \mu_a \{ t; \phi( {\tilde x}^T{\beta}), x\} \dot \mu_a\{ a_0; \phi( {\tilde x}^T{\beta}), x\}}{
	\mu\{a_0 ; \phi( {\tilde x}^T{\beta}), x\}^2  }
\\
& 
+ 2 \frac{  \mu\{t ; \phi( {\tilde x}^T{\beta}), x\} \dot \mu_a^2 \{ a_0; \phi( {\tilde x}^T{\beta}), x\} }{
	\mu\{a_0 ; \phi( {\tilde x}^T{\beta}), x\}^3}
- \frac{ \mu\{t ; \phi( {\tilde x}^T{\beta}), x\} \ddot \mu_a \{ a_0; \phi( {\tilde x}^T{\beta}), x\} }{
	\mu\{ a_0; \phi( {\tilde x}^T{\beta}), x\}^2 }
\Bigg]
\\
& \times
\frac{ S_C(a_0)}{  S_C(t)} 
f_X(x)   \{ \dot \phi( {\tilde x}^T{\beta})\}^2  {\tilde x} {\tilde x}^T dt d{x} 
+
\int  G\{ t; \phi( {\tilde x}^T{\beta}), x\}  \ddot \phi( {\tilde x}^T{\beta})  {\tilde x} {\tilde x}^T d{x}
\,,
\end{align*}
where 
$\ddot \mu_a ( t; a, x)$ and $ \ddot \phi(a) $ denote the second derivative of 
$\mu(t; a, {x})$ and $\phi(a)$ with respect to $a$.
Moreover, 
since $-D({\beta})^{-1}$ exists for all ${\beta} \in N_{\delta}$ under Condition 6, 
$D_n({\beta})^{-1}$ converges to $D({\beta})^{-1}$
uniformly for ${\beta} \in N_{\delta}$.
Then Theorem 2 follows directly from equation $(\ref{TaylorEx})$ and $(\ref{Eq_U})$.

\subsection*{Proof of Theorem 3}

Following Lemma $1$,$2$ and $4$, 
it can be proved that $(n {h_1} h_2)^{1/2} \{ M_n( {\beta}^{opt}) - M( {\beta}^{opt}) \}=O_p(1)$.
Then by  condition 5,
$U_n( {\hat \beta}^{opt})=0$
and uniform consistency of $D_n({\beta})$, 
we have
\begin{align*}
&n {h_1} h_2^3 \{ M_n({\hat \beta}^{opt}) - M( {\beta}^{opt}) \}
\\
=&
n {h_1} h_2^3 \{ M_n({\hat \beta}^{opt}) - M_n( {\beta}^{opt}) \}
+
(n {h_1} h_2^5)^{1/2}  (n {h_1} h_2)^{1/2}  \{ M_n( {\beta}^{opt}) - M( {\beta}^{opt}) \}
\\
=&
n {h_1} h_2^3 \{ M_n({\hat \beta}^{opt}) - M_n( {\beta}^{opt}) \}
+o_P(1)
\\
=&
-
\frac{1}{2} 
n {h_1} h_2^3 ( {\hat \beta}^{opt} -  {\beta}^{opt})^T 
D_n({\hat \beta}^{opt})( {\hat \beta}^{opt} -  {\beta}^{opt})
+ o_P(1)
\\
=&
\frac{1}{2} 
\{ ( n {h_1} h_2^3 )^{1/2}( {\hat \beta}^{opt} -  {\beta}^{opt})^T \}
(- D)
\{ ( n {h_1} h_2^3 )^{1/2}( {\hat \beta}^{opt} -  {\beta}^{opt}) \}
+ o_P(1)
\,,
\end{align*}
where $D=D({\beta}^{opt})$. 
As shown in Theorem 2,
$  ( n {h_1} h_2^3 )^{1/2}( {\hat \beta}^{opt} -  {\beta}^{opt}) $
converges in distribution to a mean zero normal variable with variance 
$D^{-1} \Sigma_U D^{-1}$,
thus, 
$n {h_1} h_2^3 \{ M_n({\hat \beta}^{opt}) - M({\hat \beta}^{opt}) \}$ 
converges in distribution to a random variable $Y$ 
with characteristic function 
$E[\exp(itY)] = \det ( I +  i t \Sigma_U D^{-1}  )^{-\frac{1}{2}} $, 
where $I$ is an identity matrix, 
$\det(\Sigma)$ is the determinant of $\Sigma$. 

Moreover, since 
both $\Sigma_U$ and $D$ are symmetric matrices
and $-D$ is positive definite,
by Lemma 1 in \cite{baldessari1967distribution},
$- \Sigma_U D^{-1}$ has a spectral decomposition
$
-\Sigma_U D^{-1} = \sum_{i=1}^s a_i E_i
$,
where
$a_j, j= 1,\ldots, s$ are the distinct characteristic roots of $-\Sigma_U D^{-1}$ 
and
$E_i$ are non-negative definitive matrices satisfying $E_i E_j=0, i \neq j$,
$E_i^2= E_i, i = 1,\ldots, s$.
Let $r_i$ denote the rank of $E_i$.
Since the distinct characteristic roots of $ I +  i t \Sigma_U D^{-1}$ are
$1- ita_i $ for $1\leq i \leq s$, 
we have 
$ E[\exp(itY)] = \det ( I +  i t \Sigma_U D^{-1}  )^{-\frac{1}{2}} 
= 
\prod_{i=1}^s \{ 1- 2 i ( a_i t / 2 )\}^{-r_i/2}$.
Thus, the distribution of $Y$ is the sum of $s$ independent 
chi-square distributions with degree of freedom $r_i$,
i.e.
$Y \sim  \sum_{i=1}^s a_i \chi^2(r_i)/2$.

\section*{B2.Proofs of Lemmas}

\subsection*{Proof of Lemma 1}

Lemma  $1$ consists of 
the asymptotic normality of $S_n(t; a, x)$, $\dot S_n(t; a, x)$,
$\Psi_n( a, x)$ and $\dot \Psi_n( a, x)$.

{\bf Asymptotic normality of $S_n(t; a, x)$.} 
For any fixed $(t, a, x)$ and under conditions 1--6, 
we can  obtain from Taylor expansion that
\begin{align*}
E\{ S_n(t; a, x) \}  
=&
E\Big[ I(\tilde T_i \geq t) K(\frac{X_i -x}{h_1}) K\{ \frac{g(A_i^*) - g(a)}{h_2}\} \dot g(a) \frac{1}{h_1 h_2} \Big]
 \\
 =&
E\Big[ pr(\tilde T_i \geq t \mid A_i^*,X_i ) K(\frac{X_i-x}{h_1}) K\{ \frac{g(A_i^*) - g(a)}{h_2}\} \dot g(a) \frac{1}{h_1 h_2} \Big]
 \\
 =&
E\Big[ S_T(t;A_i^*,X_i ) S_C(t) K(\frac{X_i-x}{h_1}) K\{ \frac{g(A_i^*) - g(a)}{h_2}\} \dot g(a) \frac{1}{h_1 h_2} \Big]
 \\
=&
E\Big[ \frac{\mu(t; A_i^*,X_i)}{f_{(A^*, X)}(A_i^*,X_i) } K(\frac{X_i-x}{h_1}) K\{ \frac{g(A_i^*) - g(a)}{h_2}\} \dot g(a) \frac{1}{h_1 h_2} \Big]
\\
=&
\int_{\mathcal{X}} \int_0^{a_0}
\mu(t; a', x')  K(\frac{x' -x}{h_1}) K\{ \frac{g(a') - g(a)}{h_2}\} \dot g(a) \frac{1}{h_1 h_2} d a' dx' \\
=&
\int_{-\infty}^{\infty} \int_{-\infty}^{\infty}
\frac{\mu[ t; g^{-1}\{ g(a)+v h_2\}, x+uh_1 ] }{\dot g[  g^{-1}\{ g(a)+v h_2 \} ] } \dot g(a) K(u) K(v) dv du  \\
=&
\mu(t; a, x) + \frac{1}{2} \{ B_1(t; a, x) h_1^2 + B_2(t; a, x) h_2^2 \} \kappa_{2,1} + O(h_1^4 + h_2^4) \,,
\end{align*}
where $B_1(t; a, x) =  \partial^2 \mu(t; a, x)/ ( \partial x^2)$,
\begin{align*}
B_2(t; a, x) 
=& 
\frac{ \partial^2 \mu(t; a, x)}{\partial a^2}  \frac{1}{ \{ \dot g(a) \}^2 }
- 3 \frac{\partial \mu(t; a, x)}{\partial a} \frac{ \ddot g(a)}{ \{ \dot g(a) \}^3}
+  \mu(t; a, x) \left[
\frac{ 3 \{\ddot g(a)\}^2 }{ \{ \dot g(a)\}^4} - \frac{ g^{(3)}(a) }{ \{ \dot g(a) \}^3} \right] \,,
\end{align*}
$\ddot g(a)$ is the second order derivative of $g(a)$ 
and $g^{(k)}(a)$ is the $k$th order derivative of $g(a)$ for $k = 3$ and $4$.

Similarly, we have
\begin{align*}
\text{var}\{ S_n(t; a, x) \}
=&
\frac{1}{n} \text{var} \left[ I(\tilde T_i \geq t)K(\frac{X_i-x}{h_1}) 
K\{\frac{g(A_i^*)- g(a)}{h_2}\} \dot g(a)\frac{1}{h_1 h_2}  \right] \\
=&
\frac{ \dot g(a)^2}{n h_1^2 h_2^2 } E\left[ I(\tilde T_i \geq t)  K^2( \frac{ X_i - x }{h_1}) K^2 \{ \frac{ g(A_i^*)- g(a)}{h_2}\} \right] 
 - \frac{1}{n} [E\{ S_n(t; a, x) \}]^2
 \\
=&
\frac{ \dot g(a)^2 }{n h_1 h_2 } 
\int_{-\infty}^{\infty} \int_{-\infty}^{\infty}
 \frac{\mu[ t; g^{-1}\{ g(a)+v h_2\}, x+uh_1 ]}{\dot g[  g^{-1}\{ g(a)+v h_2 \} ] } K(u)^2   K(v)^2 dv du + O( \frac{1}{n})
\\
=&
\frac{ 1 }{n h_1 h_2 } \dot g(a) \mu(t; a, x) \kappa_{0,2}^2 + o(\frac{1}{n h_1 h_2 }) \,.
\end{align*}

By the central limit theorem, when 
$n h_1 h_2 \rightarrow \infty$, 
$n h_1 h_2 (h_1^4 + h_2^4)^2  \rightarrow 0$ 
and for any fixed $(t, a, x)$, we have 
$ 
(n h_1 h_2)^{1/2}\left[ S_n(t; a, x) - \mu(t; a, x) - 
                              \{ B_1(t; a, x) h_1^2 + B_2(t; a, x) h_2^2 \} \kappa_{2,1} /2 
                            \right] 
$
converges in distribution to a normal variable with mean 0 and variance $\dot g(a) \mu(t; a, x) \kappa_{0,2}^2$.
Then the asymptotic property of $S_n(t; a, x)$ given in Lemma  $1$ holds
 if $n h_1 h_2 (h_1^2 + h_2^2)^2 \rightarrow 0$. 
 
{\bf Asymptotic normality of $ \dot S_n(t; a, x)$}.
Since $\int \dot K(u) du =0$, $\int u \dot K(u) du=-1$ and $\int u^3 \dot K(u) du=-3 \kappa_{2,1}$, we have
\begin{align*}
E\{ \dot S_n(t; a, x) \}  
=&
E\left[ I(\tilde T_i \geq t) K(\frac{X_i-x}{h_1}) \dot K\{ \frac{g(A_i^*) - g(a)}{h_2}\} 
          \frac{ \dot g(a) }{h_1 h_2^2 } \right] \\
=&
E\left[ \frac{\mu(t; A_i^*,X_i)}{f_{(A^*,X)}(A_i^*,X_i) } K(\frac{X_i-x}{h_1}) \dot K\{ \frac{g(A_i^*) - g(a)}{h_2}\} 
          \frac{\dot g(a) }{h_1 h_2^2} \right] \\
=&
\int_x \int_0^{a_0}
\mu(t; a', x')  K(\frac{x' -x}{h_1}) \dot K\{ \frac{g(a') - g(a)}{h_2}\} 
         \frac{ \dot g(a) }{h_1 h_2^2 } d a' dx' \\
=&
\frac{\dot g(a)}{h_2}
\iint \frac{\mu[ t; g^{-1}\{ g(a)+v h_2\}, x+uh_1 ] }{\dot g[  g^{-1}\{ g(a)+v h_2 \} ] }  K(u) \dot K(v)  dv du  \\
=&
- \frac{\partial}{ \partial a} \left\{ \frac{\mu(t; a, x)}{\dot g(a)} \right\}
- \frac{1}{2} \left\{ B_3(t; a, x) h_1^2 + B_4(t; a, x) h_2^2 \right\}  \kappa_{2,1} 
+ O( \frac{ h_1^5}{h_2} + h_2^4) \,,
\end{align*}
where 
\begin{align*}
B_3(t;a,x)
=&
\frac{ \partial^3 \mu(t; a, x)}{\partial a \partial x^2} \frac{1}{  \dot g(a) } 
-
\frac{ \partial^2 \mu(t; a, x)}{\partial x^2} \frac{ g^{(2)}(a)}{ \{ \dot g(a) \}^2} \,,
\\
B_4(t; a, x) 
=&
\frac{ \partial^3 \mu(t; a, x)}{\partial a^3} \frac{1}{ \{ \dot g(a) \}^3} 
-6 \frac{ \partial^2 \mu(t; a, x)}{\partial a^2} \frac{g^{(2)}(a)}{ \{ \dot g(a) \}^4} 
+ \frac{ \partial \mu(t; a, x)}{\partial a}
\left[ 
15  \frac{ \{ g^{(2)}(a)\}^2 }{ \{ \dot g(a) \}^5}
- 4  \frac{g^{(3)}(a)}{ \{ \dot g(a) \}^4}
\right]
\\
& 
+ \mu(t; a, x) 
\left[ 
- \frac{  g^{(4)}(a)  }{ \{ \dot g(a)\}^4}  
+ \frac{  10 g^{(3)}(a) g^{(2)}(a)  }{ \{ \dot g(a)\}^5}
- \frac{ 15   \{ g^{(2)}(a) \}^3}{ \{ \dot g(a) \}^6}
\right]  \,.
\end{align*}

On the other hand,
\begin{align*}
\text{var}\{ \dot S_n(t; a, x) \}
=&
\frac{1}{n} \text{var} \left [ I(\tilde T_i \geq t) K(\frac{X_i-x}{h_1}) \dot K\{ \frac{g(A_i^*)- g(a)}{h_2}\} \frac{\dot g(a)}{h_1 h_2^2}  \right] \\
=&
\frac{\dot g(a)^2}{n h_1^2 h_2^4 } E\left[ I(\tilde T_i \geq t)  K^2( \frac{ X_i - x }{h_1})  \dot K^2\{ \frac{ g(A_i^*)- g(a)}{h_2}\}  \right] 
 - \frac{1}{n} [E\{ \dot S_n(t; a, x) \}]^2
 \\
=&
\frac{\dot g(a)^2}{n h_1 h_2^3 } 
\iint
 \frac{\mu[ t; g^{-1}\{ g(a)+v h_2\}, x+uh_1 ]}{\dot g[  g^{-1}\{ g(a)+v h_2 \} ] }  K(u)^2 \dot K(v)^2 dv du + O( \frac{1}{n})
\\
=&
\frac{1}{n h_1 h_2^3 } \dot g(a) \mu(t; a, x) \kappa_{0,2} \dot \kappa_{0,2} + o(\frac{1}{n h_1 h_2^3 }) \,.
\end{align*}
By the central limit theorem, 
when $n h_1 h_2^3 \rightarrow \infty$, 
$ (n h_1 h_2)^{1/2} (h_1^5+h_2^5) \rightarrow 0 $ and for any fixed $(t, a, x)$, we have 
$$
 (n h_1 h_2^3)^{1/2}
 \left(  \dot S_n(t; a, x)+ 
          \partial[ \{ \dot g(a) \}^{-1} \mu_a(t; a, x) ]/ \partial a 
          +  \left\{ B_3(t; a, x) h_1^2 + B_4(t; a, x) h_2^2 \right\}  \kappa_{2,1} /2 
 \right) 
$$
converges in distribution to a normal variable with mean zero and variance 
$\dot g(a) \mu(t; a, x)   \kappa_{0,2}\dot\kappa_{0,2}$.

Then the asymptotic property of $\dot S_n(t; a, x)$ given in Lemma  $1$ holds
 if  $n h_1 h_2^3 (h_1^2 + h_2^2)^2 \rightarrow 0$. \\

{\bf Asymptotic normality of $\Psi_n(a, x)$ and $\dot \Psi_n(a, x)$}.

Recall that 
$
\Psi_n(a, x) = \int_{a_0}^{\tau} S_n( t; a, x) \{ S_C(t) \}^{-1} dt 
$
and
$
\dot \Psi_n(a, x) = \int_{a_0}^{\tau} \dot S_n( t; a, x) \{ S_C(t)\}^{-1} dt 
$.
Following the large sample properties of $S_n$ and $\dot S_n$, 
we have
\begin{align*}
       E\{ \Psi_n(a, x) \}  
=&
       \int_{a_0}^{\tau}  E\{ S_n( t; a, x) \} \{ S_C(t)\}^{-1} dt 
\\
=&
       \int_{a_0}^{\tau}  \mu(t; a, x)\{ S_C(t)\}^{-1} dt + 
        \frac{1}{2} \{  B_5(a, x) h_1^2 + B_6(a, x) h_2^2 \} \kappa_{2,1} + 
        O(h_1^4 + h_2^4)
\\
=&
       \Psi(a, x)+ 
       \frac{1}{2} \{  B_5(a, x) h_1^2 + B_6(a, x) h_2^2 \} \kappa_{2,1} + 
        O(h_1^4 + h_2^4)  \,,
\end{align*}
\begin{align*}
       E\{ \dot \Psi_n(a, x) \}  
=& 
        \int_{a_0}^{\tau}  E\{ \dot S_n( t; a, x) \} \{ S_C(t)\}^{-1} dt 
\\
=&
- \frac{\partial}{\partial a} \{\frac{\Psi( a, x)}{\dot g(a)} \}
- \frac{1}{2} \{ B_7(a, x) h_1^2 + B_8(a, x) h_2^2 \} \kappa_{2,1} + O(\frac{h_1^5}{h_2} + h_2^4) \,,
\\
\text{var} \{ \Psi_n(a, x)\} 
=&
\frac{1}{n} \text{var}\left[ \int_{a_0}^{\tau} \frac{ I(\tilde T_i \geq t)}{S_C(t)} dt 
               K(\frac{X_i-x}{h_1}) \dot K\{ \frac{g(A_i^*)- g(a)}{h_2}\} \frac{\dot g(a)}{h_1 h_2} \right]
\\
=&
\frac{ 1}{n } 
E\left[ 
\big \{ \int_{a_0}^{\tau} \frac{ I(\tilde T_i \geq t)}{S_C(t)} dt \big\}^2  
K^2( \frac{ X_i - x }{h_1}) K^2\{ \frac{ g(A_i^*)- g(a)}{h_2}\}  \frac{ \dot g(a)^2}{h_1^2 h_2^2 }
 \right] 
 - \frac{1}{n} E\{ \Psi_n( a, x) \}^2
 \\
 =&
\frac{ \dot g(a)^2}{n h_1^2 h_2^2 } E\left[ \frac{\omega(A_i^*,X_i)}{f_{(A^*,X)}(A_i^*,X_i)} K^2( \frac{ X_i - x }{h_1}) K^2\{ \frac{ g(A_i^*)- g(a)}{h_2}\}  \right] 
 - \frac{1}{n} E\{ \Psi_n( a, x) \}^2
 \\
=&
\frac{ \dot g(a)^2 }{n h_1 h_2 } 
\iint \frac{\omega[g^{-1}\{ g(a)+v h_2\}, x+uh_1 ]}{\dot g[  g^{-1}\{ g(a)+v h_2 \} ] } K(u)^2   K(v)^2 dv du + O( \frac{1}{n})
\\
=&
\frac{ 1 }{n h_1 h_2 } \dot g(a) \omega(a, x) \kappa_{0,2}^2 + o(\frac{1}{n h_1 h_2 }) \,,
\\
\text{var}\{ \dot \Psi_n(a, x) \}
=&
\frac{\dot g(a)^2}{n h_1^2 h_2^4 } 
E\left( \frac{ \omega(A_i^*,X_i)}{f_{(A^*,X)}(A_i^*,X_i)} 
\Big[ K( \frac{ X_i - x }{h_1}) \dot K\{ \frac{ g(A_i^*)- g(a)}{h_2}\} \Big]^2
\right)
+O( \frac{1}{n})
\\
=&
\frac{ \dot g(a)^2 }{n h_1 h_2^3 } 
\iint \frac{\omega[g^{-1}\{ g(a)+v h_2\}, x+uh_1 ]}{\dot g[  g^{-1}\{ g(a)+v h_2 \} ] } K(u)^2   \dot K(v)^2 dv du + O( \frac{1}{n})
\\
=&
\frac{1}{n h_1 h_2^3 } \dot g(a) \omega(a, x) \kappa_{0,2} \dot \kappa_{0,2} + o(\frac{1}{n h_1 h_2^3 }) \,,
 \end{align*}
where 
$B_5(a, x)= \int_{a_0}^{\tau} B_1(t; a, x) \{S_C(t)\}^{-1} dt$, 
$B_6(a, x)= \int_{a_0}^{\tau} B_2(t; a, x) \{S_C(t)\}^{-1} dt$, 
$B_7(a, x)= \int_{a_0}^{\tau} B_3(t; a, x) \{S_C(t)\}^{-1} dt$ and 
$B_8(a, x)= \int_{a_0}^{\tau} B_4(t; a, x) \{S_C(t)\}^{-1} dt$.
Similar to the proof before,
the asymptotic normality of $\Psi_n(a, x)$ and $\dot \Psi_n(a, x)$
follows directly from the central limit theorem. 

\subsection*{Proof of Lemma 2}

Note that 
\begin{align*}
c_n \left( \frac{x_n}{y_n} - \frac{x_0}{y_0} \right)
=&
  \frac{c_n (x_n-x_0)}{y_0} 
- \frac{d_n(y_n-y_0)}{y_0} \frac{x_0}{y_0}  \frac{c_n}{d_n} 
- c_n( \frac{x_n}{y_n} - \frac{x_0}{y_0}) \frac{(y_n-y_0)}{y_0} 
 \,, 
\\
c_n \left( x_n y_n -x_0y_0 \right) 
=&
c_n (x_n-x_0)y_0 +  d_n(y_n-y_0) x_0 \frac{c_n}{d_n}
   +   (y_n-y_0) c_n(x_n- x_0) \,.
\end{align*}
Lemma  $2$ follows directly from the asymptotic properties of $x_n$ and $y_n$. 

\subsection*{Proof of Lemma 3}

This lemma can be obtained by considering the form 
$\sum_{j=1}^N G(x_j) a(x_j)$ 
in jointly Gaussian random variables $G(x_j)$ with  
$a(x_j)$'s constants.
It's not hard to see that the characteristic function of
$\sum_{j=1}^N G(x_j) a(x_j)$ 
takes the form
$ \exp\{ i t \sum_{i=1}^N \sum_{j=1}^N a(x_i) r(x_i, x_j) a(x_j)^T \}$.
Then by appropriate limiting operation,
it can be obtained that the characteristic function of $Z$
is $ \exp\{ i t \iint a(x) r(x, y) a(y)^T \rm{d}x \rm{d} y \}$,
and thus Lemma  $3$ holds. 

\subsection*{Proof of Lemma 4}

Now we derive the asymptotic properties of $G_n(a, x)$. 
Rewrite
{\small
\begin{align*}
G_n(a, x)
=&
\int_{a_0}^{\tau} 
\left\{
 - \dot S_n(t; a, x) S_n(a_0; a, x) + S_n(t; a, x) \dot S_n(a_0; a, x)
\right\}
\frac{ S_C(a_0)}{S_C(t)}  dt
\frac{ \hat f_X(x) \dot g(a)}{S_n(a_0; a, x)^2}
+ R_n(a, x)
\\
=&
\left\{
     - \dot \Psi_n(a, x) S_n( a_0;a, x)
    + \Psi_n(a, x) \dot S_n( a_0; a, x)
\right\}
\frac{\hat f_X(x) \dot g(a) S_C(a_0) }{S_n( a_0;a, x)^2}
+ R_n(a, x) \,,
\end{align*}
}
where
{\small
\begin{align*}
R_n(a, x)
=& 
\int_{a_0}^{\tau} 
\left\{
 - \dot S_n(t; a, x) S_n(a_0; a, x) + S_n(t; a, x) \dot S_n(a_0; a, x)
\right\}
\left\{    \frac{\hat S_C(a_0)}{ \hat S_C(t)}-\frac{ S_C(a_0)}{ S_C(t)} 
           \right\}  dt 
\frac{ \hat f_X(x) \dot g(a)  }{S_n(a_0; a, x)^2} 
      \,. 
 \end{align*} }   
By the uniform convergence of $\hat S_C(t)$ and Lemmas  $1$ and  $2$, 
it can be calculated that $(n h_1 h_2^3)^{1/2}R_n(a, x)=o_p(1) $ 
and 
\begin{align*}
(n h_1 h_2^3)^{1/2} \{ G_n(a, x) - G(a, x) \}
 =&
 (h_1 h_2^3 / n )^{1/2} \sum_{i=1}^n 
 \left \{   \psi_{i}(a, x) - G(a, x)  \right \} +o_P(1) 
\end{align*}
with 
\begin{align*}
\psi_{i}(a, x) 
 =&
 \left\{
      - \int_{a_0}^{\tau} \frac{ I(\tilde T_i \geq t)}{S_C(t)} dt \mu( a_0; a, x) 
      + I(\tilde T_i \geq a_0)  \Psi(a, x)
 \right\} K \left (\frac{X_i-x}{h_1} \right) \dot K\left\{ \frac{g(A_i^*)-g(a) }{h_2} \right\}
 \\
 &\times
 \frac{ 1}{h_1 h_2^2} f_X(x)S_C(a_0)\big\{ \frac{  \dot g(a) }{  \mu (a_0; a, x) }\big\}^2
 \,.
 \end{align*}
Under conditions 1--6,
and with arguments similar to Lemma  $1$, 
it can be obtained that, 
$ (n h_1 h_2^3)^{1/2} \{ G_n(a, x) - G(a, x) \}$
converges in distribution to a normal variable with mean zero and variance
\begin{align*}
 \sigma^2_G(a, x)
=&
\Big\{
\frac{ \omega(a, x)}{ \mu(a_0; a, x)^2} - \frac{\Psi(a, x)^2  }{ \mu(a_0; a, x)^3} \Big\}
 \{f_X(x) S_C(a_0)\}^2 \{ \dot g(a)\}^3 \kappa_{0,2} \dot \kappa_{0,2} 
\end{align*}
for any fixed $(a, x)$.
Furthermore,
by the tightness of $G_n\{  \phi(\tilde x^T \beta),x \}$ and $G\{  \phi(\tilde x^T \beta),x \} $, 
the stochastic processes $(n h_1 h_2^3)^{1/2} [ G_n\{ \phi(\tilde x^T \beta^{opt}),x\}  - G\{ \phi(\tilde x^T \beta^{opt}),x\} ]$  
converges weakly to a mean zero Gaussian process with mean zero and covariance function
$$
r(x, y)= \lim_{n \rightarrow \infty}\text{cov}\big[ (n h_1 h_2^3)^{1/2} G_n\{ \phi(\tilde x^T \beta^{opt}), x\}, (n h_1 h_2^3)^{1/2} G_n\{\phi(\tilde y^T \beta^{opt}), y\} \big]\,.
$$

Lastly, we only need to show that $r(x, y)= 0$ for $x\neq y$. 
Given $x_1, x_2 \in \mathcal X$, 
\begin{align*}
&
\text{cov}\Big[ (n h_1 h_2^3)^{1/2} G_n\{ \phi(\tilde x_1^T \beta^{opt}), x_1\}, (n h_1 h_2^3)^{1/2} G_n\{\phi(\tilde x_2^T \beta^{opt}), x_2\} \Big]
\\
=&
n^{-1} h_1 h_2^3  \sum_{i=1}^n \sum_{j =1}^n
\text{cov} \Big[  \psi_i\{ \phi(\tilde x_1^T \beta^{opt}), x_1  \},  \psi_j\{ \phi(\tilde x_2^T \beta^{opt}), x_2  \} \Big] + o(1)
\\
=&
h_1 h_2^3 \text{cov} \Big[  \psi_i\{ \phi(\tilde x_1^T \beta^{opt}), x_1  \},  \psi_i\{ \phi(\tilde x_2^T \beta^{opt}), x_2  \} \Big] + o(1)
\\
=&
C_0(x_1,x_2) \int K(u) K(u+ \frac{x_1 - x_2}{h_1}) du 
\int \dot K(v) \dot K\{ v+\frac{\phi(\tilde x_1^T \beta^{opt}) -\phi(\tilde x_2^T \beta^{opt})}{h_2}\} dv
+ O(h_1 + h_2)  \,,
 \end{align*}
where
\begin{align*}
C_0(x_1,x_2)
=&
\Big[
\omega\{ \phi(\tilde x_1^T \beta^{opt}), x_1\} \mu\{ a_0; \phi(\tilde x_1^T \beta^{opt}), x_1\} \mu\{a_0; \phi(\tilde x_2^T \beta^{opt}), x_2\}
\\
&- \Psi^2\{ \phi(\tilde x_1^T \beta^{opt}), x_1\} \mu\{a_0; \phi(\tilde x_2^T \beta^{opt}), x_2\}
\Big]
\\
&\times 
\dot g\{ \phi(\tilde x_2^T \beta^{opt})\} \psi_0\{ \phi(\tilde x_1^T \beta^{opt}), x_1\} \psi_0\{ \phi(\tilde x_2^T \beta^{opt}), x_2\}
\end{align*}
is a bounded constant 
with $\psi_0(a, x)=\{ \mu (a_0; a, x)\}^{-2} f_X(x) \dot g(a) S_C(a_0)$. 
When $x_1=x_2$, 
$r(x, x)$ 
is the asymptotic variance of $ (n h_1 h_2^3)^{1/2} G_n\{ \phi(\tilde x^T \beta^{opt}), x\}$
which equals to $\sigma^2_G\{ \phi(\tilde x^T \beta^{opt}),x\}$.
When $x_1 \neq x_2$,
both  $ \int K(u) K\{ u+  (x_1 - x_2)/h_1\} du $
and
$ \int \dot K(v) \dot K[ v+ \{ \phi(\tilde x_1^T \beta^{opt}) -\phi(\tilde x_2^T \beta^{opt})\}/ h_2 ] dv$
will converge to zero as $h_1, h_2 \rightarrow 0 $. 
Thus $r(x_1,x_2)$ equals to zero.

\end{document}